\documentclass[usenatbib]{mn2e}

\usepackage{lscape}
\usepackage{array}
\usepackage{dcolumn}
\usepackage{graphics,epsfig}
\usepackage{txfonts}
\usepackage{times}
\usepackage{amssymb}
\usepackage{ulem}

\def\hi{H{\sc i}}
\def\wfetwo{W_0^{\lambda 2600}}

\def\wmgtwo{W_0^{\lambda 2796}}

\def\wmgone{W_0^{\lambda 2852}}

\def\mgtwo{Mg{\sc ii}}
\def\mgone{Mg{\sc i}}
\def\fetwo{Fe{\sc ii}}

\def\nhi{\rm N_{HI}}

\newcommand{\kms}{km~s$^{-1}$}
\newcommand{\cm}{cm$^{-2}$}
 
\newcommand{\dlas}{DLAs}
\newcommand{\NHI}{N_{\rm HI}}

\newcommand{\ts}{{\rm T_s}}
\newcommand{\tk}{{\rm T_k}}
\newcommand{\beq}{\begin{equation}}
\newcommand{\eeq}{\end{equation}}

\title[\hi~21cm absorption in the redshift desert]{A search for \hi~21cm absorption 
in strong Mg{\sc ii} absorbers in the redshift desert}
\author[Kanekar et al.]{N.~Kanekar$^1$\thanks{E-mail: nkanekar@aoc.nrao.edu (NK); 
xavier@ucolick.org (JXP); sarae@uvic.ca (SLE); chengalu@ncra.tifr.res.in (JNC)},
J. X. Prochaska$^2$, S. L. Ellison$^3$, J.~N.~Chengalur$^4$\\
$^1${}National Radio Astronomy Observatory, 1003 Lopezville Rd, Socorro, NM 87801, USA; \\
$^2${}UCO/Lick Observatory, UC Santa Cruz, Santa Cruz, CA 95064, USA; \\
$^3${}Department of Physics and Astronomy, University of Victoria, Victoria, B.C., V8P 1A1, Canada\\
$^4${}National Centre for Radio Astrophysics, Ganeshkhind, Pune 411007, India}

\begin{document}
\date{Received mmddyy/ accepted mmddyy}
\maketitle
\label{firstpage}

\begin{abstract}

We report results from a deep search for redshifted \hi~21cm absorption in 55~strong 
\mgtwo$\lambda$2796 absorbers (having $\wmgtwo > 0.5 \AA$) at intermediate redshifts, 
$0.58 < z_{\rm abs} < 1.70$, with the Green Bank Telescope (GBT) and the Giant 
Metrewave Radio Telescope (GMRT). Nine detections of \hi~21cm absorption were 
obtained, all at $1.17 < z_{\rm abs} < 1.68$, including three systems
reported earlier by Gupta et al. (2007). Absorption was not detected at $> 3\sigma$ 
significance in 32~other \mgtwo\ absorbers, with 26~of these providing strong upper 
limits to the \hi~21cm optical depth, $\tau_{3\sigma} < 0.013$ per $\sim 10$~\kms. 
For the latter 26~systems, the spin temperature $\ts$ of the absorber must be 
$> [800 \times f]$~K (where $f$ is the covering factor), if the \hi\ column density 
is $\ge 2 \times 10^{20}$~cm$^{-2}$, i.e. if the absorber is a damped Lyman-$\alpha$ system
(DLA). Data on the remaining 13~systems of the sample were affected by radio frequency 
interference and were hence not useful. 


Two of the \mgtwo\ absorbers, at $z_{\rm abs} \sim 1.4106$ towards 2003$-$025 and at 
$z_{\rm abs} \sim 0.9115$ towards 2149+212, are known DLAs. We detect \hi~21cm absorption 
towards 2003$-$025 with the GMRT and estimate the spin temperature of the DLA to be 
$\ts = [(905 \pm 380) \times f]$~K. Conversely, the GBT observations of 2149+212 
resulted in a non-detection of \hi~21cm absorption, yielding the $3\sigma$ 
limit $\ts > [2700 \times f]$~K. 

Excluding ``associated'' systems (within $3000$~\kms\ of the quasar redshift), 
the detection rate of \hi~21cm absorption in 
strong \mgtwo$\lambda$2796 absorbers is $x_{\rm 21,MgII} ({\bar z} = 1.1) = 
25^{+11}_{-8}$\%, at a $3\sigma$ optical depth sensitivity of $\sim 0.013$ per 10~\kms.
Comparing the detection rates of \hi~21cm and damped Lyman-$\alpha$ absorption 
in strong \mgtwo\ absorber samples yields a detection rate of \hi~21cm absorption 
in DLAs of $x_{\rm 21,DLA} ({\bar z} = 1.1) = (73 \pm 27)$\%, consistent with 
the detection rate in low-$z$ DLAs. Since \hi~21cm absorption
arises in cold neutral gas, this indicates that most gas-rich galaxies contain 
significant fractions of cold \hi\ by $z \sim 1$. 


Finally, we use the observed detection rate of \hi~21cm absorption in 
\mgtwo\ absorbers to infer the cosmological mass density of neutral gas in 
DLAs, assuming that (1)~the average \hi\ column density in our \hi~21cm 
absorber sample is the same as that measured by Rao et al. (2006) in their 
DLA sample, and (2)~the detection rate of \hi~21cm absorption in the DLAs of 
our \mgtwo\ sample is the same as that in known DLAs at $0.09 < z_{\rm abs} < 3.45$. 
We obtain $\Omega_{\rm GAS} \sim (0.55^{+0.42}_{-0.22}) \times 10^{-3}$, at 
${\bar z} \sim 1.1$, slightly lower than, but consistent with, the value 
obtained by Rao et al. (2006) from their DLA survey at similar redshifts.


\end{abstract}


\section{Introduction}
\label{intro}


Much of what is known about gas-rich galaxies at high redshifts comes from
studies of the highest \hi\ column density systems detected in the spectra of 
distant quasars, the damped Lyman-$\alpha$ absorbers [DLAs, with $\nhi \ge 
2.0 \times 10^{20}$~\cm; see, e.g., \citet{wolfe05} for a recent review].
Unlike objects such as Lyman-break or sub-mm galaxies obtained from emission (i.e. 
flux-limited) surveys, DLA samples are selected from absorption surveys and are 
hence not biased towards the high luminosity end of the galaxy distribution. Studies 
of DLAs over a range of redshifts thus provide direct observational constraints 
on the evolution of the predecessors of today's galaxies, and the gas therein.

Current DLA samples are strongly biased towards high redshifts ($z_{\rm abs} > 1.7$, 
for which the Lyman-$\alpha$ line redshifts into optical wavebands), as it has not 
been possible to carry out large blind DLA surveys with space-based facilities. There are
only $\sim 50$~known DLAs at $z_{\rm abs} < 1.7$ (e.g. \citealt{rao06}; 
hereafter RTN06), while more than a thousand have been found at $z_{\rm abs} \gtrsim 2$, 
primarily from searches based on the Sloan Digital Sky Survey (SDSS; 
\citealt{prochaska04,prochaska05}
\footnote{http://www.ucolick.org/$\sim$xavier/SDSSDLA/index.html}).
Optical imaging studies have found low redshift DLAs to be associated with a mixture of 
galaxy types, including dwarfs, low surface brightness systems and luminous spiral disks 
(e.g. \citealp{lebrun97,rao03,chen03}).  However, at high redshifts, $z \gtrsim 1$, 
their small angular separation from the much brighter background quasars has meant 
that it is very difficult to directly image the absorbing galaxies.  As a result,
detailed studies of high-$z$ DLAs have mostly been limited to high-resolution absorption 
spectroscopy. While such studies have provided much information on physical 
conditions in the absorbers, they have also given rise to some conundrums, partly, no 
doubt, due to the mix of morphologies, impact parameters and orientations 
in the high-$z$ DLA population. For example, while DLA metallicities, 
measured in $\sim 150$~absorbers 
(including $\sim 25$ at $z_{\rm abs} < 1.7$; \citealt{akerman05,meiring06,prochaska07}),
do show evidence for an increase with decreasing redshift 
\citep{kulkarni02,prochaska03a,kulkarni05}, low metallicities (one-tenth solar 
or lower) are the norm at all redshifts (e.g. 
\citealt{pettini99,kulkarni05,prochaska07}). Contrary to expectations from most 
models of chemical evolution (e.g. \citealp{pei99}), the extrapolated mean DLA metallicity 
at $z = 0$ is well below solar values (although see \citealt{zwaan05}). Similarly, 
while it has been argued that typical abundance patterns in DLAs indicate star formation 
histories similar to those of dwarf galaxies (e.g.  \citealt{dessauges07}), it 
is difficult to explain the large observed velocity widths ($\gtrsim  100$~\kms) 
by absorption in individual dwarfs \citep{prochaska97}. These and other problems have 
meant that, despite more than two decades of systematic study, the nature of, and 
physical conditions in, high $z$ DLAs, as well as their redshift evolution, remain 
issues of much controversy. 


For DLAs lying towards radio-loud background quasars, absorption studies in the 
redshifted \hi~21cm line can be used to derive additional information on physical 
conditions in the absorber [see \citet{kanekar04} for a review]. Combining the \hi\ 
column density of the DLA (determined from the Lyman-$\alpha$ line) and the integrated 
\hi~21cm optical depth yields the column-density-weighted harmonic mean spin temperature 
$\ts$ of the neutral hydrogen, which gives the distribution of the gas between the warm 
and cold phases. The spin temperature is one of the few direct probes of physical 
conditions in the neutral gas in DLAs; most optical techniques infer conditions 
in the \hi\ from observations of the low-ionization metal lines, which are assumed 
to be associated with the neutral hydrogen (e.g. \citealt{wolfe05}). Further, most
optical resonance transitions (including Lyman-$\alpha$) are quite insensitive to 
physical conditions like density and temperature.

Low-$z$ DLAs show a mix of low and high spin temperatures (see Fig.~3 of \citealt{kanekar03}), 
while DLAs at $z_{\rm abs} \gtrsim 1.7$ typically have high spin temperatures, 
$\ts \gtrsim 1000$~K (e.g. \citealt{carilli96,chengalur00,kanekar03}). So far, 
only one DLA at $z_{\rm abs} \gtrsim 1.7$ has been found to have $\ts < 500$~K \citep{york07}. 
Further, the fraction of detections of \hi~21cm absorption is significantly higher at 
$z_{\rm abs} < 0.7$ ($\sim 85$\%) than at $z_{\rm abs} > 1.7$ [$\sim 33$\%; Kanekar et 
al. (\textit{in prep.})].
Note that searches for \hi~21cm absorption in DLAs are usually carried out to a sensitivity 
limit in spin temperature (usually $\gtrsim 800$~K; e.g. \citealt{kanekar03}), thus taking 
into consideration the \hi\ column density of the DLA, rather than to a fixed \hi~21cm 
optical depth sensitivity. 

The redshift dependence of the spin temperature suggests an 
evolution in the temperature distribution of the DLAs, due to either evolving physical 
conditions within the DLA galaxies or changes in the nature of the galaxies that 
typically give rise to DLAs at a given redshift. In the local Universe, large spiral 
galaxies (including the Milky Way) tend to have low spin temperatures ($\ts \lesssim 300$~K; 
e.g. \citealp{braun92}), indicating large cold gas fractions. On the other hand, warm 
gas has been found to dominate the temperature distribution in dwarfs (e.g. 
\citealp{young97}). A similar trend between spin temperature and galaxy type has been 
seen in low-$z$ DLAs, wherein absorbers associated with luminous disk galaxies have 
$\ts \lesssim 300$~K, while absorbers associated with low luminosity galaxies have 
$\ts \gtrsim 800$~K \citep{chengalur00}. These results are suggestive of a scenario 
in which high-$z$ DLAs are typically smaller systems like dwarfs, while low-$z$ DLAs 
arise in both low and high luminosity galaxies \citep{kanekar01a}.

The spin temperature of DLAs thus appears to be one of the few tracers of physical conditions 
in DLAs that shows evidence for evolution with redshift and has hence been a subject of much interest 
and controversy (e.g. \citealt{kanekar03,wolfe03b,curran05}). A detailed understanding 
of the evolution of the spin temperature in DLAs has been hampered by the paucity of 
DLAs detected towards radio-loud quasars that are suitable for \hi~21cm absorption studies. 
The situation has now improved at high redshifts, $z_{\rm abs} \gtrsim 2$, primarily due 
to recent surveys for DLAs towards samples of quasars selected on the basis of 
their radio flux density \citep{ellison01,jorgenson06,ellison08}. However, there is 
yet a critical shortage of $\ts$ measurements in the ``redshift desert'', 
$0.7 < z_{\rm abs} < 1.7$, as almost no DLAs have been found here towards 
radio-loud quasars. In fact, prior to this work, there was not a single 
spin temperature estimate in a ``classical'' DLA (i.e. with 
$\nhi \ge 2 \times 10^{20}$~\cm) in the redshift range $0.9 < z_{\rm abs} < 1.7$, 
which encompasses more than a quarter of the history of the Universe.  Unfortunately, 
it is precisely at these redshifts that the star formation activity in the 
Universe shows strong evolution, either peaking here or flattening, and then 
decreasing significantly to lower redshifts (e.g. \citealt{madau96,hopkins04}). 
This redshift range is also likely to see significant evolutionary effects in 
the neutral gas, such as the build-up of the cold phase of \hi\ and hence, the 
onset of stronger \hi~21cm absorption and the transition to the high detection 
fraction and the low spin temperatures seen at low redshifts.

Besides probing physical conditions in the absorbing galaxies, detections of \hi~21cm 
absorption are also important as a means of probing evolution in the fundamental
constants [see \citet{kanekar08b} for a recent review]. A comparison between the absorption 
redshifts in the \hi~21cm and optical resonance transitions allows a probe of changes in three 
fundamental constants, the fine structure constant $\alpha$, the proton-electron
mass ratio $\mu \equiv m_p/m_e$ and the proton g-factor $g_p$ \citep{wolfe76}. Given that the \hi~21cm 
and optical lines need not arise in the same gas, statistically large absorption 
samples must be used for the comparison, to ensure that any observed redshift offsets
are not dominated by local systematic effects. The redshift desert $ 0.7 < z_{\rm abs} < 1.7$ is the 
obvious place to attempt to set up a large sample of \hi~21cm absorbers, if the fraction of 
cool gas increases significantly here. Note that the lookback time at $z \sim 1.7$ is 
$\sim 9.8$~Gyrs, implying that the comparison between \hi~21cm and optical redshifts out 
to this redshift would probe $\sim 70$\% of the age of the Universe.


The weakness of the \hi~21cm transition, combined with observational issues such as radio frequency 
interference (RFI) at low frequencies, makes it very difficult to directly carry out a blind 
survey for \hi~21cm absorption with current radio telescopes (e.g. \citealt{kanekar04}). In 
addition, almost no DLAs are known at $0.7 < z_{\rm abs} < 1.7$ towards radio-loud quasars and 
it would be difficult to carry out space-based surveys for such DLAs. However, RTN06  found 
that systems with \mgtwo$\lambda$2796 and \fetwo$\lambda$2600 rest equivalent widths larger 
than $0.5 \AA$ have a $\sim (36 \pm 6)$~\% probability of being DLAs (see also \citealt{rao00}). 
These transitions are detectable with ground-based optical telescopes at $z_{\rm abs} \gtrsim 
0.15$, implying that it should be possible to set up new samples of \hi~21cm absorbers by 
targetting such ``strong \mgtwo\ absorbers'' towards radio-loud quasars. Large samples 
of such \mgtwo\ absorbers ($\sim 10^4$~systems) have been found in the SDSS, at $z_{\rm abs} 
\gtrsim 0.4$ (e.g.  \citealt{prochter06}), while the Complete Optical and Radio Absorption Line Systems 
(CORALS) survey \citep{ellison04} obtained a smaller sample of such absorbers, but towards a 
radio-loud quasar sample, well-suited for follow-up \hi~21cm absorption studies. We have used 
the Giant Metrewave Radio Telescope (GMRT) and the Green Bank Telescope (GBT) to carry out a deep 
search for redshifted \hi~21cm absorption from a large sample of strong \mgtwo\ absorbers in the 
redshift desert, selected from the above surveys. A similar program, carried out in parallel with ours 
on the GMRT, has recently detected three new \hi~21cm absorbers at these redshifts \citep{gupta07}. 
The first results of our survey, based on GMRT and GBT observations of 55~absorbers, are 
presented in this paper.


\section{The \mgtwo\ absorber sample}
\label{sec:sample}

\begin{table}
\begin{center}
\begin{tabular}{|c|c|c|c|c|c|c|}
\hline
 QSO & $z_{\rm em}$ & R & $z_{\rm abs}$ & $W_0^{\lambda 2796}$ & $W_0^{\lambda 2600}$ & $W_0^{\lambda 2852}$ \\
     &              & mag &               & $\AA$              & $\AA$              &  $\AA$                 \\
\hline
0017+154 & 2.009  & 17.7$^\dagger$& 1.6260$^a$& 1.44 & 0.76 & 0.48  \\
0039$-$407 & 2.478  & 19.7$^\dagger$&  0.8483$^b$ & 2.35 & 2.09 & $-$  \\
0014+813 & 3.366  & 16.5 &  1.1109$^c$ & 0.85 & 0.53 & 0.20  \\
0014+813 & 3.366  & 16.5 &  1.1125$^c$ & 2.49 & 2.14 & 0.40  \\
0105$-$008 & 1.374  & 17.7 &  1.3710$^a$ & 0.58 & 0.38 & 0.22  \\
0109+176 & 2.157  & 18.0 &  0.8392$^d$ & 1.75 & 1.09 & $<0.2$  \\
0237$-$233 & 2.223  & 16.6 &  1.3647$^e$ & 2.05 & 0.84 & $<0.16$  \\
0237$-$233 & 2.223  & 16.6 &  1.6724$^j$ & 1.31 & 0.57 & 0.47  \\
0240$-$060 & 1.805  & 18.7$^\dagger$ &  0.5810$^b$ & 1.44 & $-$ & 0.11  \\
0240$-$060 & 1.805  & 18.7$^\dagger$ &  0.7550$^b$& 1.65 & 1.25 & 0.63  \\
0244$-$128 & 2.201  & 18.4$^\dagger$ &  0.8282$^b$ & 1.77 & 1.23 & 0.45  \\
0311+430   & 2.870  & 21.5$^\dagger$ &  1.068$^k$ & 3.12 & 2.31 & $-$ \\
0409$-$045 & 1.684  & 19.9 &  0.8797$^a$ & 1.81 & 1.29 & $<0.25$  \\
0445+097   & 2.108  & 19.6 &  0.8392$^d$ & 3.17 & 1.97 & 0.91  \\
0458$-$020 & 2.286  & 19.0$^\dagger$ &  1.5605$^b$ & 0.94 & 0.75 & $<0.21$  \\
0642+449 & 3.396  & 18.5$^\ast$ &  1.2468$^f$ & 0.57 & 0.50 & $-$  \\
0741+294 & 1.184  & 17.1 & 1.0625$^a$ & 1.19 & 0.73 & 0.46  \\
0801+303 & 1.451  & 18.3 &  1.1908$^a$ & 1.45 & 0.97 & 0.30  \\
0804+499 & 1.436  & 19.0 &  1.4071$^a$ & 1.29 & 0.70 & 0.26  \\
0812+332 & 2.426  & 19.4 &  0.8518$^a$ & 2.36 & 0.98 & 0.60  \\
0821+394 & 1.216  & 18.5 &  1.0545$^a$ & 1.56 & 0.88 & 0.17  \\
0829+425 & 1.051  & 18.9 &  1.0459$^a$ & 1.72 & 1.17 & 0.65  \\
0955+476 & 1.882  & 18.9 &  1.0291$^a$ & 2.44 & 1.63 & $<0.17$ \\
0957+003 & 0.905  & 19.3 &  0.6722$^a$ & 1.85 & 1.32 & 0.33  \\
1005$-$333 & 1.837  & 18.0$^\dagger$ &  1.3734$^b$ & 0.93 & 0.84 & $-$ \\
1012+022 & 1.375  & 17.3 &  0.7632$^a$ & 1.60 & 0.76 & 0.29  \\
1011+280 & 0.899  & 18.6$^\ast$ &  0.8895$^g$ & $>2$ & 1.46 & $-$  \\
1116+128 & 2.129  & 18.9 &  0.5163$^a$ & 2.48 & 0.61 & $<0.15$  \\
1116+128 & 2.129  & 18.9 &  0.6346$^a$ & 0.92 & $-$ & 0.40  \\
1136+408 & 2.366  & 20.0 &  1.3702$^a$ & 1.61 & 0.82 & $<0.13$  \\
1142+052 & 1.345  & 19.5 &  1.3431$^a$ & 2.21 & 1.32 & 1.05  \\
1200+068 & 2.182  & 19.8$^\ddagger$&  0.862$^h$ & 5.57 & 3.75 & 2.98  \\
1204+399 & 1.518  & 19.0 &  1.3254$^a$ & 1.22 & 0.61 & $<0.12$  \\
1210+134 & 1.139  & 18.4 &  0.7717$^a$ & 1.26 & 0.75 & 0.32  \\
1222+438 & 1.075  & 19.7 &  0.7033$^a$ & 1.51 & 1.29 & $<0.24$  \\
1226+105 & 2.307  & 19.1 &  0.9382$^a$ & 1.69 & 0.78 & 0.33  \\
1318$-$263 & 2.027  & 21.3$^\dagger$ &  1.1080$^b$ & 1.38 & 0.61 & 0.49  \\
1324$-$047 & 1.882  & 19.8$^\dagger$&  0.7850$^b$ & 2.58 & 1.77 & 0.73  \\
1343+386 & 1.852  & 18.4 &  0.8076$^a$ & 1.50 & 1.03 & $<0.13$  \\
1402$-$012 & 2.518  & 18.0$^\dagger$&  0.8901$^b$ & 1.21 & 0.99 & 0.19 \\
1430$-$178 & 2.331  & 19.4$^\dagger$&  1.3269$^a$& 0.60 & 0.45 & 0.32  \\
1602+241 & 2.531  & 19.0 &  1.5246$^a$& 1.50 & 1.23 & 0.72 \\
1611+343 & 1.397  & 17.5 &  0.6672$^a$& 1.36 & 0.58 & 0.15  \\
1625+262 & 1.656  & 18.7 &  1.0156$^a$ & 1.43 & 0.73 & 0.22  \\
1629+120 & 1.792  & 18.4$^\ast$ &  0.9004$^d$ & 1.06 & 0.63 & $-$  \\
1634+213 & 1.802  & 19.6 &  0.8001$^a$ & 2.60 & 1.56 & 0.73 \\
1701+593 & 1.798  & 19.6 &  0.7238$^a$ & 1.86 & 1.06 & 0.45  \\
2003$-$025 & 1.457  & 19.0$^\ast$ &  1.2116$^e$ & 2.65 & 1.27 & $<0.31$ \\
2003$-$025 & 1.457  & 19.0$^\ast$&  1.4106$^e$ & 0.74 & 0.34 & $-$  \\
2149+212 & 1.538  & 19.0$^\ast$ &  0.9115$^d$ & 0.72 & 0.95 & 0.34 \\
2149+212 & 1.538  & 19.0$^\ast$ &  1.0017$^d$ & 2.46 & 1.0 & $<0.15$\\
2149$-$307 & 2.345  & 18.4$^\dagger$&  1.0904$^b$ & 1.45 & 0.78 & $<0.11$ \\
2149$-$307 & 2.345  & 18.4$^\dagger$ &  1.6996$^b$ & $-$ & $-$ & $-$  \\
2337$-$011 & 2.085  & 17.8 &  1.3606$^i$ & 1.55 & 1.21 & 0.54  \\
2355$-$106 & 1.639  & 19.3 &  1.1727$^a$ & 1.66 & 1.14 & 0.50  \\
\hline
\end{tabular}
\end{center}
\vskip -0.05in
\caption{The absorption systems: optical data. Typical errors on the rest equivalent widths 
are $\sim 0.05 - 0.1 \AA$. Notes: $^\dagger${}B-magnitude, from the APM catalogue or 
\citet{jackson02}; $^\ast${}V-magnitude, from RTN06 or the NASA/IPAC Extragalactic Database; 
$^\ddagger${}g-magnitude, from the SDSS; 
References for redshifts and equivalent widths: 
(a)~SDSS DR-5 catalogue; (b)~\citet{ellison04}; (c)~\citet{sargent88}; 
(d)~\citet{barthel90}; (e)~\citet{aldcroft94}; (f)~\citet{sargent89}; (g)~\citet{peterson78}; 
(h)~\citet{wild05}; (i)~\citet{khare04}; (j)~\citet{steidel92}; (k)~\citet{york07}.}
\label{tab:optobs}
\end{table}

We used \mgtwo\ absorber samples drawn from the CORALS survey \citep{ellison04}, Data Releases~(DRs)~3-5 
of the SDSS (e.g. \citealt{prochter06}) and the literature (e.g. \citealp{sargent88,barthel90,aldcroft94}) 
to select our targets. The primary selection criterion was that the system be a ``strong'' \mgtwo\ absorber,
with $W_0^{\lambda 2796} \ge 0.5 \AA$ (where $W_0^{\lambda 2796}$ is the \mgtwo$\lambda$2796 rest equivalent 
width), lying towards a background quasar of ``sufficient'' radio flux density at the redshifted 
\hi~21cm line frequency to allow a sensitive search for \hi~21cm absorption. A secondary criterion, 
that the rest equivalent width in the \fetwo$\lambda$2600 transition also be large, $\wfetwo \ge 0.5 \AA$, 
was also typically used, although a few systems with $\wfetwo < 0.5 \AA$ or without information on $\wfetwo$ 
were also observed. We thus have two effective sub-samples:\\
(1)~Systems with $\wmgtwo \ge 0.5 \AA$ and $\wfetwo \ge 0.5 \AA$.\\
(2)~Systems with $\wmgtwo \ge 0.5 \AA$ and either $\wfetwo < 0.5 \AA$ or no information on 
the $\wfetwo$ transition.

The best optical depth sensitivities achieved in present searches for \hi~21cm absorption 
in DLAs are $\tau_{3\sigma} \sim 0.01$ per $\sim 10$~\kms\ (e.g. \citealp{kanekar06,york07}). 
For DLAs, with N$_{\rm HI} \ge 2 \times 10^{20}$~\cm, a non-detection at this sensitivity would yield
the $3\sigma$ limit $[\ts/f] \ge 1000$~K. In other words, only DLAs with high spin temperatures 
and/or low covering factors\footnote{The covering factor $f$ gives the fraction of quasar radio flux 
density occulted by the foreground absorber.} would remain undetected at such a sensitivity. 
Note that fewer than $5$\% of Galactic sightlines have $[\ts/f] > 1000$~K (see Fig.~2 of \citealp{braun92}).

Our search for \hi~21cm absorption was hence aimed at achieving an \hi~21cm optical depth 
sensitivity of $\tau_{3\sigma} \sim 0.01$ per $\sim 10$~\kms\ for all non-detections, so 
as to detect all DLAs with $[\ts/f] \lesssim 1000$~K. Absorbers with high \hi\ column 
densities or low spin temperatures would be detected at higher significance. 

The practical effect of the above sensitivity criterion was that the background quasars were 
chosen to have flux densities $S_{\nu} \gtrsim 300$~mJy at the redshifted \hi~21cm line 
frequency, to ensure a high optical depth sensitivity in a reasonable integration time 
($\lesssim 10$~hours). In most cases, the quasar flux densities had not been measured at 
(or near) the expected \hi~21cm line frequency; we hence estimated the flux density at 
the line frequency by interpolating between measurements at higher and lower frequencies, 
using radio surveys such as the 1.4~GHz NRAO VLA Sky Survey (NVSS; \citealp{condon98}), 
the 1.4~GHz VLA FIRST Survey (Faint Images of the Radio Sky at Twenty 
Centimeter; \citealp{becker95}), the 327~MHz Westerbork Northern Sky Survey (WENSS; 
\citealp{rengelink97}) or the 365~MHz Texas Survey \citep{douglas96}.

A few quasars with lower flux densities, but having foreground \mgtwo\ absorbers 
with $W_0^{\lambda 2796} \ge 0.5 \AA$, 
were also observed, especially when data towards the primary targets were affected by RFI. 
We also observed one absorber, at $z_{\rm abs} \sim 1.6996$ towards 2149$-$307 (Ellison \& Lopez, 
\textit{in prep.}), where the \mgtwo$\lambda$2796 transition has not been observed, 
but which shows strong metal lines, and is hence likely to be a strong \mgtwo\ absorber. 

Our final absorber sample consists of 55~systems, at $0.58 < z_{\rm abs} < 1.70$. Details of
the absorbers and their background quasars are provided in Table~\ref{tab:optobs}, whose columns contain 
(1)~the quasar name, (2)~the quasar redshift, (3)~the quasar R-magnitude (but see the notes to the table), 
(4)~the \mgtwo$\lambda$2796 absorption redshift (or metal-line redshift, for the $z_{\rm abs} \sim 1.6996$ 
absorber towards 2149$-$307), (5)~the \mgtwo$\lambda$2796 rest equivalent width, $W_0^{\lambda 2796}$, 
(6)~the \fetwo$\lambda$2600 rest equivalent width, $W_0^{\lambda 2600}$, and (7)~the \mgone$\lambda$2852 
rest equivalent width, $W_0^{\lambda 2852}$. Most absorbers of the sample have $W_0^{\lambda 2600} 
\ge 0.5 \AA$, while typical errors on the rest equivalent widths are $\sim 0.05 - 0.1 \AA$.  We 
emphasize that our \mgtwo\ absorber sample is a heterogeneous one, drawn from different surveys 
with very different selection criteria.

\section{Observations and data analysis}
\label{sec:obs}

The GMRT and GBT searches for redshifted \hi~21cm absorption were carried out between January~2002
and September~2008. 14 systems at $1.12 < z_{\rm abs} < 1.54$, for which the redshifted \hi~21cm 
absorption frequency lies in the GMRT 610~MHz band ($\sim 560 - 670$~MHz) were observed with the GMRT 
(proposals 07NKa02, 10NKa01, 10NKa02, 11NKa02, 12NKa01), due to its better RFI environment and 
superior spectral baselines; the remaining 41 targets were observed with the PF1-600, 
PF1-800 or PF2 receivers of the GBT (proposals 6A-026, 6C-048 and 7B-007).

\begin{table*}
\begin{center}
\begin{tabular}{|c|c|c|c|c|c|c|c|c|c|c|c|c|c|} 
\hline
 QSO & $z_{\rm abs}$ & Tel. & $\nu_{\rm obs}$ & BW & Resn. & Time  & $S_\nu$ & RMS$^a$ & $\int \tau {\rm d}V$ & $\ts/f$~$^\dagger$ & $\NHI$~$^\dagger$ & $\alpha$ & \\
     &               &           &  MHz            & MHz& \kms  & Hrs. & Jy      & mJy     & \kms  & K & $\times 10^{20}$~\cm & & \\
     &               &           &        &      &       &       &         &         & &  & & & \\
\hline
0017+154 & 1.6260 & GBT & 540.90 & 1.25 & 1.4 & 0.8 & 7.3 & 27.2 & $<0.045$ & $>2500$ & $<0.8$ & $-1.08$ & Y \\
0039$-$407 & 0.8483 & GBT & 768.49 & 2.5 & 1.9 & 3.5 & 0.11 & 3.7 & $<0.49$ & $>225$ & $<8.9$ & $+0.50$ & N \\
0014+813 & 1.1109 & GBT & 672.89 & 2.5 & 2.2 & 0.7 & RFI & -- & -- & -- & -- & $+0.00$ & R \\
0014+813 & 1.1125 & GBT & 672.38 & 2.5 & 2.2 & 0.7 & RFI & -- & -- & -- & -- & $+0.00$ & R \\
0105$-$008 & 1.3710 & GMRT & 599.07 & 1.0 & 2.0 & 3.5 & 1.26 & 3.3 & $0.995\pm0.023$ & $110\pm3$ & $18.14\pm0.41$ & $-0.32$ & A \\ 
0109+176 & 0.8392 & GBT & 772.30 & 2.5 & 1.9 & 4.2 & 1.5 & 4.3 & $<0.039$ & $>2855$ & $<0.7$ & $-0.97$ & Y \\
0237$-$233 & 1.3647 & GMRT & 600.67 & 1.0 & 3.9$^b$ & 12 & 5.53 & 2.5 & $<0.014$ & $>6665$ & $<0.3$ & $+0.55$ & Y \\
0237$-$233 & 1.6724 & GBT & 531.51 & 0.625 & 0.69 & 1.5 & 6.7 & 55.9 & $0.076\pm0.016$ & $1440\pm300$ & $1.39\pm0.29$ & $+0.55$ & Y \\
0240$-$060 & 0.5810 & GBT & 898.42 & 12.5$^c$ & 0.51 & 4.3 & 0.66 & 4.5 & $<0.075$ & $>1430$ & $<1.4$ & $+0.27$ & Y \\
0240$-$060 & 0.7550 & GBT & 809.35 & 12.5$^c$ & 0.57 & 4.3 & 0.65 & 3.7 & $<0.061$ & $>1820$ & $<1.1$ & $+0.27$ & Y \\
0244$-$128 & 0.8282 & GBT & 776.94 & 1.25 & 0.94 & 2.2 & 0.55 & 6.5 & $<0.13$ & $>835$ & $<2.4$ & $-0.32$ & Y \\
0311+430 & 1.068 & GBT & 686.85 & 1.25 & 1.1 & 2.3 & 4.55 & 10.3 & $<0.03$ & $>3335$ & $<0.6$ & $-1.06$ & Y \\
0409$-$045 & 0.8797 & GBT & 755.66 & 2.5 & 1.9 & 2.5 & 0.41 & 4.5 & $<0.14$ & $>800$ & $<2.5$ & $-0.79$ & Y \\
0445+097 & 0.8392 & GBT & 772.30 & 2.5 & 1.9 & 1.8 & 2.0 & 5.9 & $<0.054$ & $>2000$ & $<1.0$ & $-0.74$ & Y \\
0458$-$020 & 1.5605 & GBT & 554.74 & 1.25 & 1.3 & 4.5 & 2.2 & 8.2 & $0.161\pm0.015$ & $685\pm65$ & $2.93\pm0.28$ & $-0.07$ & Y \\
0642+449 & 1.2468 & GMRT & 632.19 & 1.0 & 7.8 & 6.0 & 0.35 & 1.0 & $<0.09$ & $>1250$ & $<1.6$ & $+0.04$ & Y \\
0741+294 & 1.0625 & GBT & 688.68 & 2.5 & 2.1 & 1.7 & 0.59 & 16.8 & $<0.56$ & $>200$ & $<10$ & $-0.62$ & N \\
0801+303 & 1.1908 & GMRT & 648.35 & 1.0 & 3.6$^b$ & 5.0 & 2.07 & 2.2 & $0.305\pm0.025$ & $360\pm30$ & $5.56\pm0.46$ & $-0.74$ & Y \\
0804+499 & 1.4071 & GMRT & 590.09 & 1.0 & 4.0 & 3.3 & 0.68 & 2.8 & $<0.13$ & $>835$ & $<2.4$ & $-0.97$ & Y \\
0812+332 & 0.8518 & GBT & 767.04 & 2.5 & 1.9 & 2.0 & 0.76 & 4.9 & $<0.11$ & $>1000$ & $<2.0$ & $-0.62$ & Y \\
0821+394 & 1.0545 & GBT & 691.36 & 1.25 & 1.1 & 0.5 & 5.2 & 20.0 & $<0.053$ & $>2000$ & $<1.0$ & $-0.67$ & Y \\
0829+425 & 1.0459 & GBT & 694.27 & 2.5 & 2.1 & 1.2 & RFI & -- & -- & -- & -- & $-0.60$ & R \\
0955+476 & 1.0291 & GBT & 700.02 & 1.25 & 1.1 & 2.0 & 1.4 & 11.5 & $<0.12$ & $>950$ & $<2.1$ & $+0.48$ & Y \\
0957+003 & 0.6722 & GBT & 849.42 & 2.5 & 1.7 & 0.6 & 2.2 & 9.8 & $<0.06$ & $>1820$ & $<1.1$ & $-0.90$ & Y \\
1005$-$333 & 1.3734 & GBT & 598.47 & 1.25 & 1.2 & 0.8 & 1.8 & 17.4 & $<0.11$ & $>950$ & $<2.1$ & $-0.82$ & Y \\
1012+022 & 0.7632 & GBT & 805.58 & 12.5$^c$ & 0.29 & 0.7 & 2.2 & 17.9 & $<0.06$ & $>1665$ & $<1.2$ & $-0.82$ & Y \\
1011+280 & 0.8895 & GBT & 751.74 & 12.5$^c$ & 0.30 & 1.2 & RFI & -- & -- & -- & -- & $-0.87$ & R \\
1116+128 & 0.5163 & GBT & 936.76 & 12.5$^c$ & 0.25 & 0.1 & RFI & -- & -- & -- & -- & $-0.90$ & R \\
1116+128 & 0.6346 & GBT & 868.96 & 12.5$^c$ & 0.27 & 0.1 & RFI & -- & -- & -- & -- & $-0.90$ & R \\
1136+408 & 1.3702 & GMRT & 599.28 & 1.0 & 3.9 & 2.3 & 0.49 & 2.6 & $<0.13$ & $>910$ & $<2.2$ & $-0.10$ & Y \\
1142+052 & 1.3431 & GMRT & 606.21 & 1.0 & 3.9 & 6.5 & 1.01 & 1.3 & $0.557\pm0.030$ & $195\pm10$ & $10.15\pm0.55$ & $-0.55$ & A \\
1200+068 & 0.862 & GBT & 762.84 & 2.5$^d$ & 1.9 & 2.2 & 0.79 & 3.8 & $<0.08$ & $>1430$ & $<1.4$ & $-0.45$ & Y \\
1204+399 & 1.3254 & GMRT & 610.82 & 1.0 & 3.8 & 3.0 & 0.38 & 2.7 & $<0.23$ & $>475$ & $<4.2$ & $-0.30$ & N \\
1210+134 & 0.7717 & GBT & 801.72 & 1.25 & 0.91 & 1.8 & 2.2 & 10.8 & $<0.052$ & $>2220$ & $<0.9$ & $-0.46$ & Y \\
1222+438 & 0.7033 & GBT & 833.91 & 2.5 & 1.8 & 2.2 & 0.19 & 4.8 & $<0.42$ & $>260$ & $<7.7$ & $-0.41$ & N \\
1226+105 & 0.9382 & GBT & 732.85 & 2.5 & 2.0 & 1.0 & RFI & -- & -- & -- & -- & $-0.82$ & R \\
1318$-$263 & 1.1080 & GBT & 673.82 & 1.25 & 1.1 & 0.7 & RFI & -- & -- & -- & -- & $-0.20$ & R \\
1324$-$047 & 0.7850 & GBT & 795.75 & 2.5 & 1.8 & 2.2 & 0.14 & 4.8 & $<0.48$ & $>225$ & $<8.8$ & $-0.57$ & N \\
1343+386 & 0.8076 & GBT & 785.80 & 1.25 & 0.93 & 1.3 & 1.7 & 7.4 & $<0.06$ & $>1820$ & $<1.1$ & $-0.52$ & Y \\
1402$-$012 & 0.8901 & GBT & 751.50 & 1.25 & 0.97 & 2.0 & 0.87 & 6.5 & $<0.09$ & $>1175$ & $<1.7$ & $+0.33$ & Y \\
1430$-$178 & 1.3269 & GMRT & 610.43 & 1.0 & 3.7 & 12 & 1.05 & 1.0 & $0.127\pm0.022$ & $860\pm150$ & $2.32\pm0.40$ & $-0.17$ & Y \\
1602+241 & 1.5246 & GMRT & 562.63 & 1.0 & 4.2 & 2.3 & 0.63 & 12.0 & $<0.6$ & $>180$ & $<11$ & $+0.23$ & N \\
1611+343 & 0.6672 & GBT & 851.97 & 1.25 & 0.86 & 0.3 & 5.7 & 26 & $<0.09$ & $>1250$ & $<1.6$ & $-0.63$ & Y \\
1625+262 & 1.0156 & GBT & 704.71 & 2.5 & 2.1 & 1.0 & RFI & -- & -- & -- & -- & $-0.94$ & R \\
1629+120 & 0.9004 & GBT & 747.43 & 2.5 & 2.0 & 1.2 & RFI & -- & -- & -- & -- & $-0.53$ & R \\
1634+213 & 0.8001 & GBT & 789.07 & 2.5 & 1.9 & 0.8 & RFI & -- & -- & -- & -- & $-0.32$ & R \\
1701+593 & 0.7238 & GBT & 824.00 & 2.5 & 1.8 & 2.0 & 0.76 & 4.5 & $<0.1$ & $>1175$ & $<1.7$ & $-0.89$ & Y \\
2003$-$025 & 1.2116 & GMRT & 642.25 & 1.0 & 7.2 & 8.0 & 3.4 & 2.4 & $<0.022$ & $>5000^d$ & $<0.4^d$ & $-0.48$ & Y \\
2003$-$025 & 1.4106 & GMRT & 589.23 & 1.0 & 4.0 & 15 & 3.7 & 2.8 & $0.210\pm0.011$ & $520\pm25$ & $3.83\pm0.20$ & $-0.48$ & Y \\
2149+212 & 0.9115 & GBT & 743.08 & 1.25 & 1.0 & 1.3 & 2.8 & 10.9 & $<0.052$ & $>2220$ & $<0.9$ & $-0.79$ & Y \\
2149+212 & 1.0017 & GBT & 709.60 & 2.5 & 2.1 & 1.8 & RFI & -- & -- & -- & -- & $-0.79$ & R \\
2149$-$307 & 1.0904 & GBT & 679.49 & 1.25 & 1.1 & 0.5 & RFI & -- & -- & -- & -- & $+0.21$ & R \\
2149$-$307 & 1.6996 & GBT & 526.15 & 1.25 & 1.4 & 2.5 & 1.6 & 13.9 & $<0.12$ & $>910$ & $<2.2$ & $+0.21$ & M \\
2337$-$011 & 1.3606 & GMRT & 601.71 & 0.5 & 1.9 & 15 & 0.063 & 1.9 & $2.04\pm0.13$ & $55\pm2$ & $37.2\pm2.4$ & $+0.73$ & Y \\
2355$-$106 & 1.1727 & GMRT & 653.75 & 1.0 & 1.8$^b$ & 4.0 & 0.42 & 2.5 & $0.256\pm0.062$ & $430\pm55$ & $4.67\pm0.62$ & $+0.62$ & Y \\
\hline
\end{tabular}
\end{center}
\caption{Observing details and results. Notes to the table: $^\dagger${}The quoted values of $\ts/f$ and $\NHI$ in 
columns~(11) and (12) assume, respectively, an \hi\ column density of $2 \times 10^{20}$~\cm\ and $\ts/f = 1000$~K; 
see the main text for discussion.  $^a${}RMS noise at the Hanning-smoothed 
resolution of column~(6). $^b${}The spectrum was not Hanning-smoothed.  $^c${}These observations 
used the Auto-Correlation Spectrometer as the 
Spectral Processor was not available. $^d${}The observations of 1200+068 used two settings, with the final spectrum 
(details listed) obtained by smoothing to the coarser resolution and averaging. }
\label{tab:obs}
\end{table*}

\subsection{GMRT observations and data analysis}

The GMRT observations used the 30-station FX correlator as the backend, with two circular 
polarizations and a bandwidth of 0.5 or 1~MHz sub-divided into either 128 or 256~channels, yielding a 
velocity resolution of $\sim 2-8$~\kms. The standard calibrators 3C147, 3C286, or 3C48, were 
used to calibrate the flux density scale and usually also to determine the shape of the passband. 
Observations of target sources were interleaved (at $\sim 45$-minute intervals) with short observations 
of nearby phase calibrators to obtain an initial estimate of the antenna gains. Our experience with 
the GMRT indicates that the flux density calibration is reliable to better than $\sim 15$\% 
in this observing mode.

The GMRT data were analysed in ``classic'' AIPS, using standard procedures. After initial 
editing of corrupted data (e.g. due to ``dead'' antennas, correlator errors, RFI, etc), 
the antenna gains and bandpass shapes were determined using the calibrator data. A 
continuum image was then made of each target field (after averaging line-free 
channels to increase the signal-to-noise ratio) and self-calibration procedures used to 
obtain a better estimate of the antenna gains.  Three or four iterations of phase 
self-calibration, followed by one or two of amplitude~\&~phase self-calibration (along 
with further editing of bad data), were typically found necessary to obtain convergent 
antenna gains; the source flux density was then measured from the final image. The 
continuum emission was next subtracted out from the calibrated visibilities, using the 
task UVSUB (followed, in some cases, by the task UVLIN, to remove any residual emission). 
For all antenna baselines, the residual visibilities were then inspected in the 
time-frequency plane (using the task SPFLG) and any channel-dependent RFI was edited out. 
Finally, the edited residual visibilities were imaged in all channels to produce a spectral 
cube. The target sources were usually unresolved (or, in a few cases, marginally resolved) 
by the GMRT synthesized beam and a single spectrum was hence extracted in each case from 
the final spectral cube at the source location. Typically, a second-order baseline was 
subtracted from this, using line-free channels, to produce the final spectrum for each source.

\subsection{GBT observations and data analysis}
\label{sec:gbt}

The GBT observations (in projects 6A-026, 6C-048 and 7B-007) usually used the Spectral 
Processor as the backend, due to its 
high dynamic range, with two linear polarizations. The observing bandwidths were $1.25$ or 
$2.5$~MHz, sub-divided into 1024 channels, yielding spectral resolutions of $\sim 1-2$~\kms\ 
after Hanning smoothing. Five of the observing sessions (in one of which the two \mgtwo\ 
absorbers towards 0240$-$060 were observed simultaneously) used the Auto-Correlation 
Spectrometer as the backend, as the Spectral Processor was not available. These had a 
bandwidth of 12.5~MHz, sub-divided into 32768 channels, yielding a spectral resolution 
of $\sim 0.3$~\kms\ after Hanning smoothing (with 16384 channels and $\sim 0.6$~\kms\ for 
each spectrum towards 0240$-$060).  All GBT observations were carried out in position-switched, 
total power mode, with online system temperature measurements obtained by firing a noise diode.

The GBT data were analysed in {\sc DISH}, the {\sc AIPS++} single-dish package. Much editing was 
usually necessary to remove data corrupted by RFI. Despite this, data on 13 sources could 
not be salvaged, due to strong RFI at or near the observing frequency. 
For the remaining sources, after the initial editing and calibration, the continuum flux density 
was first measured using RFI- and line-free channels. A first- or second-order baseline (determined 
by inspecting the individual spectra) was then fitted to each record and subtracted out during the 
process of calibration. The residual data were then averaged together to obtain the final spectrum
for each source. In the case of multiple observing epochs for a single source (sometimes with
different observational parameters), the spectra were first smoothed to the same spectral resolution, 
then interpolated onto the same heliocentric frequency scale, and finally averaged, with appropriate 
weights derived from the noise values. A low-order polynomial (upto second order) was then subtracted 
from this, using line- and RFI-free channels, to produce the final spectrum for each absorber.

\section{Spectra and results}
\label{sec:spectra}


For optically-thin \hi~21cm absorption against a radio-loud quasar, the \hi\ column density,
${\rm N_{HI}}$, \hi~21cm optical depth, $\tau$, and spin temperature, $\ts$, are related 
by the expression (e.g. \citealp{rohlfs06})
\begin{equation}
\rm{N_{HI}} = 1.823 \times 10^{18} \times \left[\ts/f\right] \times \int \tau \:{\rm d}V ,
\label{eqn:hi}
\end{equation}
where N$_{\rm HI}$ is in \cm, $\ts$ in K, and the integral is over velocity, in \kms. For the general
case of multi-phase gas along the line of sight, $\ts$ is the column-density-weighted harmonic mean 
of the spin temperatures of the different phases. The covering factor $f$, the fraction of quasar radio
flux density obscured by the foreground absorber, can be estimated by comparing the flux density 
measured at high spatial resolution [e.g. with very long baseline interferometry (VLBI)
studies] with the total flux density of the quasar at the redshifted \hi~21cm line frequency (e.g. 
\citealp{briggs83,kanekar07,kanekar09a}). While VLBI studies at frequencies $< 1.4$~GHz
are not available in the literature for most of the quasars of our sample, higher-frequency VLBI 
observations (when available) will be used to estimate $f$ and hence, to infer the \hi\ column 
densities of the absorbers, using equation~(\ref{eqn:hi}), for an assumed spin temperature of 
$\ts = 1000$~K.

The observational details and results are summarized in Table~\ref{tab:obs}; the columns here contain 
(1)~the quasar name, (2)~the \mgtwo$\lambda$2796 absorption redshift $z_{\rm abs}$, (3)~the telescope 
used for the observations, 
(4)~the redshifted \hi~21cm frequency, in MHz, (5)~the observing bandwidth, in MHz, (6)~the velocity 
resolution, after Hanning smoothing, in \kms, (7)~the on-source integration time, in hours, 
(8)~the source flux density, in Jy, (9)~the root-mean-square (RMS) noise, in mJy, (10)~the 
integrated \hi~21cm optical depth, $\int \tau \rm{d}V$, in \kms, or, for non-detections, 
$3\sigma$ upper limits to the integrated \hi~21cm optical depth, assuming a Gaussian profile with a 
full width at half maximum (FWHM) of $10$~\kms, (11)~the ratio $[\ts/f]$ derived from these observations, 
or $3\sigma$ lower limits to this quantity, assuming the DLA threshold \hi\ column density of 
$2 \times 10^{20}$~\cm, (12)~the \hi\ column density inferred from the present 
observations, in units of $[\ts/1000] \times (1/f) \times 10^{20}$~\cm\ (see equation~\ref{eqn:hi}), 
(13)~the quasar spectral index, $\alpha$, defined by $S_{\nu_1}/S_{\nu_2} = (\nu_1/\nu_2)^\alpha$ 
(for most sources of the present sample, $\nu_1 = 1.4$~GHz, and $\nu_2 = 365$~MHz), and (14)~a letter
indicating whether the absorber was included in the final sample (Y), or, if not, the reason for the 
exclusion, weak \hi~21cm optical depth limit (N), RFI (R), associated \mgtwo\ absorption (A), 
or unknown rest \mgtwo$\lambda$2796 equivalent width (M).

\begin{figure*}
\centering
\epsfig{file=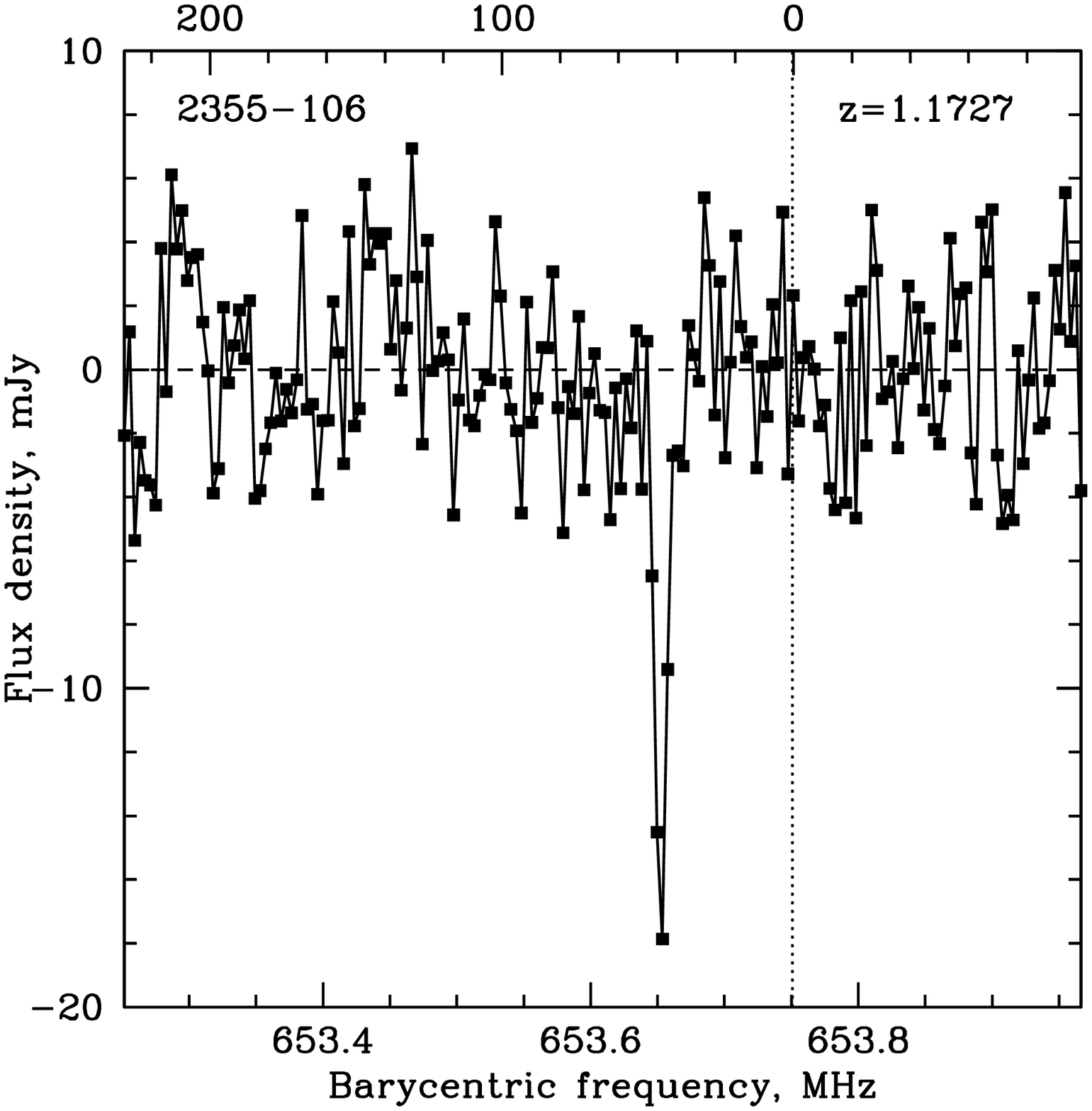,height=2.2truein,width=2.2truein}
\epsfig{file=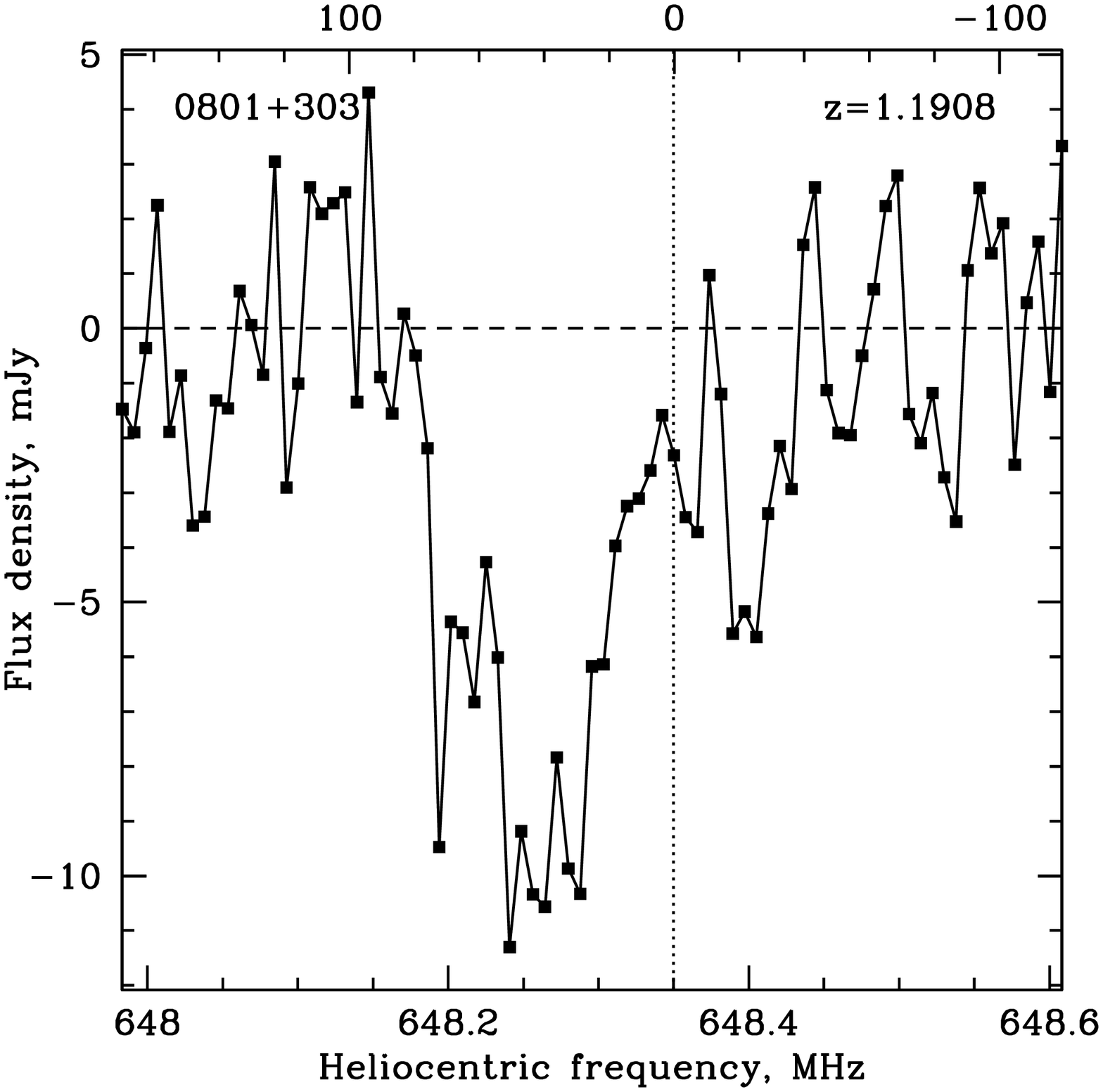,height=2.2truein,width=2.2truein}
\epsfig{file=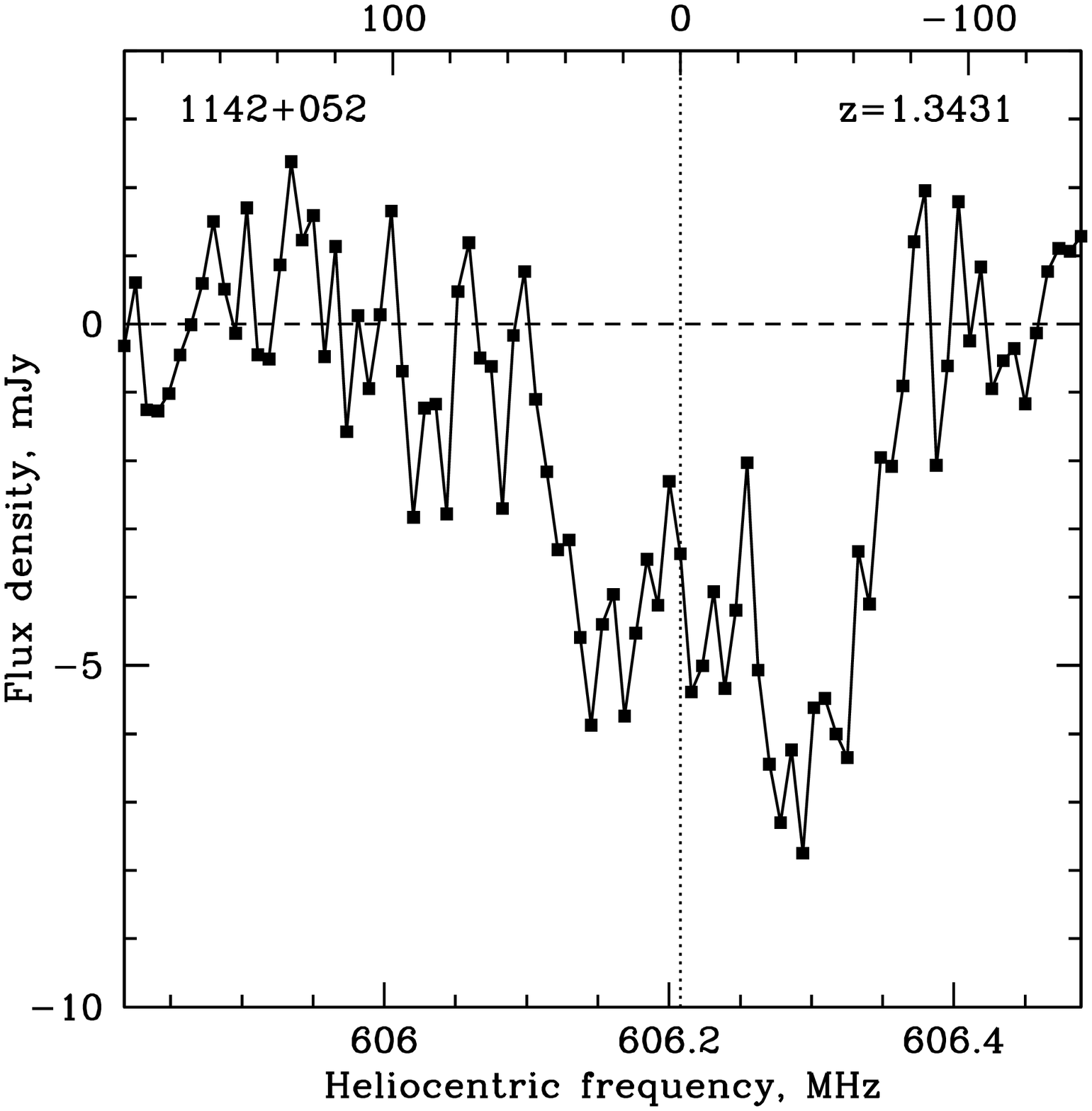,height=2.2truein,width=2.2truein}
\epsfig{file=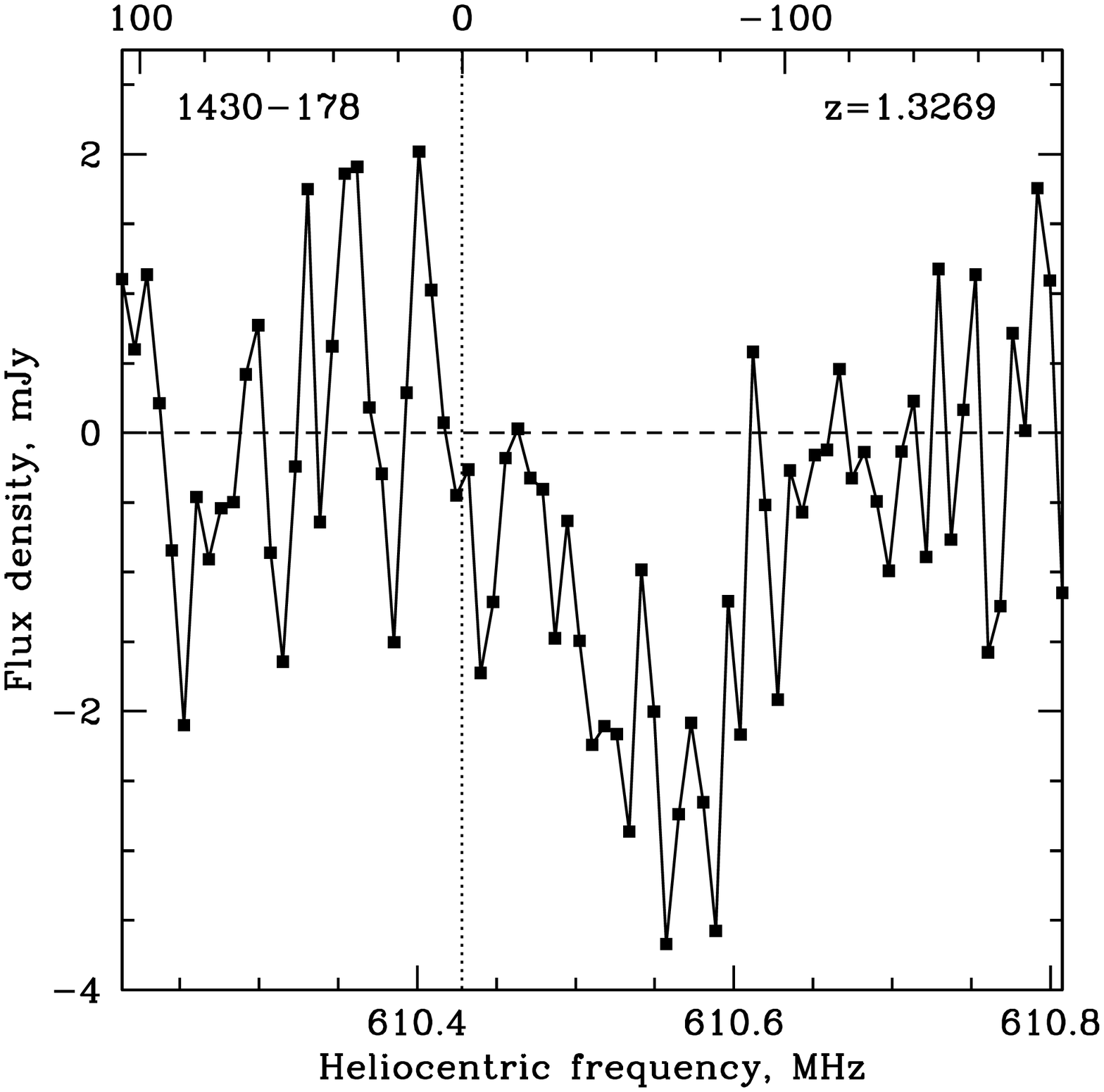,height=2.2truein,width=2.2truein}
\epsfig{file=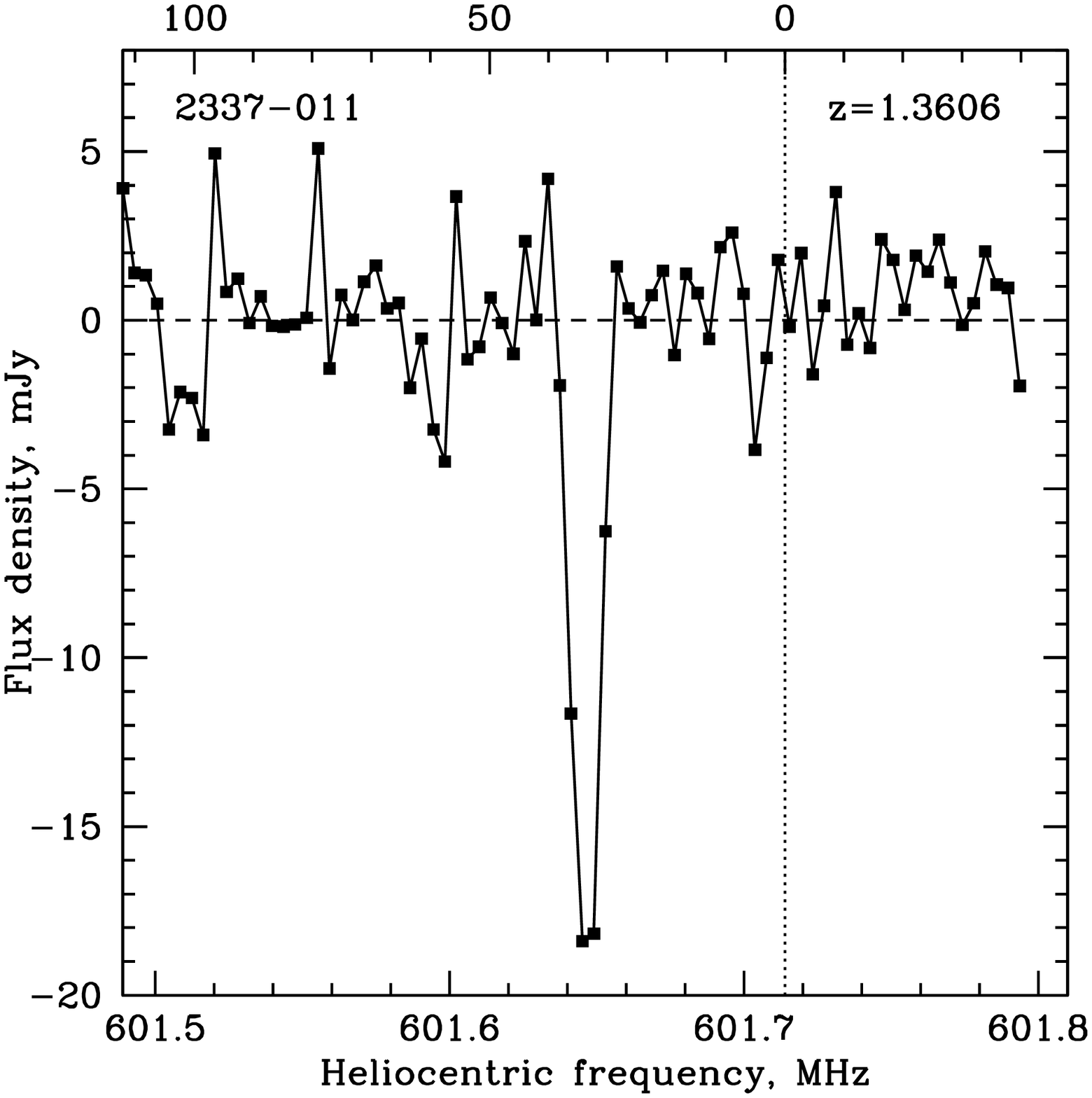,height=2.2truein,width=2.2truein}
\epsfig{file=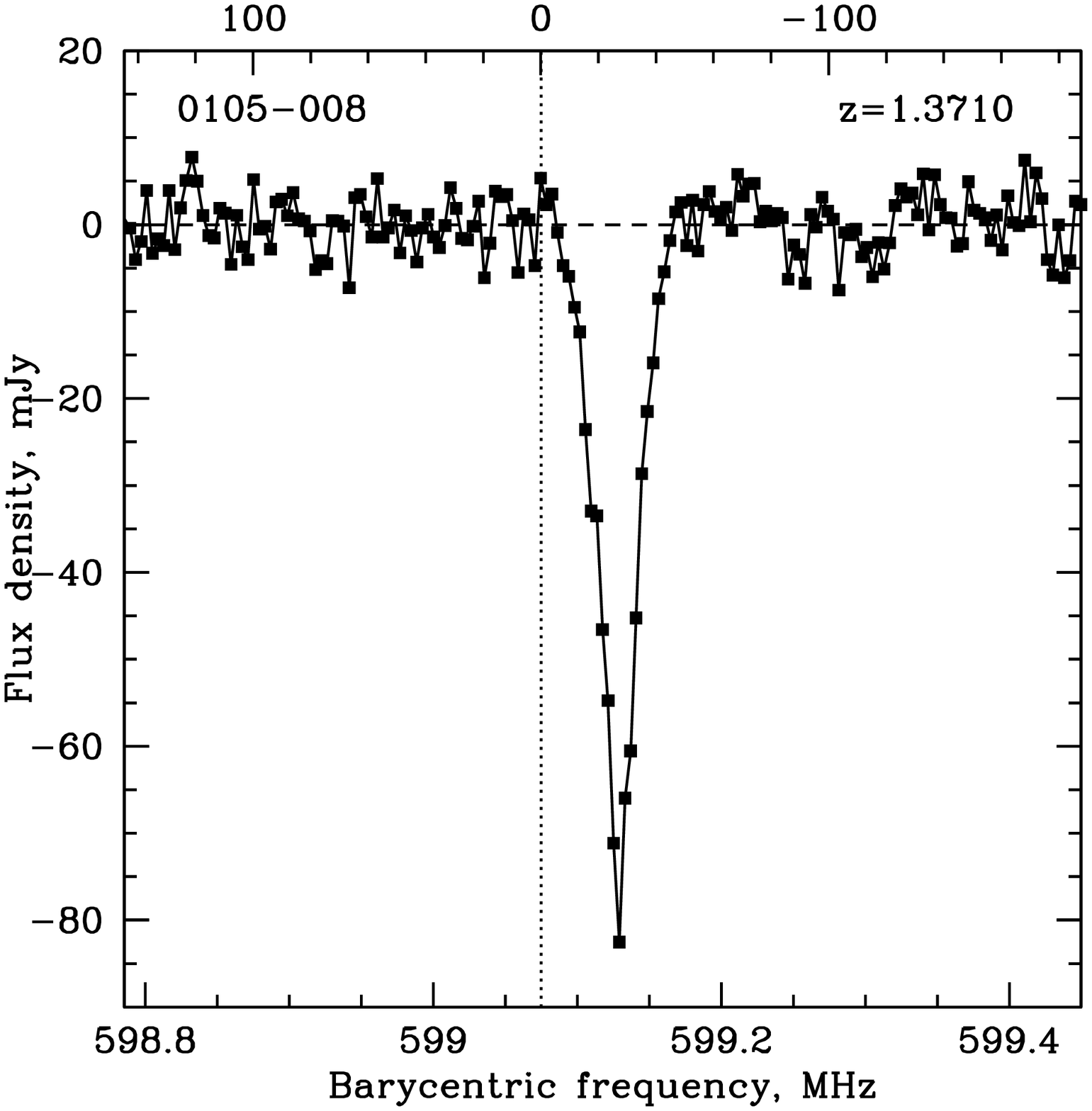,height=2.2truein,width=2.2truein}
\epsfig{file=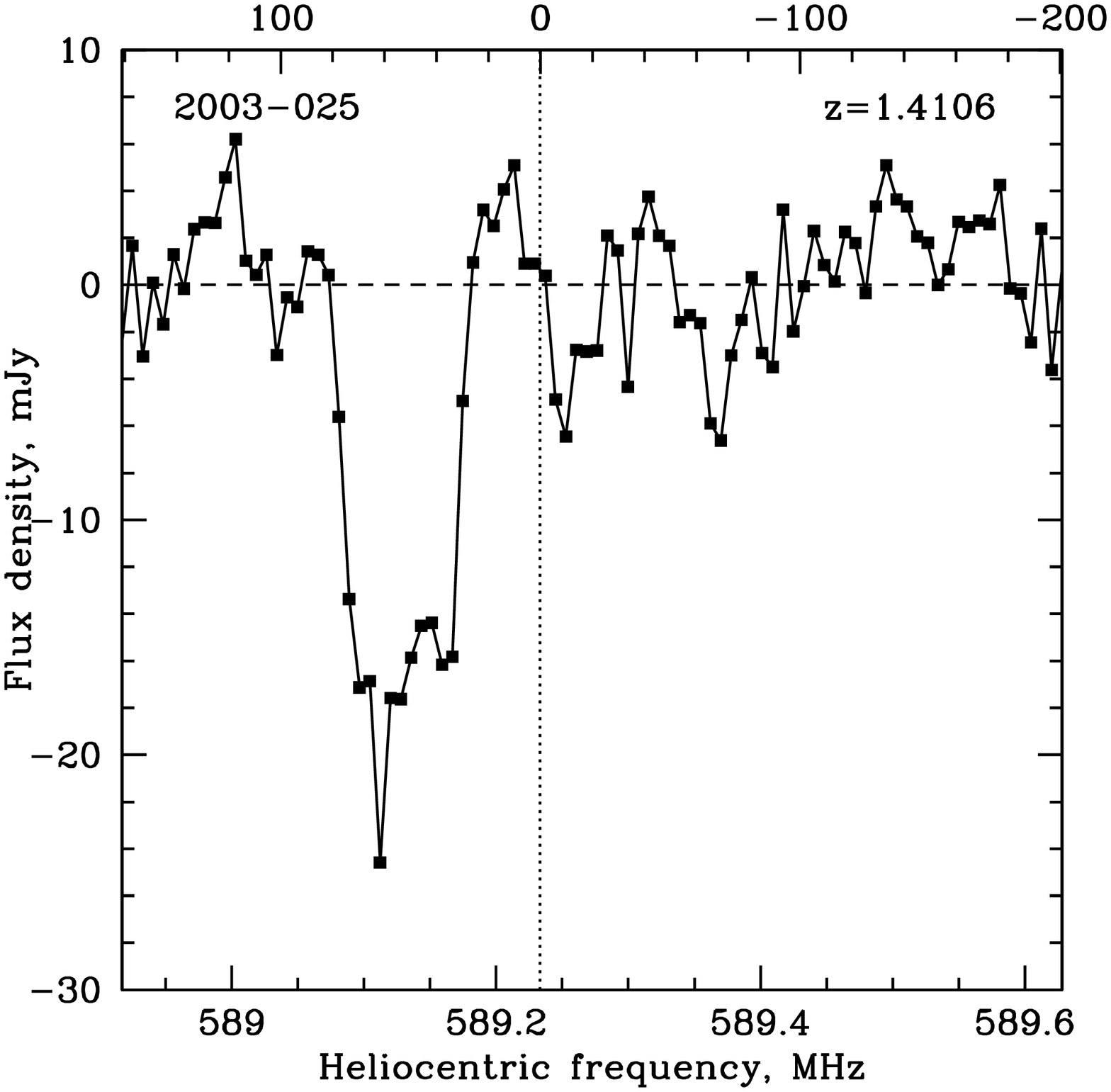,height=2.2truein,width=2.2truein}
\epsfig{file=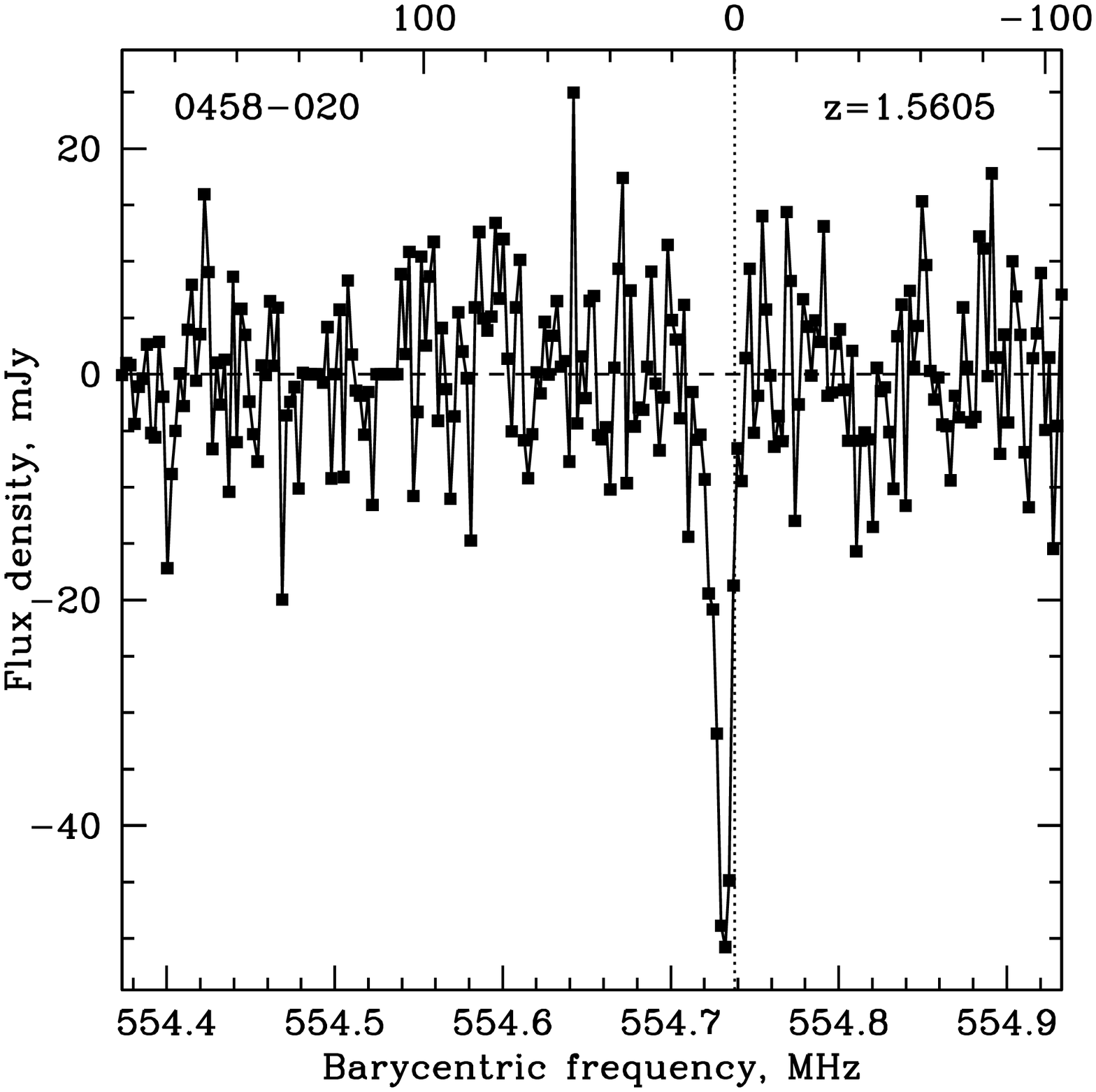,height=2.2truein,width=2.2truein}
\epsfig{file=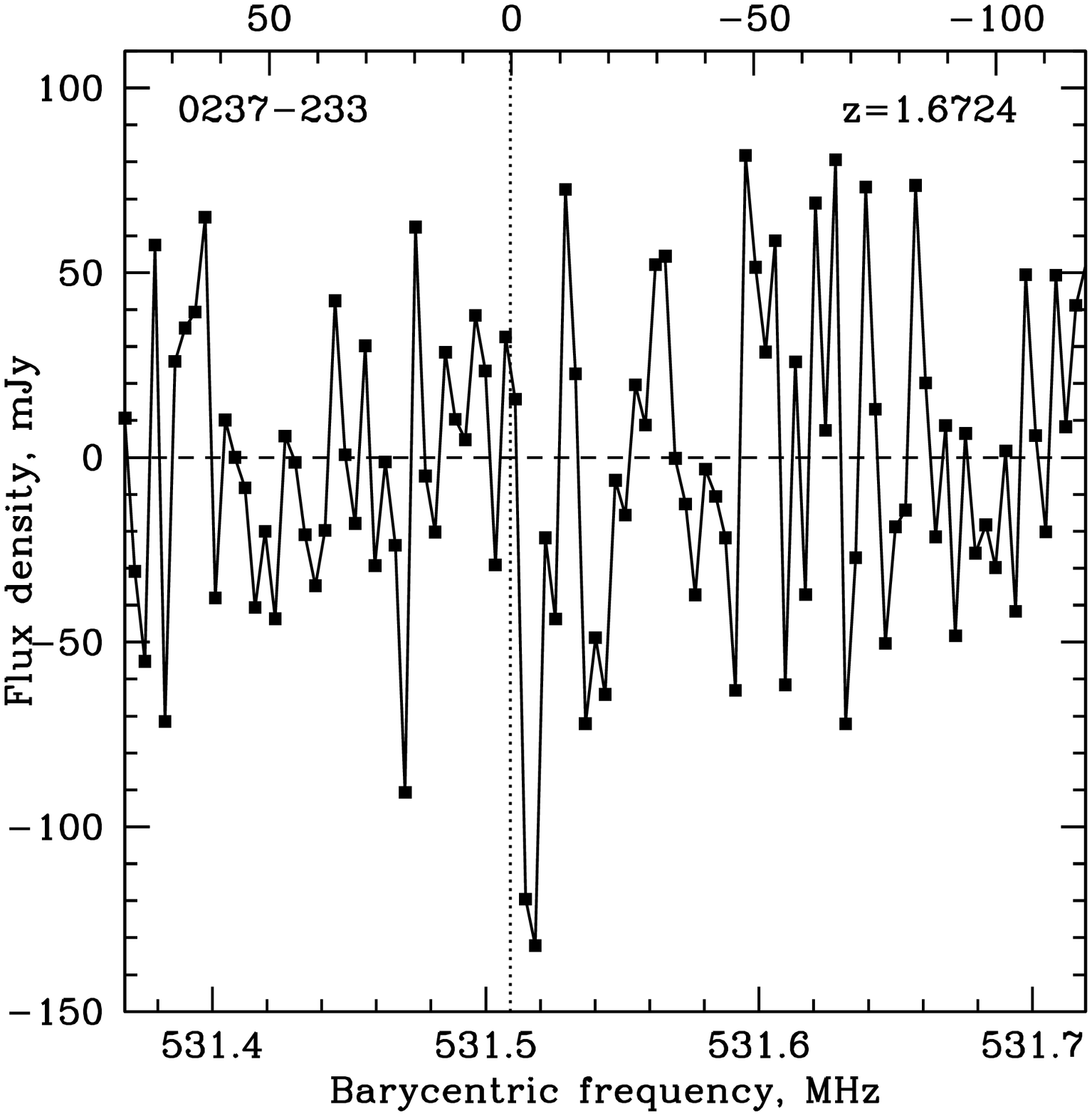,height=2.2truein,width=2.2truein}
\caption{The GMRT and GBT spectra for the eight confirmed, and one tentative,
detections of \hi~21cm absorption, in order of increasing redshift. The 
\mgtwo$\lambda$2796 redshift is listed at the top right corner of each 
panel, and the expected \hi~21cm line frequency indicated by the vertical 
dashed line. The velocity scale relative to the \mgtwo$\lambda$2796 redshift
is indicated on the top axis of each panel. }
\label{fig:detect}
\end{figure*}

Eight definite detections (all with $> 5\sigma$ significance) of redshifted \hi~21cm 
absorption, and one tentative detection (with $\sim 4.6 \sigma$ significance), were 
obtained, at $1.17 \lesssim z_{\rm abs} \lesssim 1.68$.  Seven of these were with the 
GMRT, three of which had been earlier independently found by \citet{gupta07}, while 
two were with the GBT.  All candidate \hi~21cm absorbers were observed on multiple epochs 
to ensure that the spectral feature did not arise due to intermittent RFI. In most cases, 
the expected doppler shift in the topocentric line frequency (due to the motion of the 
Earth) was detected between the different observing epochs, confirming that the feature is 
likely to arise from an astronomical source. For the weak absorption feature at 
$z_{\rm abs} \sim 1.3269$ towards 1430$-$178 (GMRT) and the tentative detection at 
$\sim 1.6724$ towards 0237$-$233, we have not detected the doppler shift, as the expected 
shift was very small ($< 5$~\kms) between the different observing sessions. However, the 
\hi~21cm line in the former system was detected at the same heliocentric frequency on two 
observing epochs separated by $\sim 2$ years, and is hence unlikely to arise due to RFI. 
The nine \hi~21cm absorption profiles are shown in Fig.~\ref{fig:detect}, in order of 
increasing absorption redshift, with individual systems discussed in the next section.

As mentioned earlier, the GBT spectra of 13~sources, mostly at $z_{\rm abs} \sim 1$, were 
affected by strong RFI and did not prove useful; these are indicated by the label ``RFI'' in 
column~(8) of Table~\ref{tab:obs}. No evidence for absorption was seen at (or near) the 
expected redshifted \hi~21cm frequency in the spectra of 32~sources, yielding (mostly) 
strong upper limits on the \hi~21cm optical depth and the \hi\ column density [see 
columns~(10) and (11) of Table~\ref{tab:obs}]. The spectra of the non-detections are 
shown in Fig.~\ref{fig:nondetect}, in order of increasing absorption redshift.

After excluding the 13~systems affected by RFI and the $z_{\rm abs} \sim 1.6996$ 
absorber towards 2149$-$307 (in which the \mgtwo$\lambda$2796 transition has not so far 
been observed), our sample consists of 41~strong \mgtwo$\lambda$2796 absorbers, at 
$0.58 < z_{\rm abs} < 1.68$.  Our observing goal was to achieve an optical depth 
sensitivity of $\tau_{3\sigma} \sim 0.01$ per $\sim 10$~\kms\ for all non-detections.
In practice, we either detected \hi~21cm absorption or achieved $\tau_{3\sigma} < 
0.013$ per $\sim 10$~\kms\ for 35~of the above 41~targets. For six absorbers, our
optical depth sensitivity was only $\tau_{3\sigma} \sim 0.02 - 0.06$, typically because 
the background quasar has a low flux density at the redshifted \hi~21cm line frequency 
or because a significant fraction of the data were affected by RFI. We will exclude 
these six systems from later statistical analysis due to the lower sensitivity of 
their \hi~21cm spectra, so as to obtain a uniform optical depth criterion,
$\tau_{3\sigma} < 0.013$. In addition, two systems (towards 0105$-$008 and 1142+052) 
lie within $3000$~\kms\ of the background quasar and are thus ``associated'' absorbers.
Following the general practice in the literature (e.g. \citealp{wolfe86,ellison01}),
these are excluded from our statistical sample, as conditions in the absorbing 
gas could be affected by the presence of the nearby active galactic nucleus. These criteria 
imply that all ``non-associated'' DLAs with $[\ts/f] \lesssim 800$~K that are present 
in our statistical sample of 33~systems would have been detected in \hi~21cm 
absorption. Of course, absorbers with higher \hi\ column densities or lower spin 
temperatures would have been detected at higher significance.

\section{Individual systems}
\label{sec:detect}

In this section, we discuss physical conditions in a few absorbers of interest, the 
nine detections of \hi~21cm absorption and a non-detection in the known DLA towards 2149+212 (RTN06).

\begin{enumerate}
\item{0105$-$008, $z_{\rm abs} \sim 1.3710$: The absorber redshift is only $\sim 660$~\kms\ blueward 
of the quasar redshift ($z_{\rm em} \sim 1.374$, from the SDSS), making this an ``associated'' absorber; 
it also has a relatively low \fetwo$\lambda$2600 rest equivalent width, $\wfetwo = (0.38 \pm 0.05) \AA$. 
\hi~21cm absorption from this system was originally detected by \citet{gupta07}. Our new GMRT 
spectrum has a higher velocity resolution than the detection spectrum, and a slightly higher 
sensitivity. The peak optical depth in our spectrum matches that of \citet{gupta07}.
However, unlike them, we find that a single Gaussian component does not provide a good fit to the 
spectrum; two components, with FWHMs of $\sim (10.2 \pm 2.2)$~\kms\ and $\sim (20.4 \pm 2.0)$~\kms, 
are needed to obtain residuals consistent with noise. The kinetic temperatures of the 
two components inferred from these FWHMs are $\tk \le (2275 \pm 980)$~K and $\tk \le 
(9095 \pm 1785)$~K.\footnote{Note that $\tk \approx \ts$, for a single cold cloud (e.g. \citealp{liszt01}). 
However, the kinetic temperature of an individual cold cloud in a DLA would typically be lower 
than the estimated absorber spin temperature, due to the presence of warm \hi\ along the
line of sight.} Only $\sim 30$\% (170~mJy) of the total 2.3~GHz quasar flux density ($\sim 0.58$~Jy) is 
recovered in a Very Long Baseline Array (VLBA) 2.3~GHz image [from the VLBA Calibrator Survey 
(VCS); \citealp{beasley02}], indicating that significant flux density is present in extended 
emission. The covering factor is hence likely to be low, $< 0.3$ at the redshifted \hi~21cm line 
frequency. The \hi~21cm column density of the absorber thus appears quite high, 
N$_{\rm HI} \sim (6.05 \pm 0.14) \times \left[\ts/1000\right] (0.3/f) \times 10^{21}$~\cm. }
\item{0458$-$020, $z_{\rm abs} \sim 1.5605$: The GBT \hi~21cm profile is well fit by a single
Gaussian component, with FWHM$\sim (6.48 \pm 0.76)$~\kms; the upper limit to the gas 
kinetic temperature is thus $(915 \pm 215)$~K. \citet{briggs89} used VLBI studies at 327, 467 and 
608~MHz to find that the central core-jet structure has a flux density of $\sim 1.8$~Jy at 
608~MHz on scales $< 0.1''$, with a spectral index of $\alpha \sim +1.7$ between 467 and 608~MHz. 
This implies a compact flux density of $\sim 1.55$~Jy at the redshifted \hi~21cm line 
frequency of 554.7~MHz, giving a covering factor $f > 0.7$. The \hi\ column density of the 
absorber is thus N$_{\rm HI} \sim (4.19 \pm 0.40) \times \left[\ts/1000\right] 
(0.7/f) \times 10^{20}$~\cm. This line of sight also contains another \hi~21cm absorber, 
at $z_{\rm abs} \sim 2.03945$ \citep{wolfe85}.}
\item{0801+303, $z_{\rm abs} \sim 1.1908$: \citet{gupta07} detected \hi~21cm absorption from this 
absorber, but found differences between spectra taken at different epochs (probably due to low-level 
RFI), and were hence unable to confirm the presence of wide absorption or determine the velocity width 
of the line. We clearly detect wide \hi~21cm absorption, obtaining a total velocity spread (full width 
between nulls, FWBN) of $\sim 125$~\kms. The covering factor is again likely to be low: 
\citet{kunert02} find that only 456~mJy of the 1.6~GHz flux density is associated with the compact core, 
while the total 1.4~GHz flux density was measured to be 1.27~Jy in the NVSS. While both the core
and the extended emission have steep spectra \citep{kunert02}, the covering factor at the redshifted \hi~21cm 
line frequency of 648.35~MHz is likely to be lower than the 1.4~GHz core fraction, i.e. $f < 0.35$.
The \hi\ column density is N$_{\rm HI} = (1.59 \pm 0.13) \times \left[\ts/1000\right] (0.35/f) 
\times 10^{21}$~\cm.  }
\item{1142+052, $z_{\rm abs} \sim 1.3431$: The \hi~21cm (and \mgtwo) absorption arise within
$\sim 40$~\kms\ of the QSO redshift $(z_{\rm em} \sim 1.3425)$, implying that this is also an 
``associated'' system; the absorption may originate in the quasar host galaxy. The \hi~21cm 
profile is quite wide, with FWBN$\sim 145$~\kms. Again, the absorber covering factor is 
likely to be low at the redshifted \hi~21cm line frequency; only $\sim 30$\% of the total 2.3~GHz 
quasar flux density ($\sim 0.6$~Jy) is recovered in the VCS image \citep{beasley02}. The 
\hi\ column density is N$_{\rm HI} = (3.38 \pm 0.18) \times \left[\ts/1000\right] (0.3/f) 
\times 10^{21}$~\cm.}
\item{1430$-$178, $z_{\rm abs} \sim 1.3269$: This is the weakest absorption line of our sample, with peak 
line depth $\sim 4$~mJy. The \hi~21cm absorption is again quite wide, with FWBN~$\sim 115$~\kms. The 
2.3~GHz VCS image finds a quasar flux density of $\sim 0.46$~Jy, $\sim 55$\% of the total 2.3~GHz
flux density \citep{beasley02}. The quasar also has a flat spectrum between 5~GHz and 408~MHz, suggesting
that it is likely to be core-dominated. The covering factor at the redshifted \hi~21cm frequency is thus
likely to be similar to that at 2.3~GHz, i.e. $f \sim 0.55$. The total \hi\ column density is then
N$_{\rm HI} = (4.21 \pm 0.73) \times \left[\ts/1000\right] (0.55/f) \times 10^{20}$~\cm.}
\item{2003$-$025, $z_{\rm abs} \sim 1.4106$: The \hi\ column density of this absorber has been measured 
to be log~N$_{\rm HI} = (20.54^{+0.15}_{-0.24}) \times 10^{20}$~\cm, from the damped Lyman-$\alpha$ 
absorption profile (RTN06). The integrated \hi~21cm optical depth of $(0.210 \pm 0.011)$~\kms\ 
then gives a spin temperature of $\ts = (905 \pm 380) \times f$~K, where the error is
dominated by that on the \hi\ column density. Unfortunately, no VLBI observations of the background 
quasar are available in the literature; it is thus not possible to estimate the covering factor. 
We note that the quasar is unresolved by the GMRT synthesized beam (of angular size $\sim 7''.1 
\times 5''.3$) and is hence likely to be compact. This absorber has a low \fetwo$\lambda$2600 
rest equivalent width, $\wfetwo = (0.34 \pm 0.08) \AA$.}
\item{2149+212, $z_{\rm abs} \sim 1.1727$: This is another known DLA, with an \hi\ column 
density of log~N$_{\rm HI} = 20.70^{+0.08}_{-0.10}$ (RTN06). Equation~\ref{eqn:hi} then 
yields $\ts > (2700 \times f)$~K. No VLBI information on the quasar is available in the 
literature and it is hence not possible to estimate its covering factor. }
\item{2337$-$011, $z_{\rm abs} \sim 1.3606$: This absorber has the highest peak \hi~21cm 
optical depth of our sample, $\tau_{\rm peak} \sim 0.35$. The \hi~21cm line is quite narrow, 
and is well fit by a single Gaussian of FWHM$= (5.74 \pm 0.54)$~\kms; this implies 
an upper limit to the kinetic temperature of $(720 \pm 135)$~K. The background quasar has 
a highly inverted spectrum (spectral index $\alpha \sim +0.73$), and is clearly 
core-dominated. The covering factor is thus likely to be close to unity, especially 
in view of the high peak optical depth; formally, $f > 0.35$. The \hi\ column density 
of the absorber is N$_{\rm HI} = (3.72 \pm 0.24) \times \left[\ts/1000\right] 
(1/f) \times 10^{21}$~\cm, among the highest in our sample.}
\item{2355$-$106, $z_{\rm abs} \sim 1.1727$: The \hi~21cm absorption in this system was 
originally detected by \citet{gupta07}; our new GMRT spectrum has similar sensitivity and 
velocity resolution. As noted by \citet{gupta07}, a single Gaussian provides a good fit 
to this profile; we obtain a peak optical depth of $(0.043 \pm 0.005)$ and 
FWHM$=(5.03 \pm 0.73)$~\kms, slightly deeper and narrower than that of \citet{gupta07}, 
but consistent within the errors. \citet{gupta07} argue for a high covering factor, based 
on the compactness of the source in a 5~GHz VLBA image \citep{fomalont00}. Similarly, a 
2.3~GHz VCS image recovers $\sim 90$\% of the total 2.3~GHz quasar flux density 
\citep{beasley02}, and the source has a highly inverted spectrum, indicating that it is 
core-dominated. The covering factor is thus likely to be $\sim 0.9$ at the redshifted 
\hi~21cm line frequency, yielding an \hi\ column density of N$_{\rm HI} = (5.18 \pm 0.19) 
\times \left[\ts/1000\right] (0.9/f) \times 10^{20}$~\cm.} 
\item{0237$-$233, $z_{\rm abs} \sim 1.6724$: The absorption feature was seen in data from 
two separate GBT observing sessions in November~2006, close to the expected redshifted \hi~21cm 
line frequency from the \mgtwo\ redshift. Although attempts were made to confirm the line in
August 2008 with the GBT, the RFI environment at these frequencies was found to have significantly 
deteriorated and it was not possible to obtain useful data; we hence list this as a tentative 
detection. \citet{srianand07} discuss high-frequency ($2.3-8.4$~GHz) VLBI studies of this 
source from the literature (e.g. \citealp{fomalont00}) and conclude that most of the flux 
density arises from two components, separated by $\sim 10$~mas. The lowest frequency at 
which a VLBI study has been carried out is 1.6~GHz, where \citet{hodges84} found no evidence 
for extended structure, obtaining a size of $\sim 6 \times 3$~mas. The covering 
factor is thus likely to be high, $f \sim 1$. The tentative detection of \hi~21cm 
absorption in this system yields the lowest \hi\ column density of all our detections, 
N$_{\rm HI} \sim (1.39 \pm 0.29) \times \left[\ts/1000\right](1/f) \times 10^{20}$~\cm; 
note that the absorber must have a high spin temperature, if it is indeed a DLA.}
\end{enumerate}

\begin{figure*}
\begin{centering}
\epsfig{file=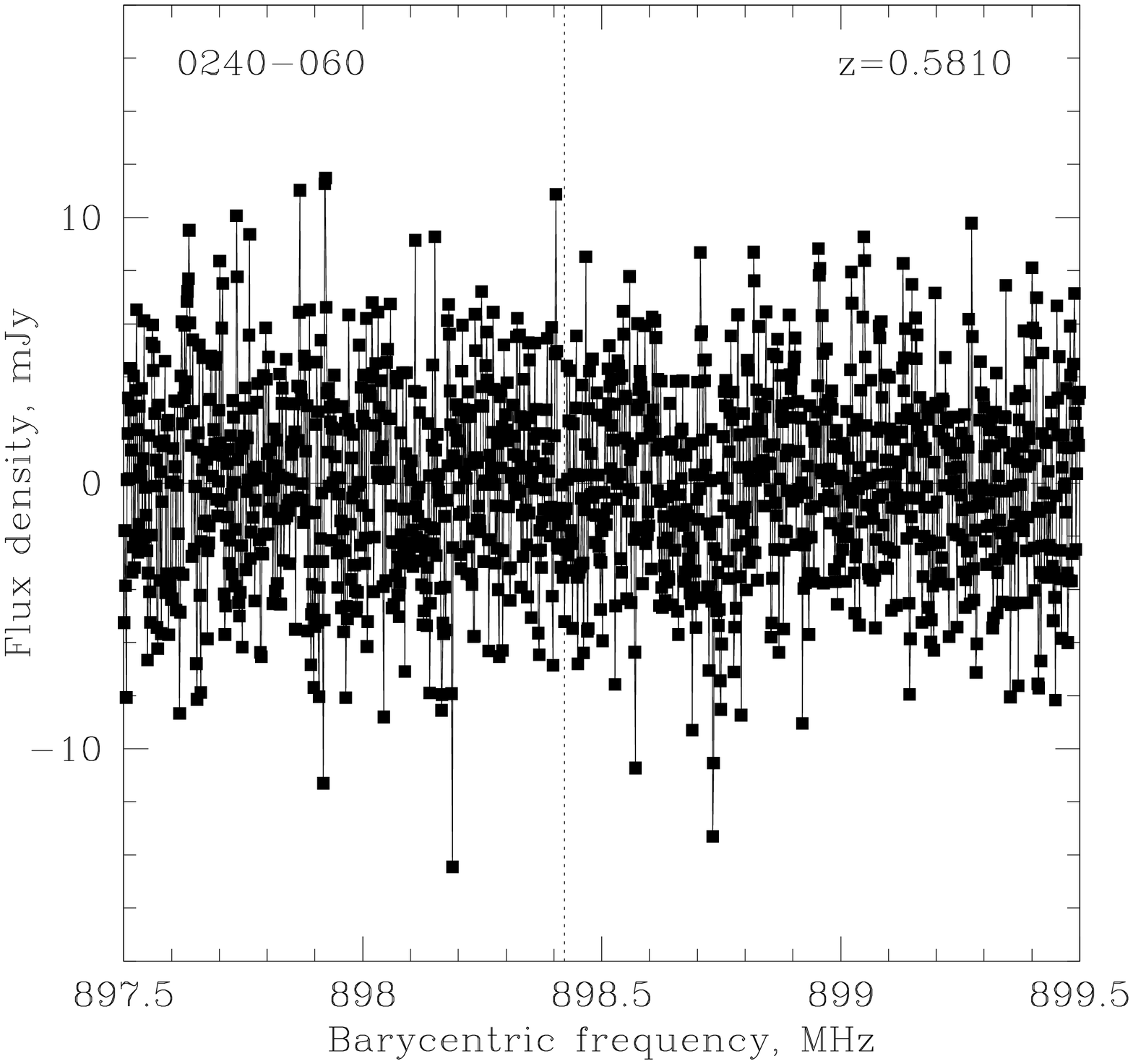,height=1.65in}
\epsfig{file=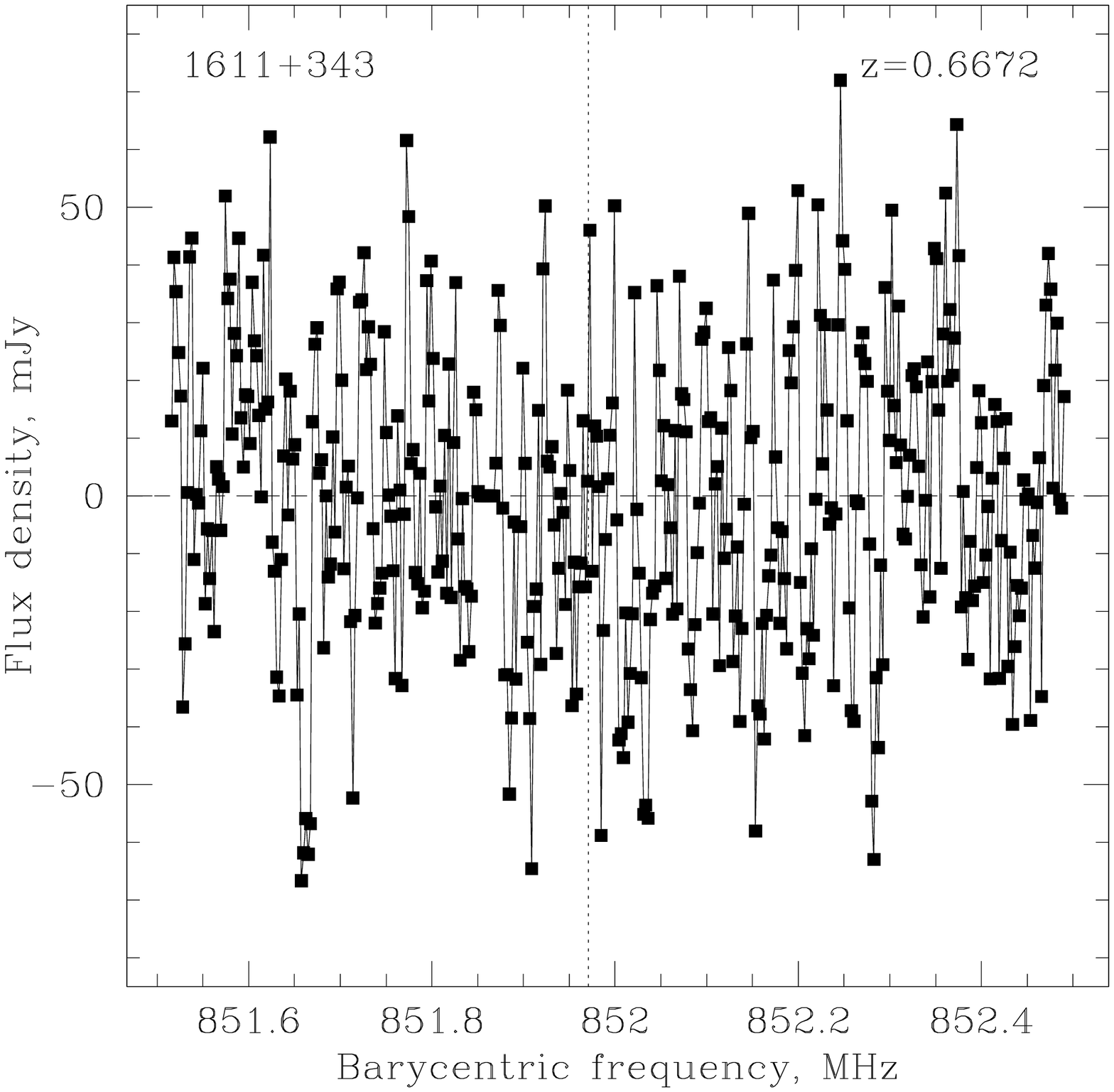,height=1.65in}
\epsfig{file=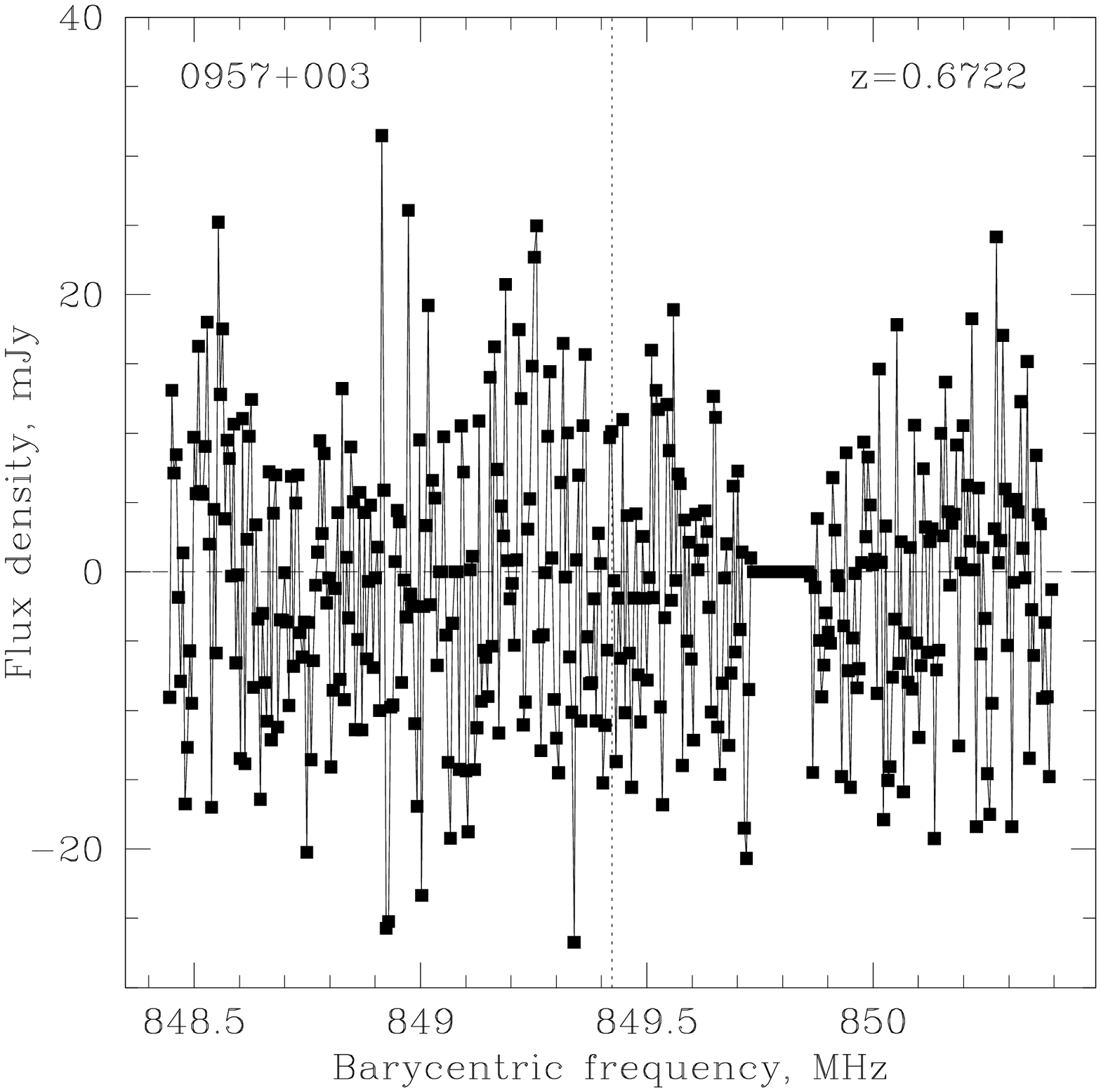,height=1.65in}
\epsfig{file=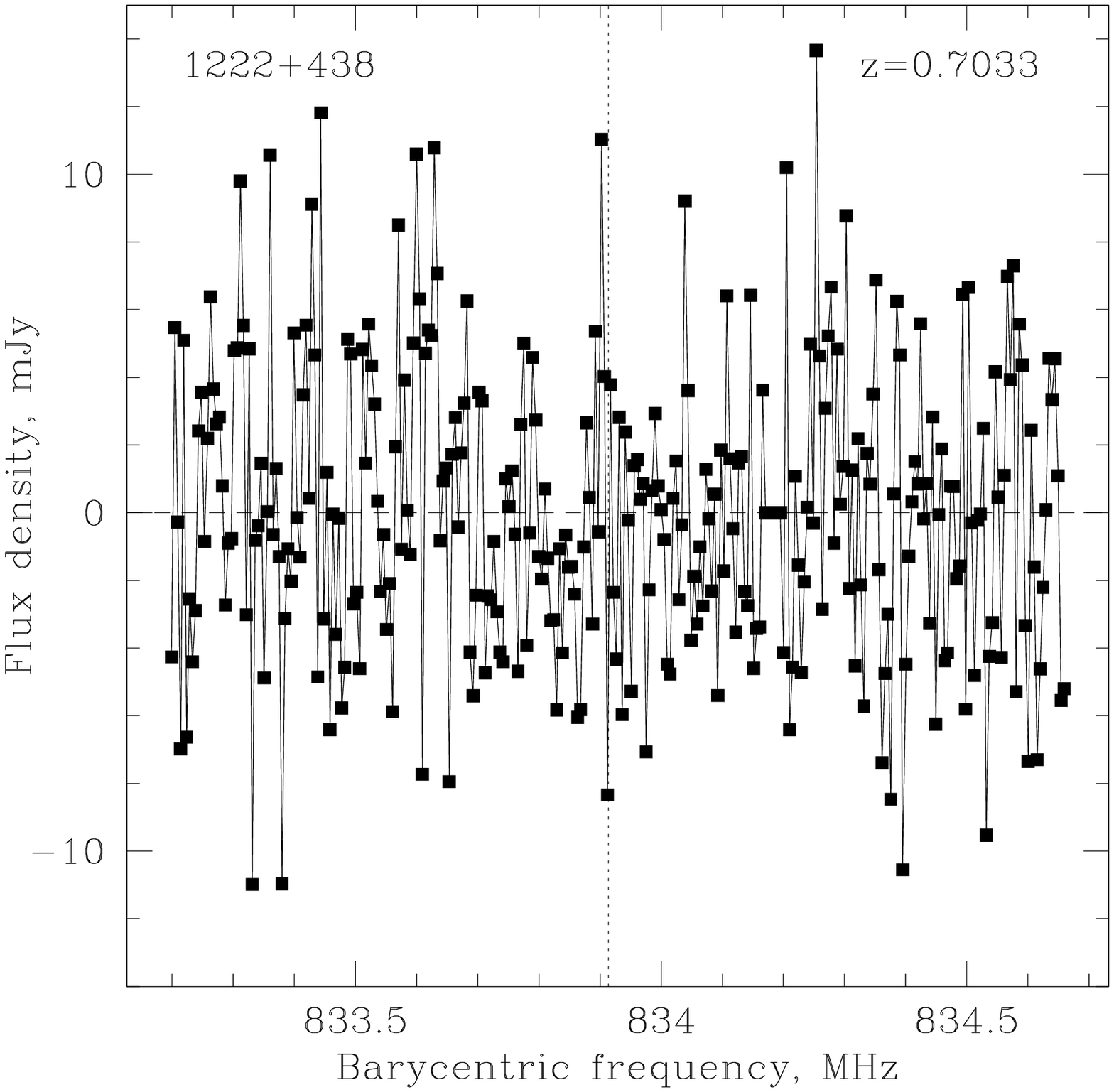,height=1.65in}
\epsfig{file=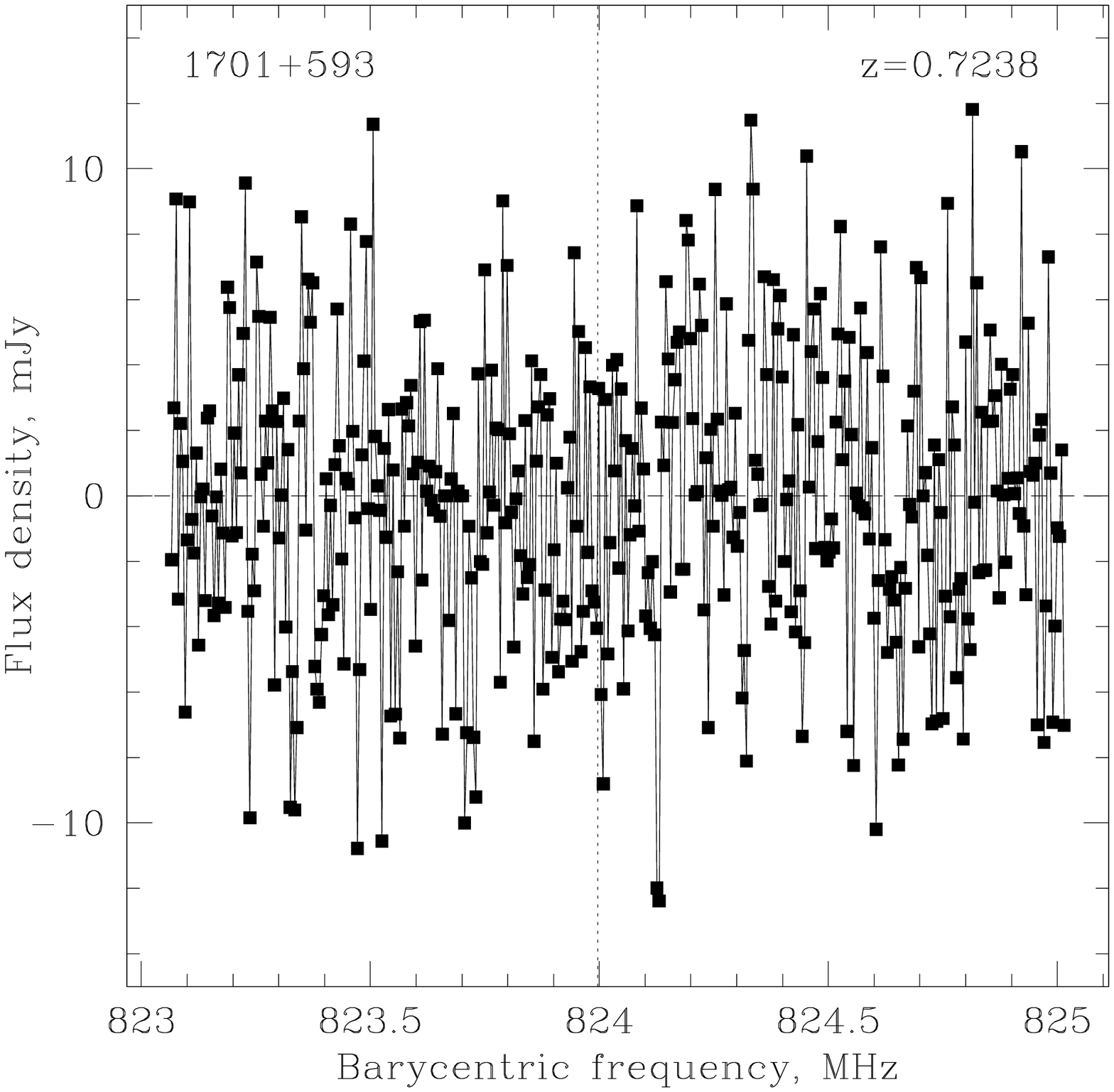,height=1.65in}
\epsfig{file=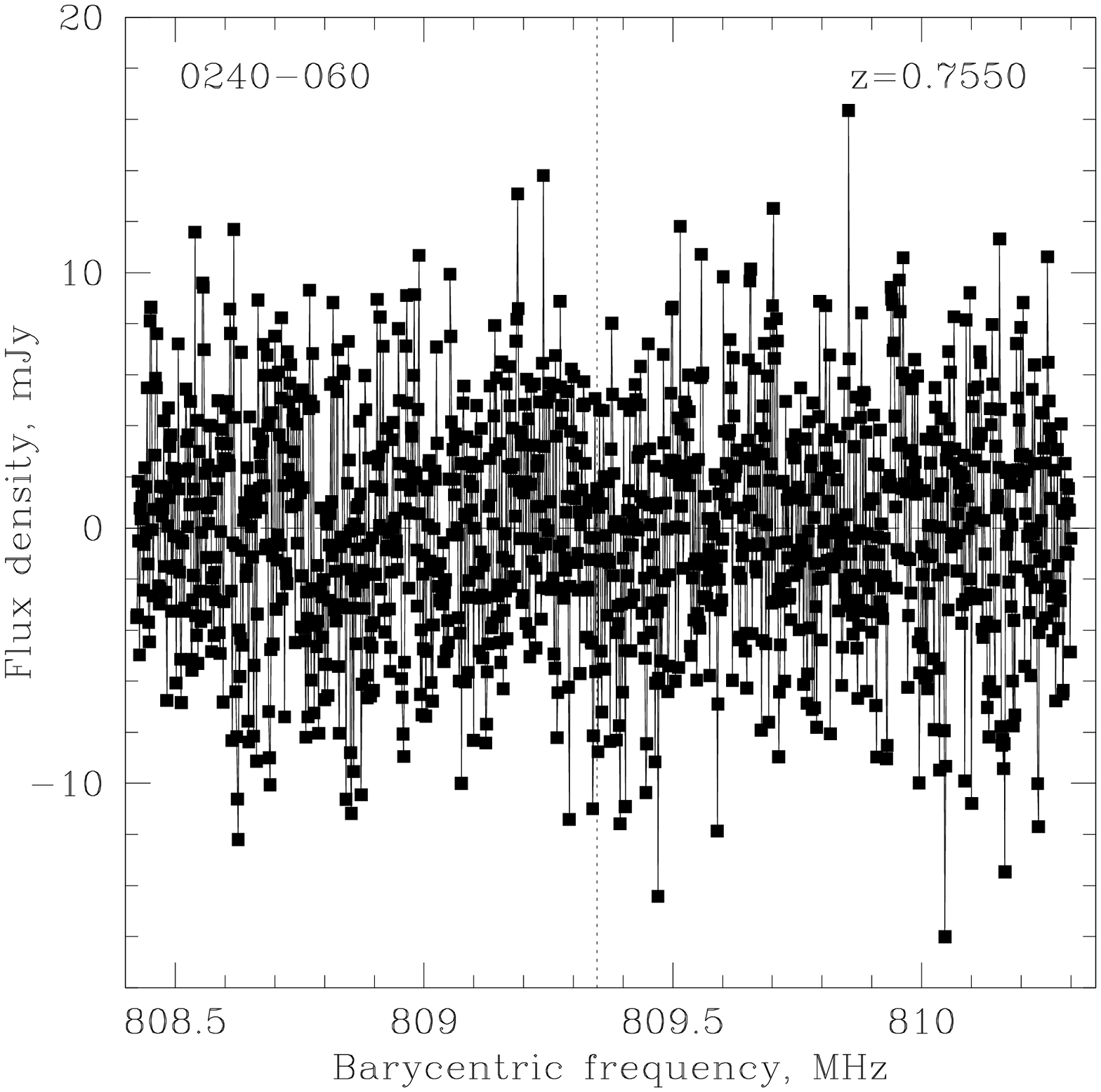,height=1.65in}
\epsfig{file=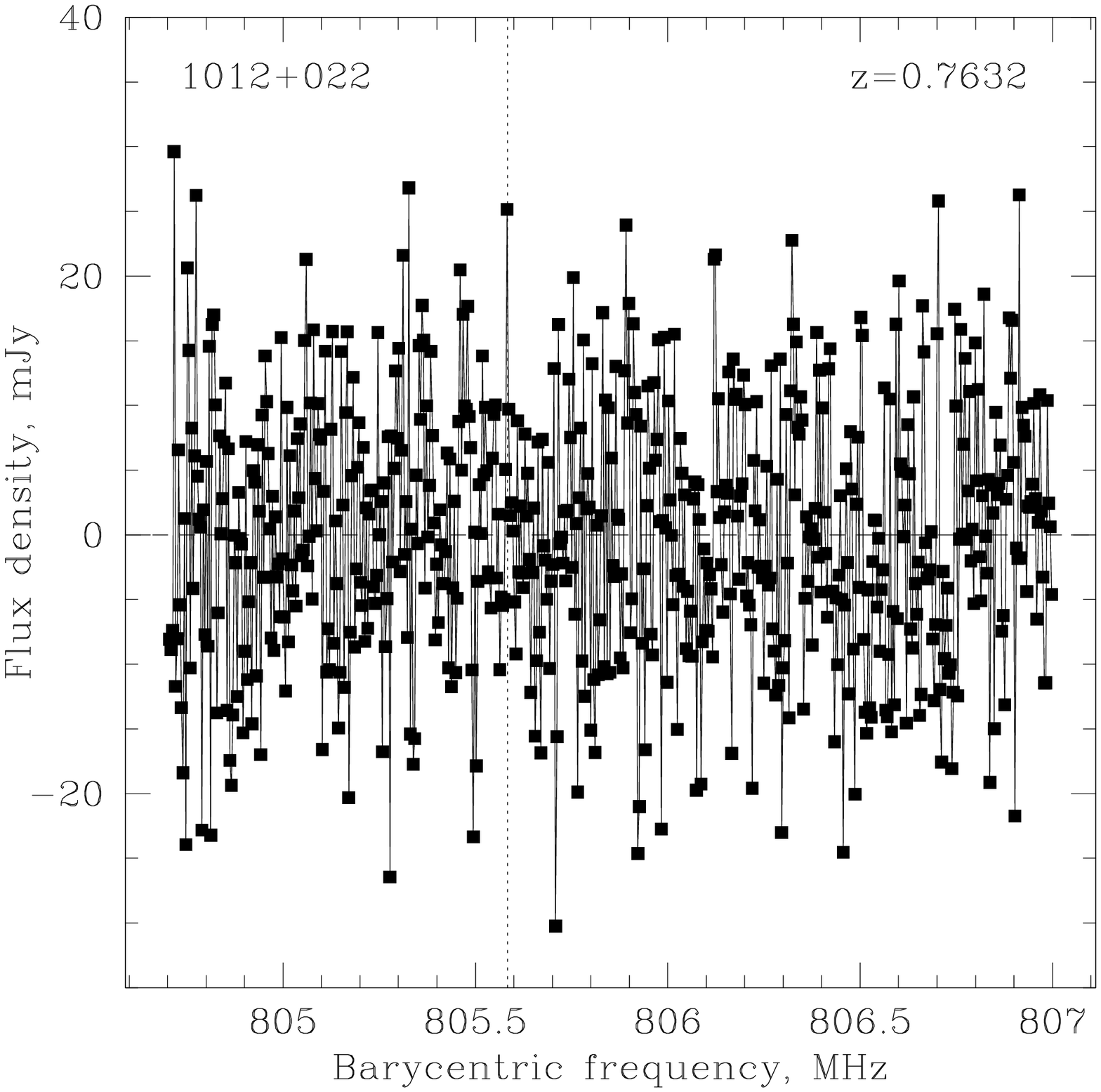,height=1.65in}
\epsfig{file=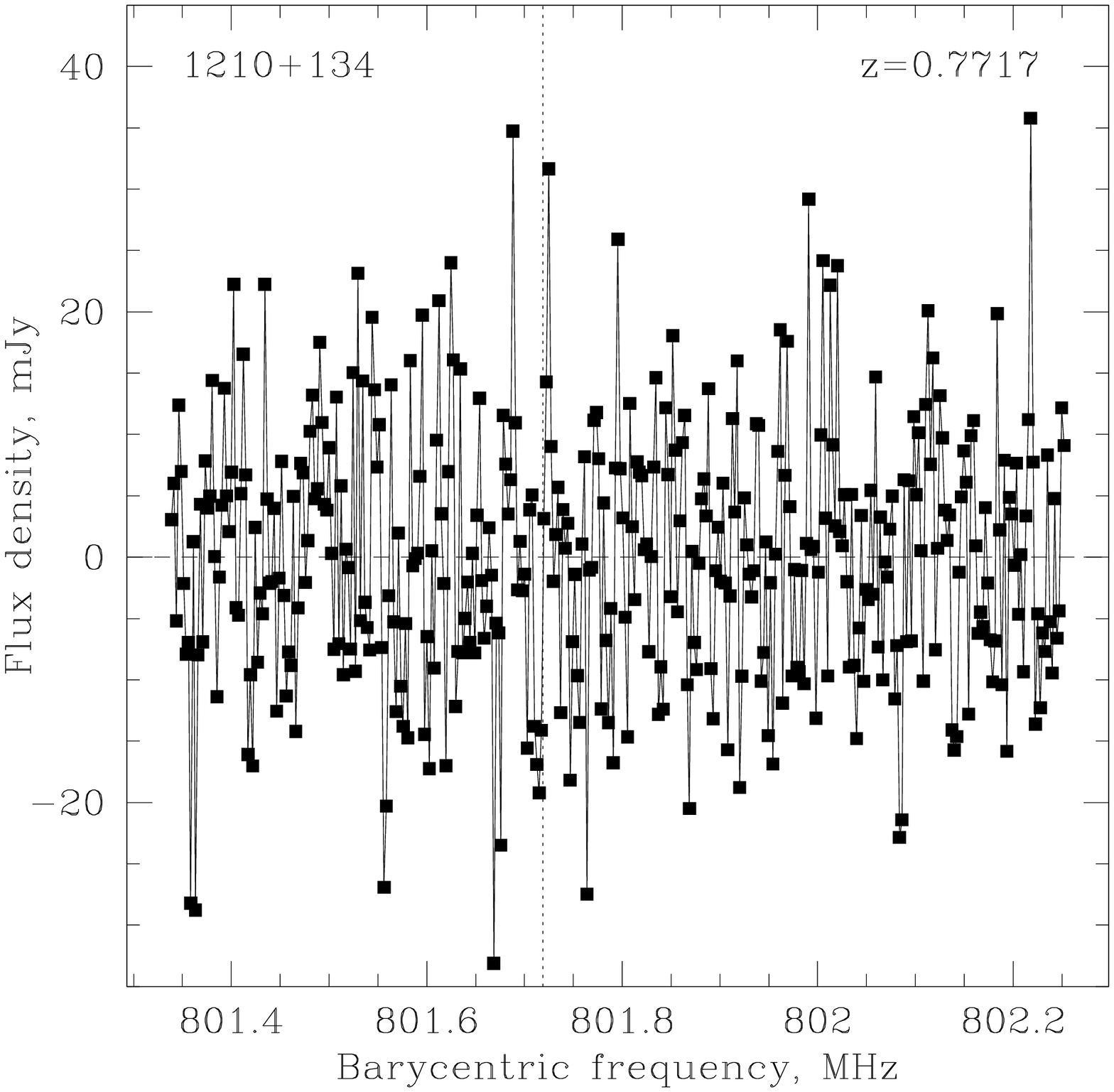,height=1.65in}
\epsfig{file=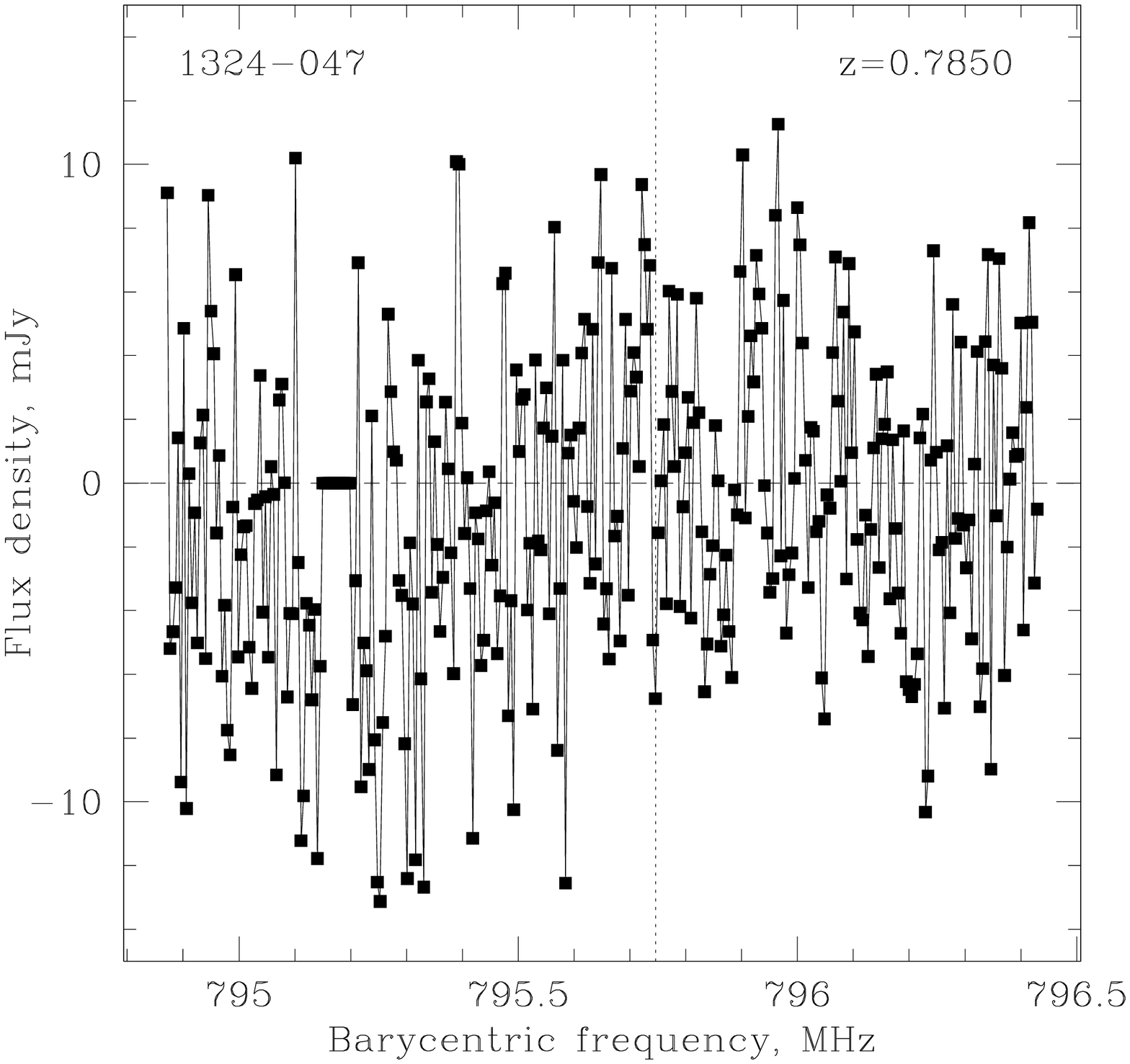,height=1.65in}
\epsfig{file=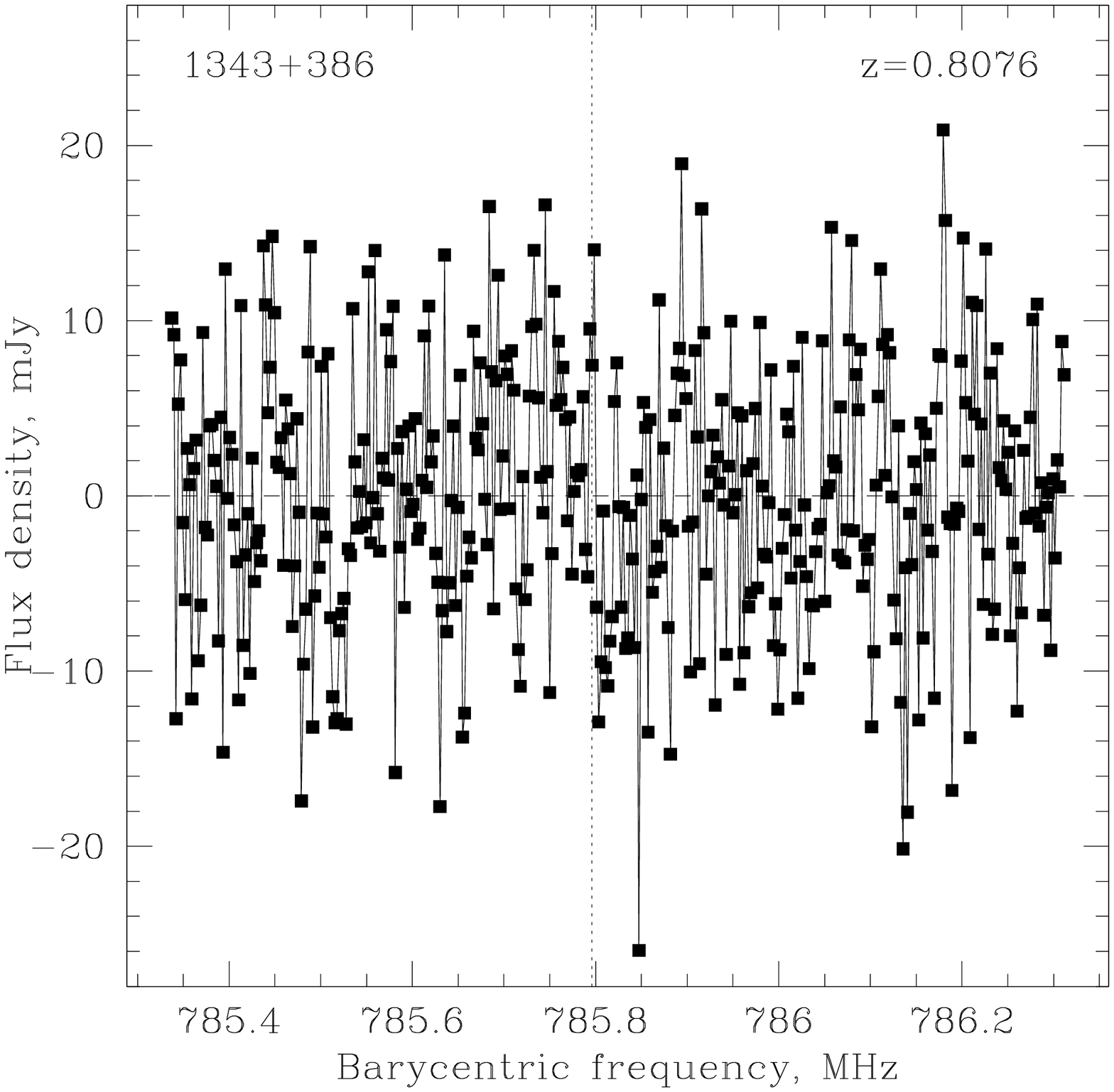,height=1.65in}
\epsfig{file=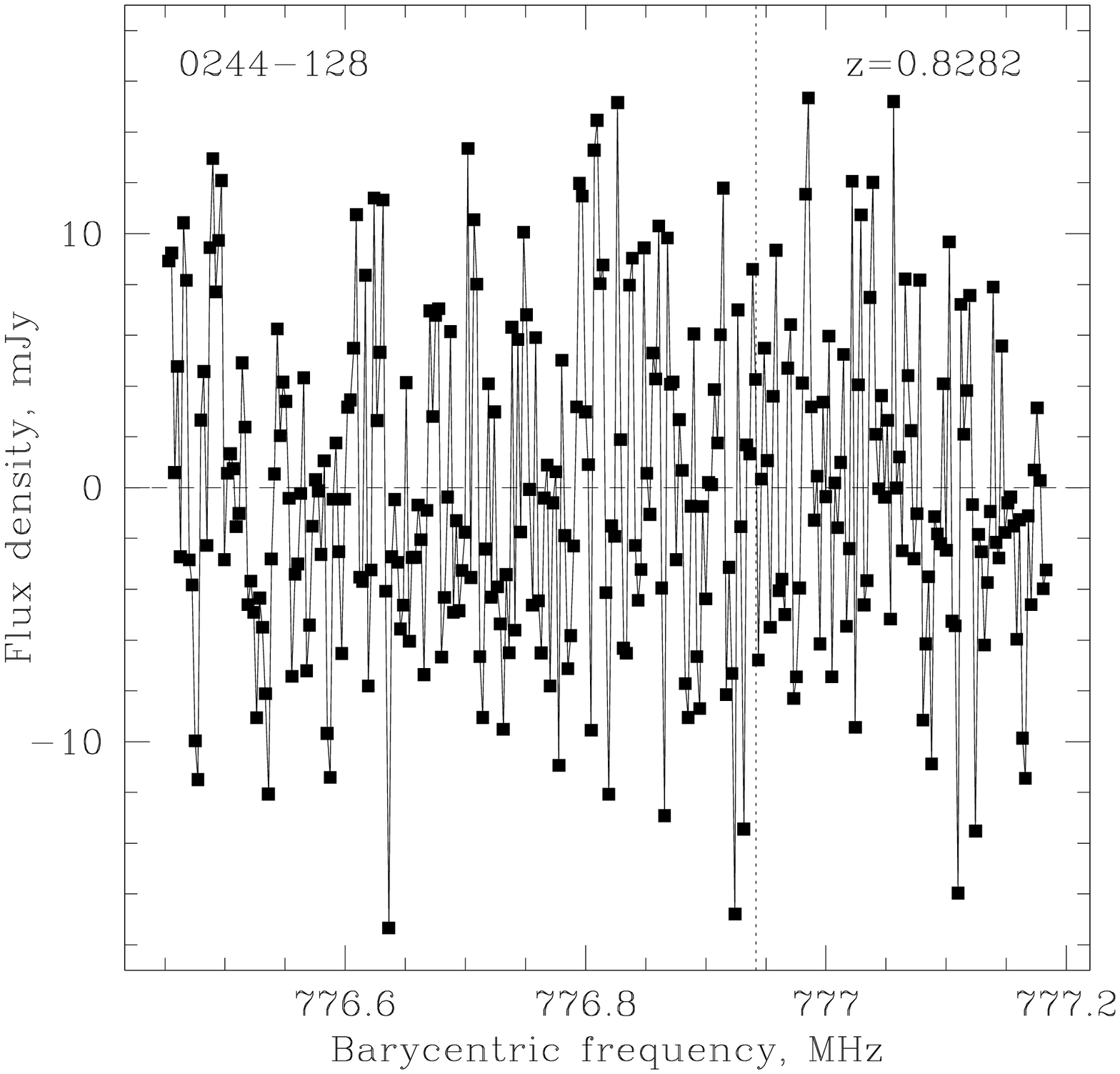,height=1.65in}
\epsfig{file=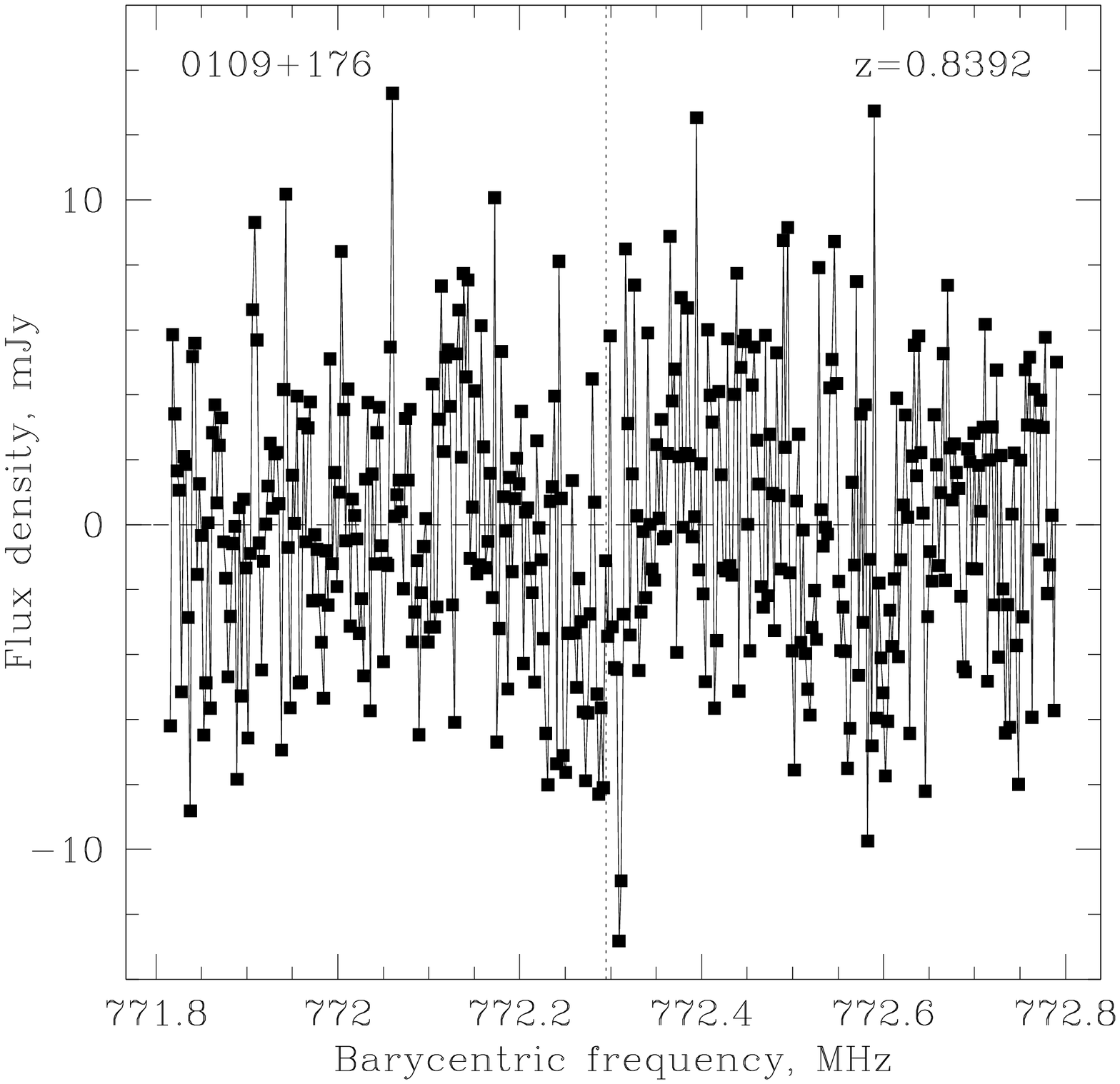,height=1.65in}
\epsfig{file=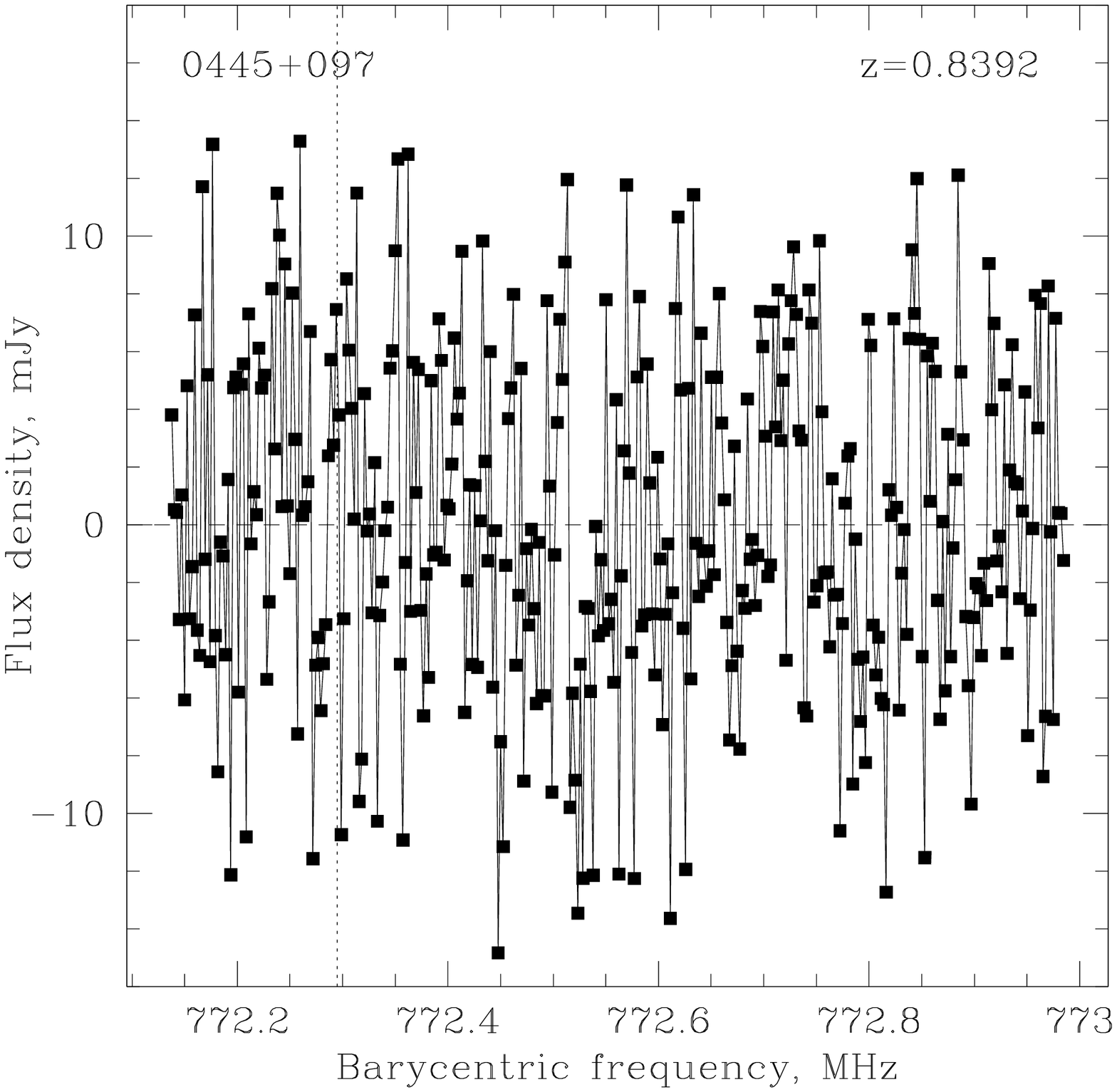,height=1.65in}
\epsfig{file=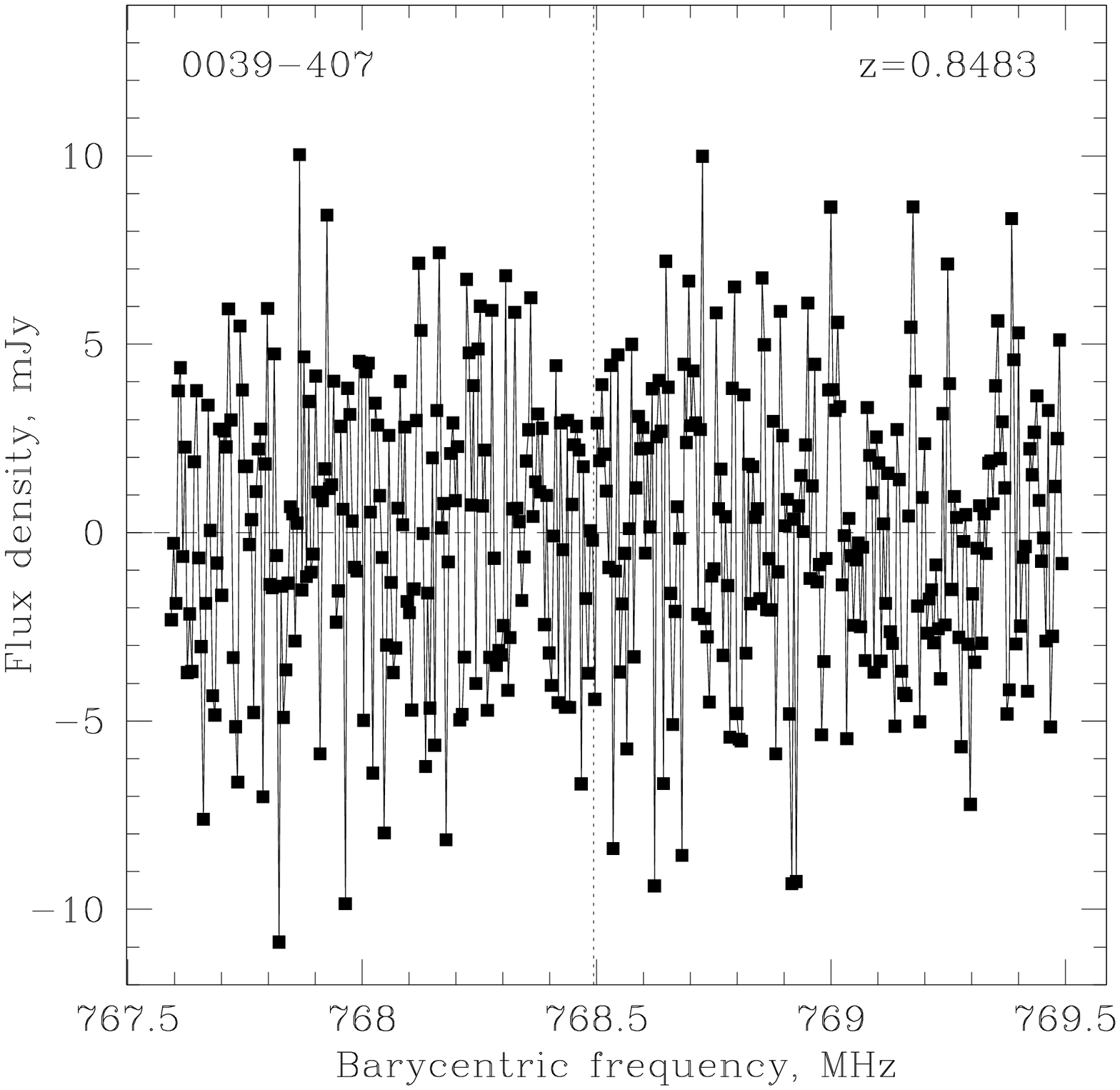,height=1.65in}
\epsfig{file=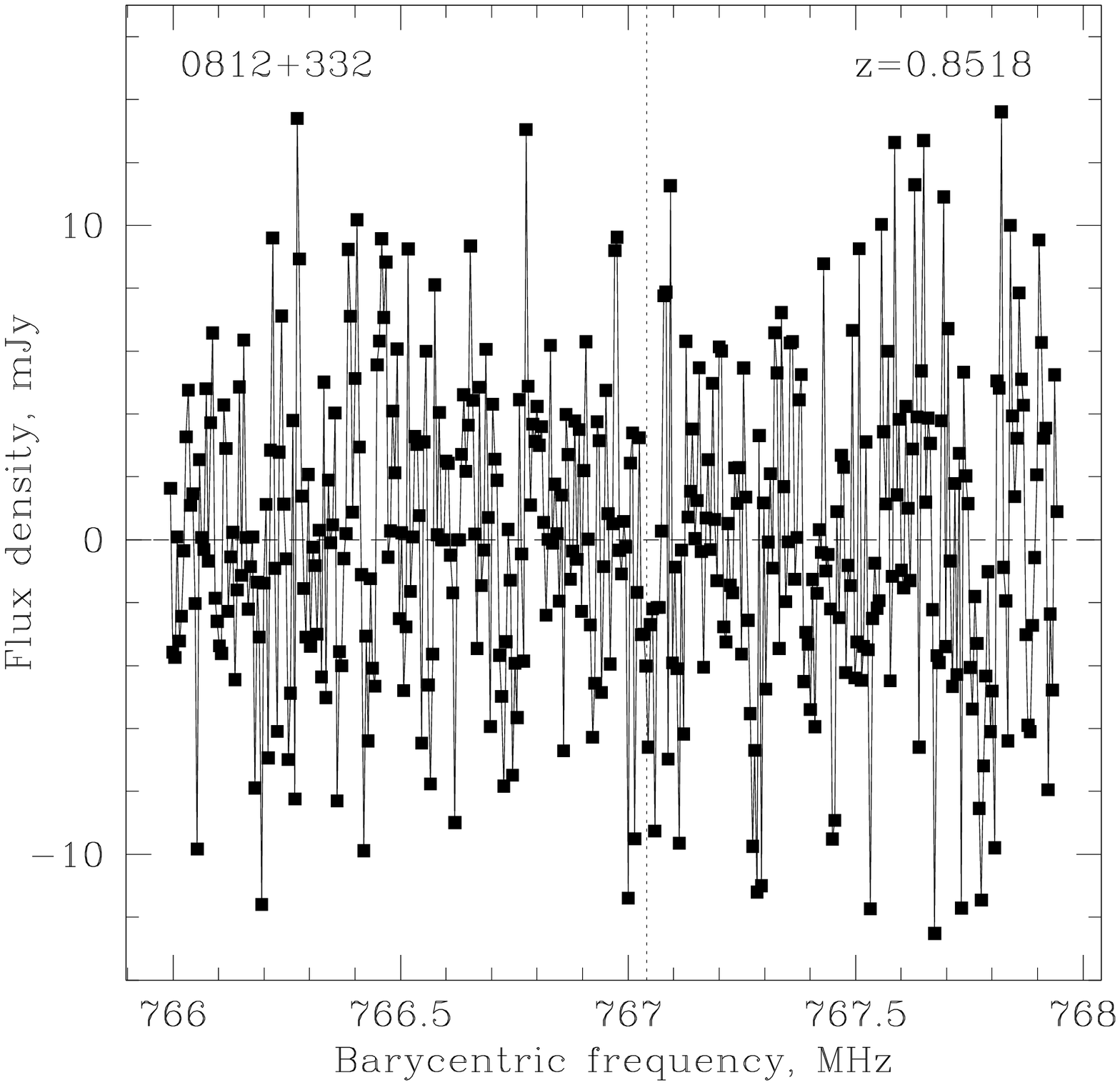,height=1.65in}
\epsfig{file=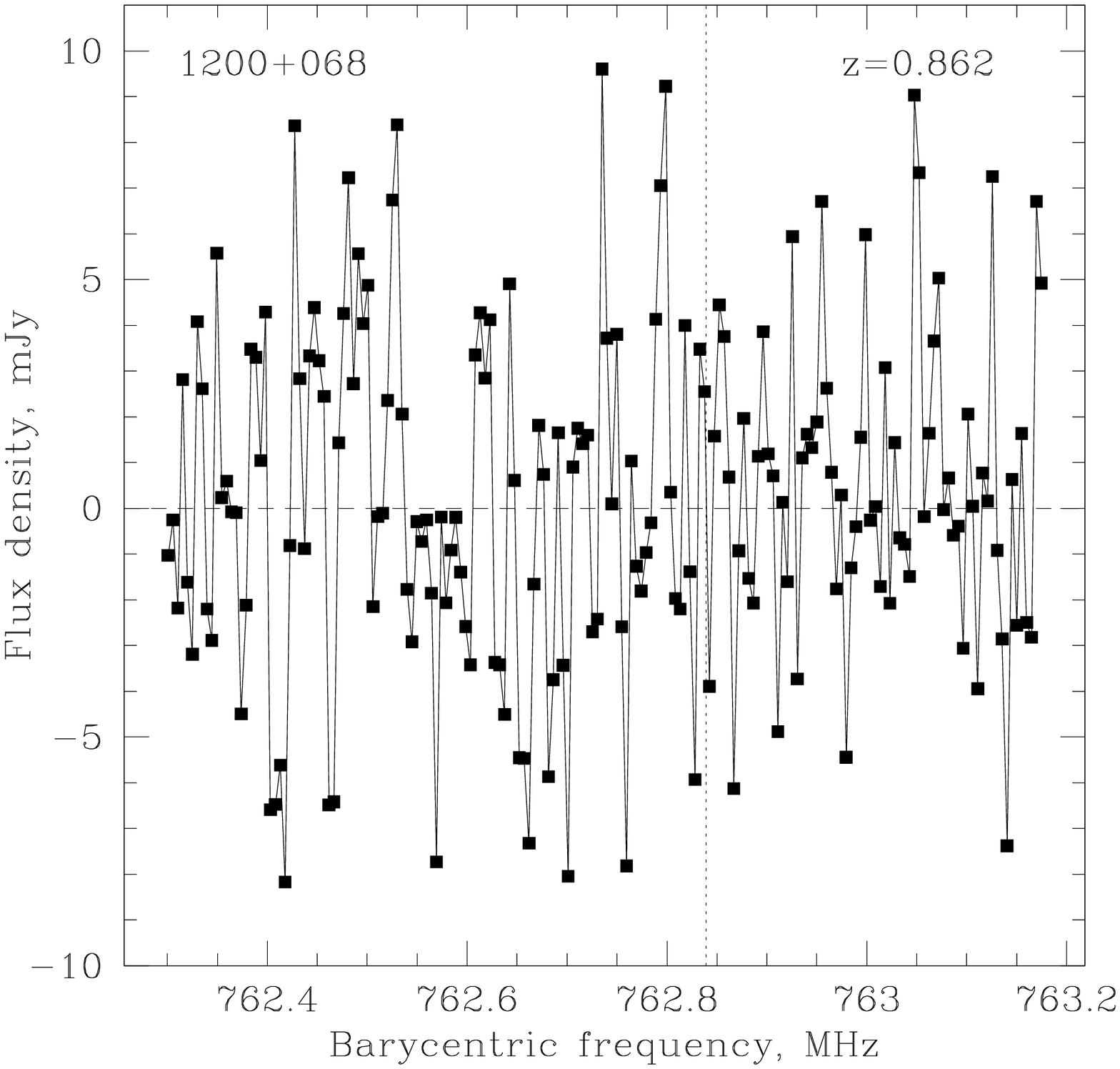,height=1.65in}
\epsfig{file=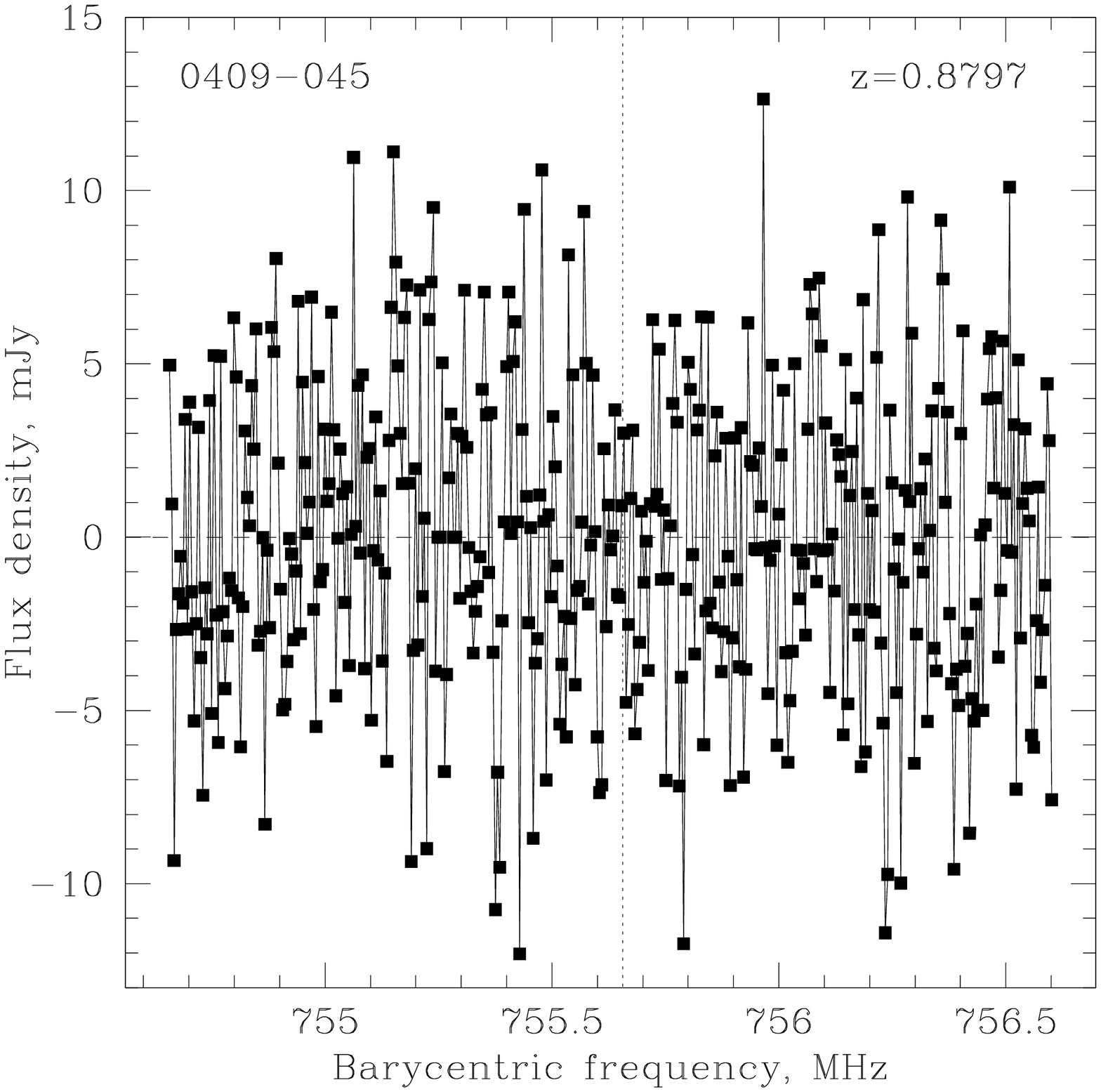,height=1.65in}
\epsfig{file=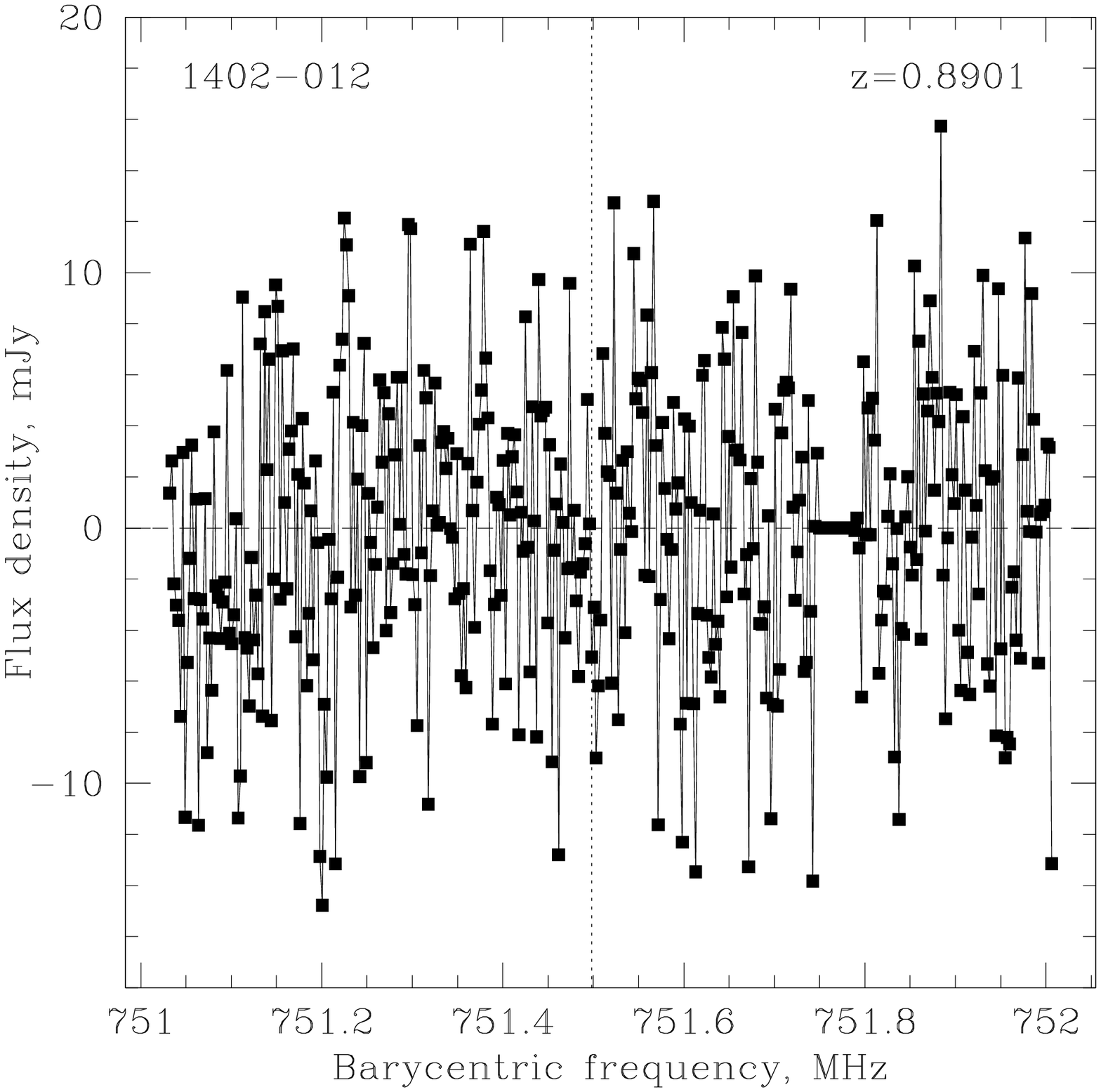,height=1.65in}
\epsfig{file=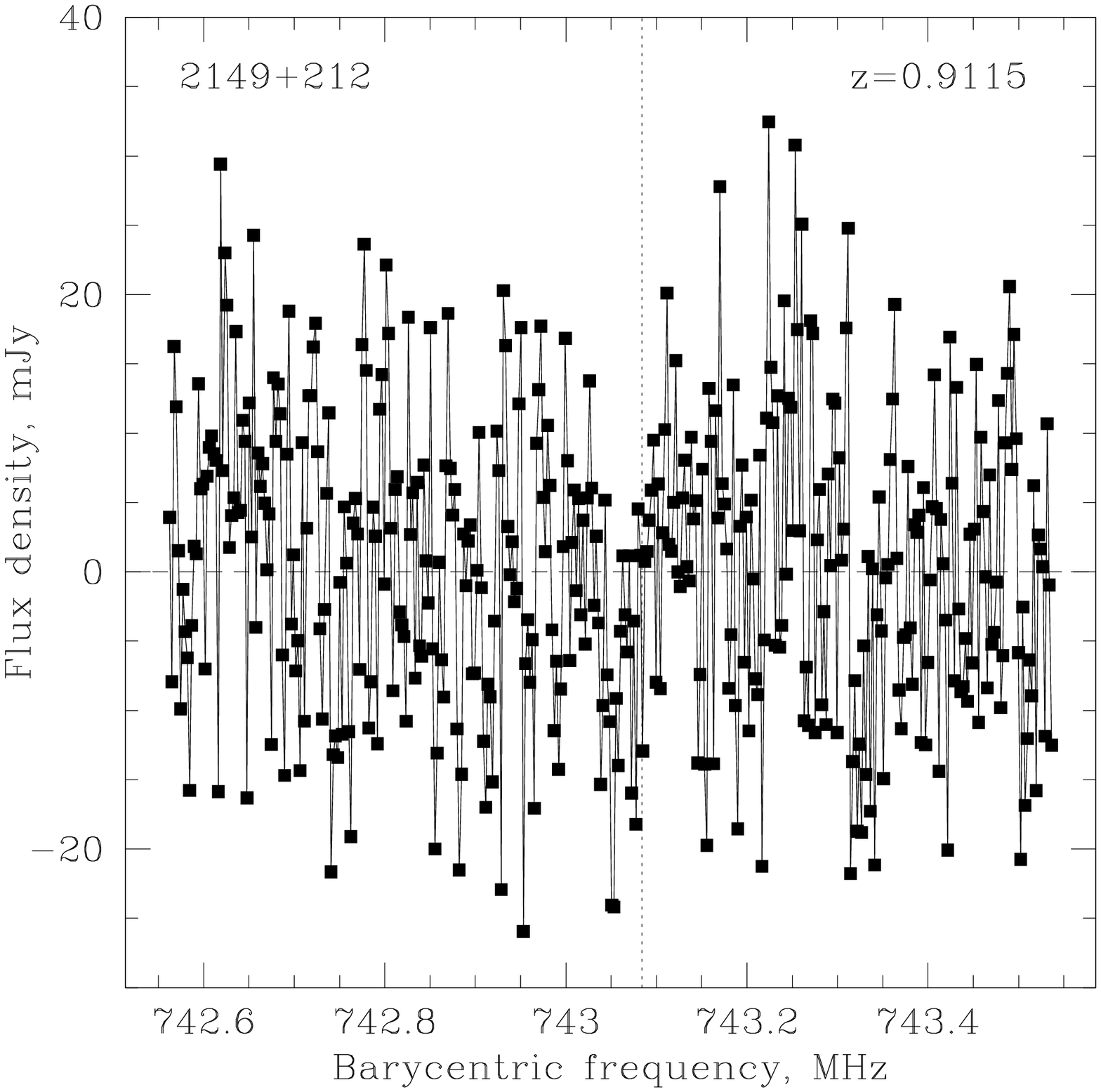,height=1.65in}
\epsfig{file=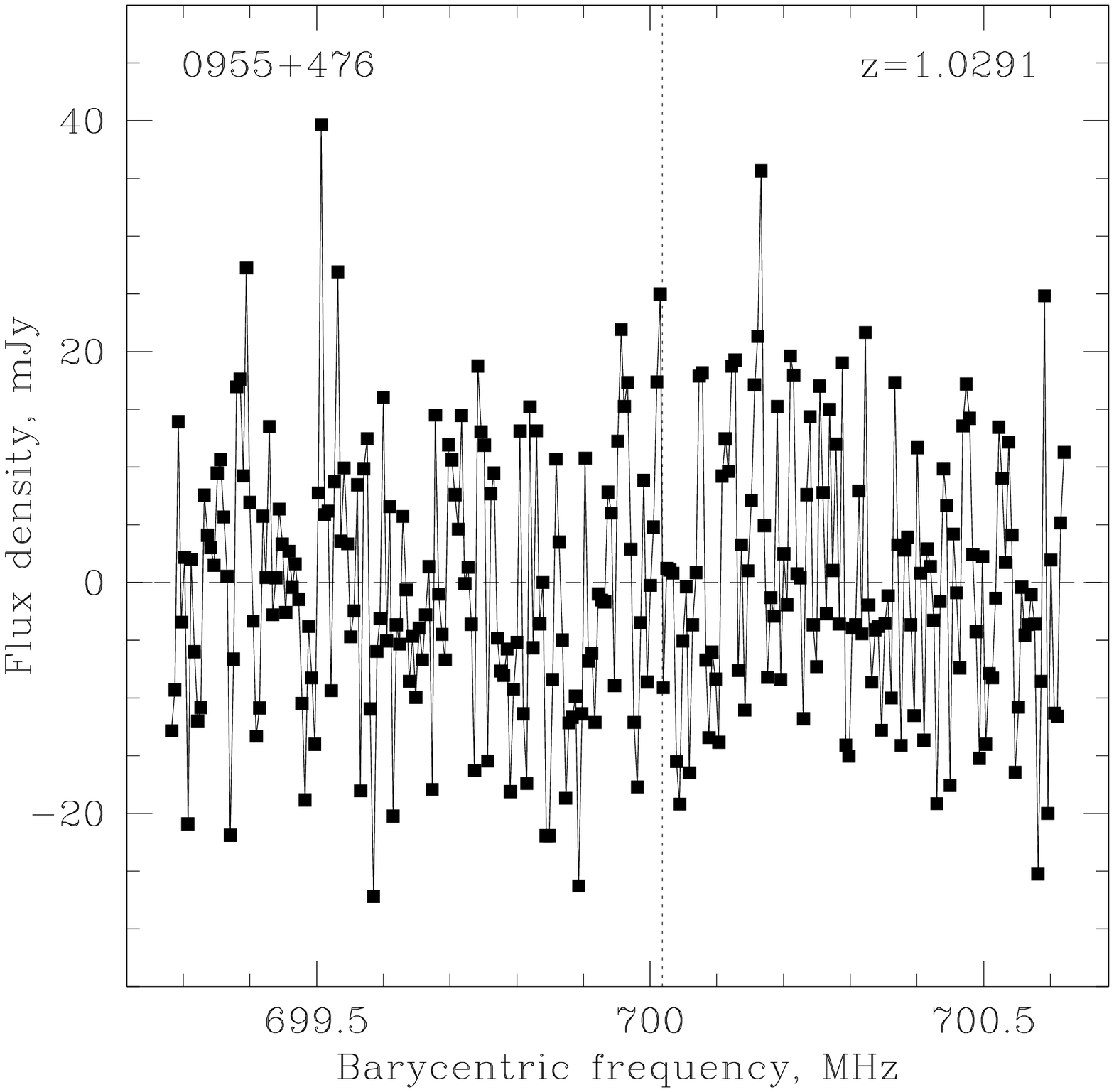,height=1.65in}
\caption{The GMRT and GBT spectra for the 35 non-detections of \hi~21cm absorption, 
in order of increasing \mgtwo$\lambda$2796 absorption redshift. The expected \hi~21cm 
line frequency, based on the \mgtwo$\lambda$2796 redshift, is indicated by the 
dashed line in each panel.}
\label{fig:nondetect}
\end{centering}
\end{figure*}

\setcounter{figure}{1}
\begin{figure*}
\begin{centering}
\epsfig{file=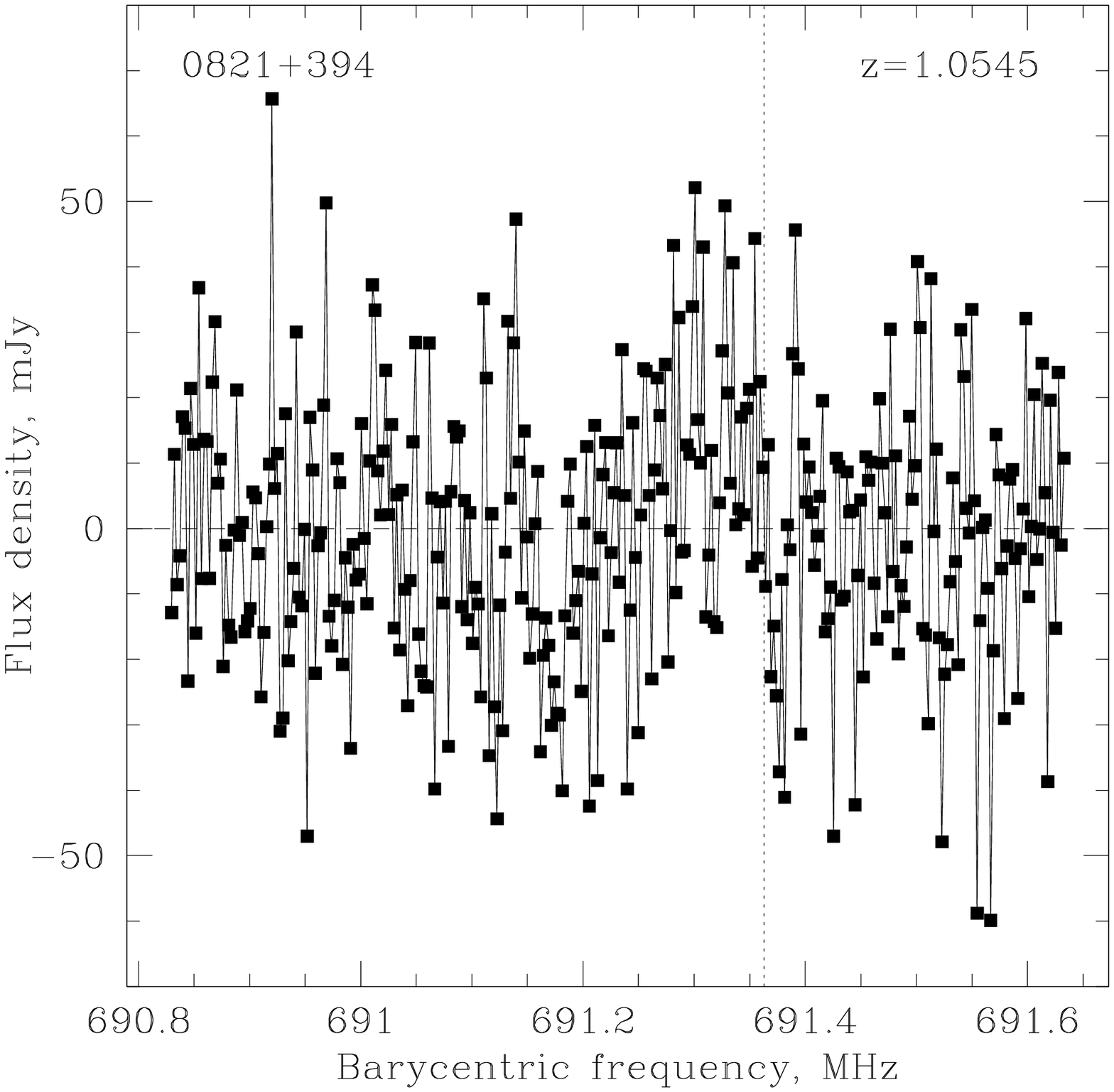,height=1.65in}
\epsfig{file=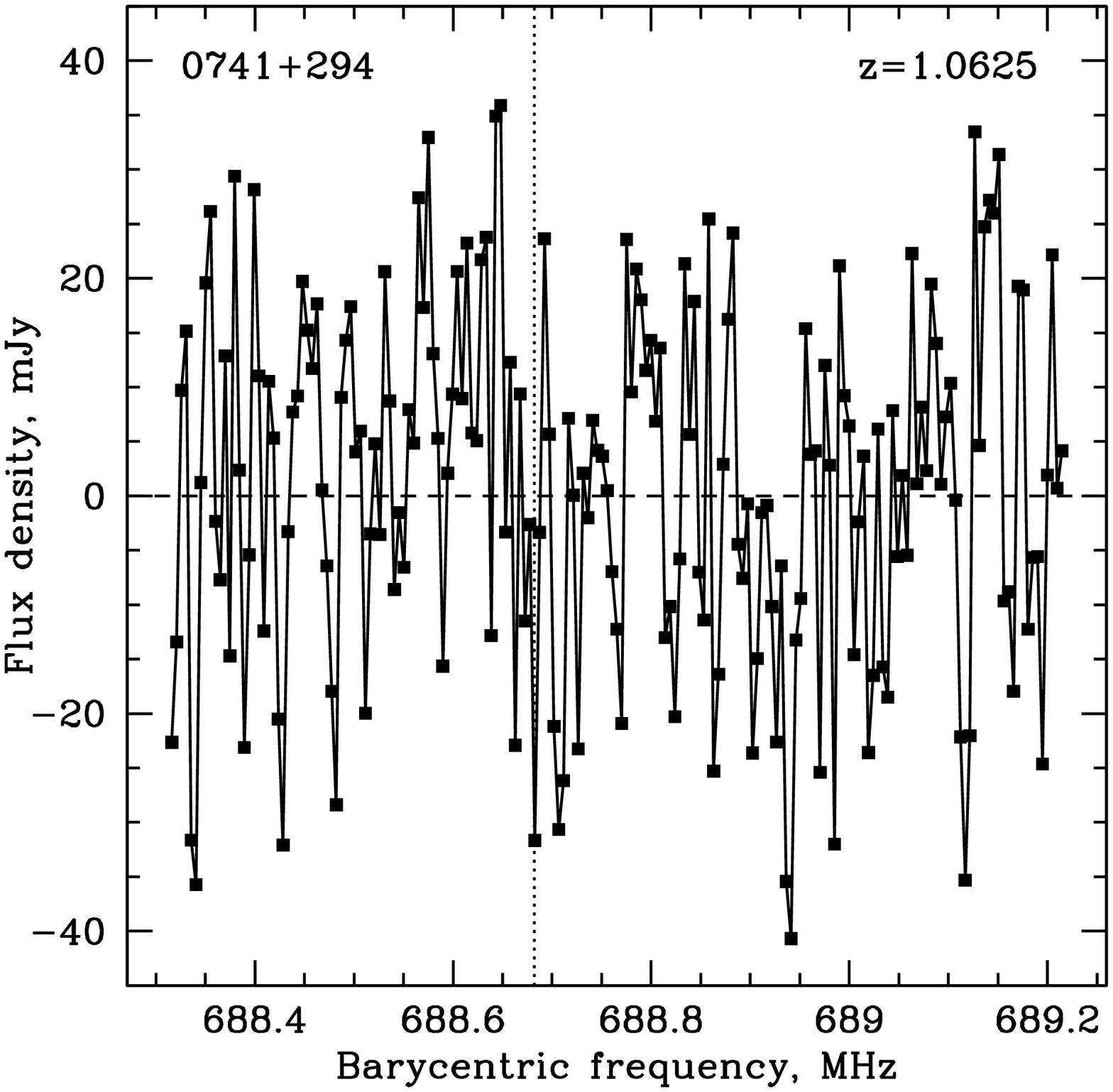,height=1.65in}
\epsfig{file=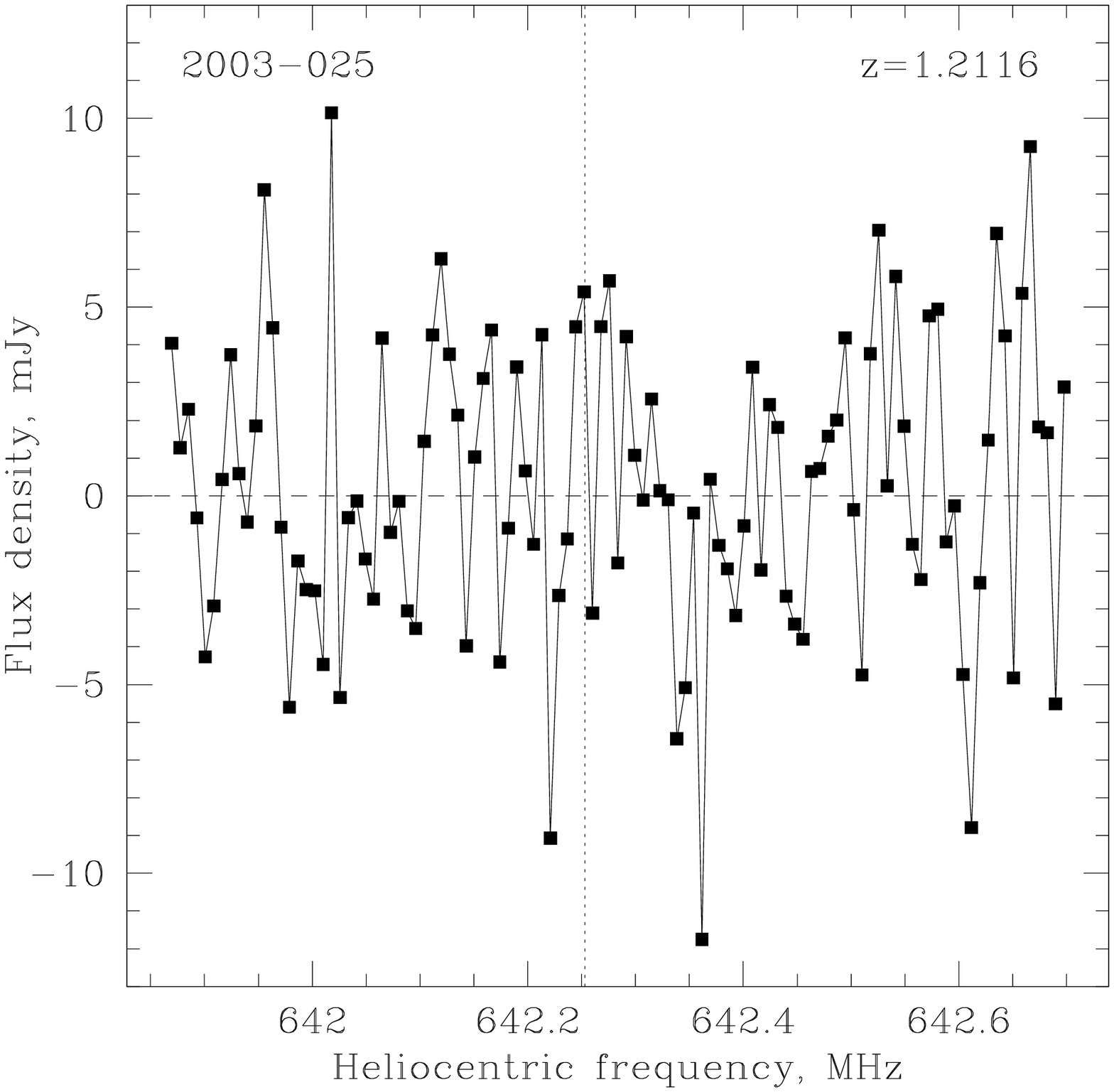,height=1.65in}
\epsfig{file=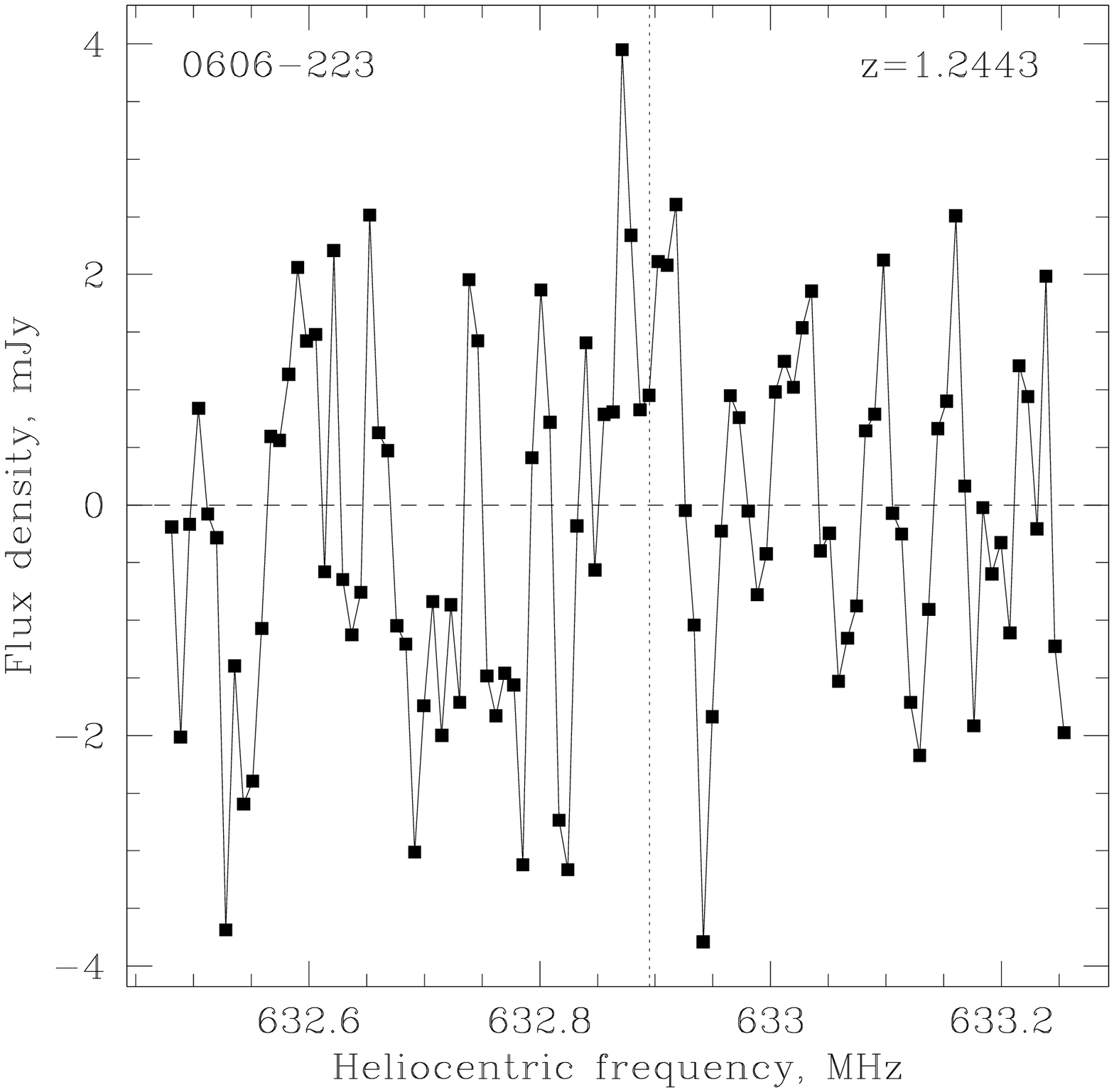,height=1.65in}
\epsfig{file=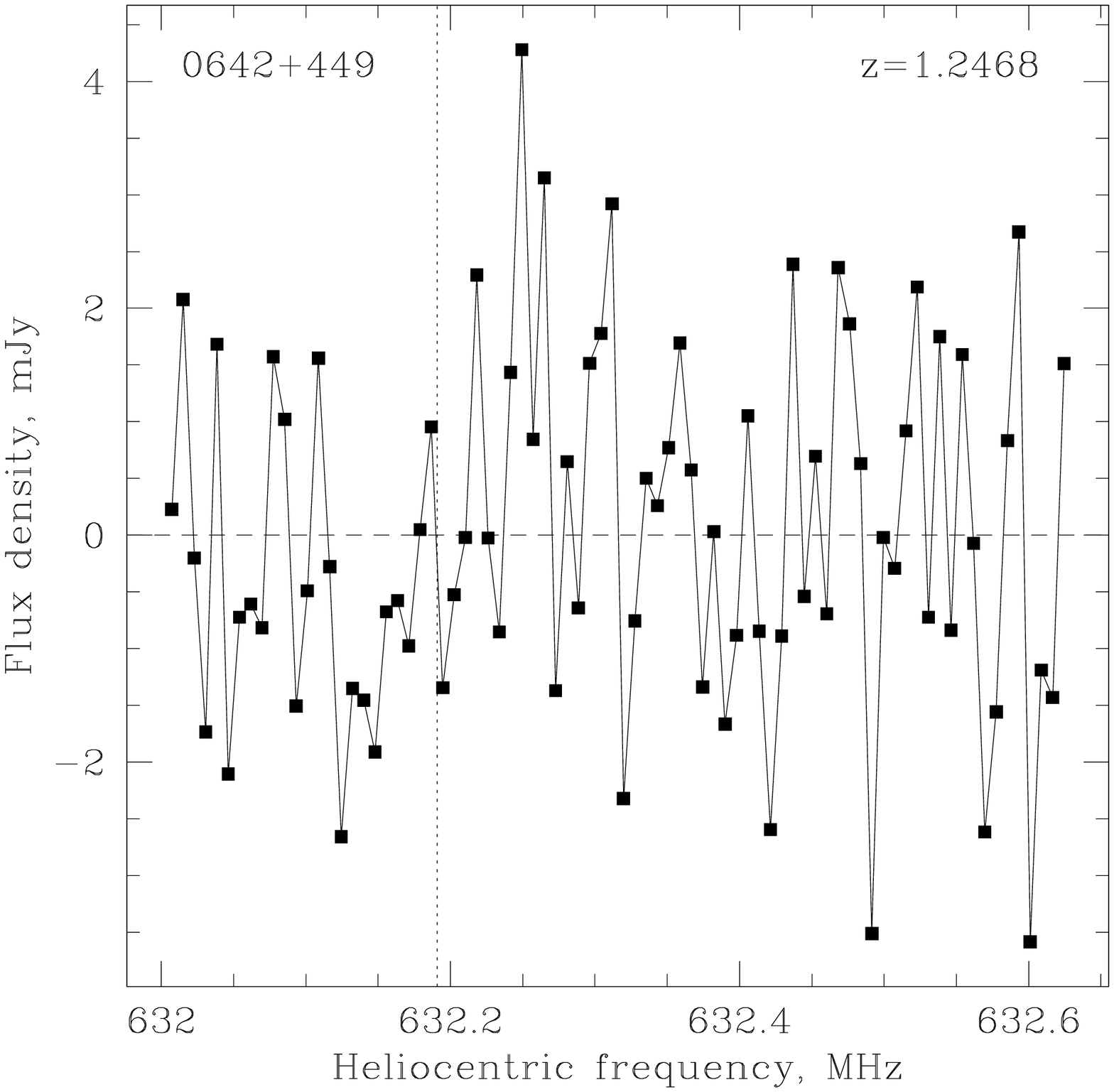,height=1.65in}
\epsfig{file=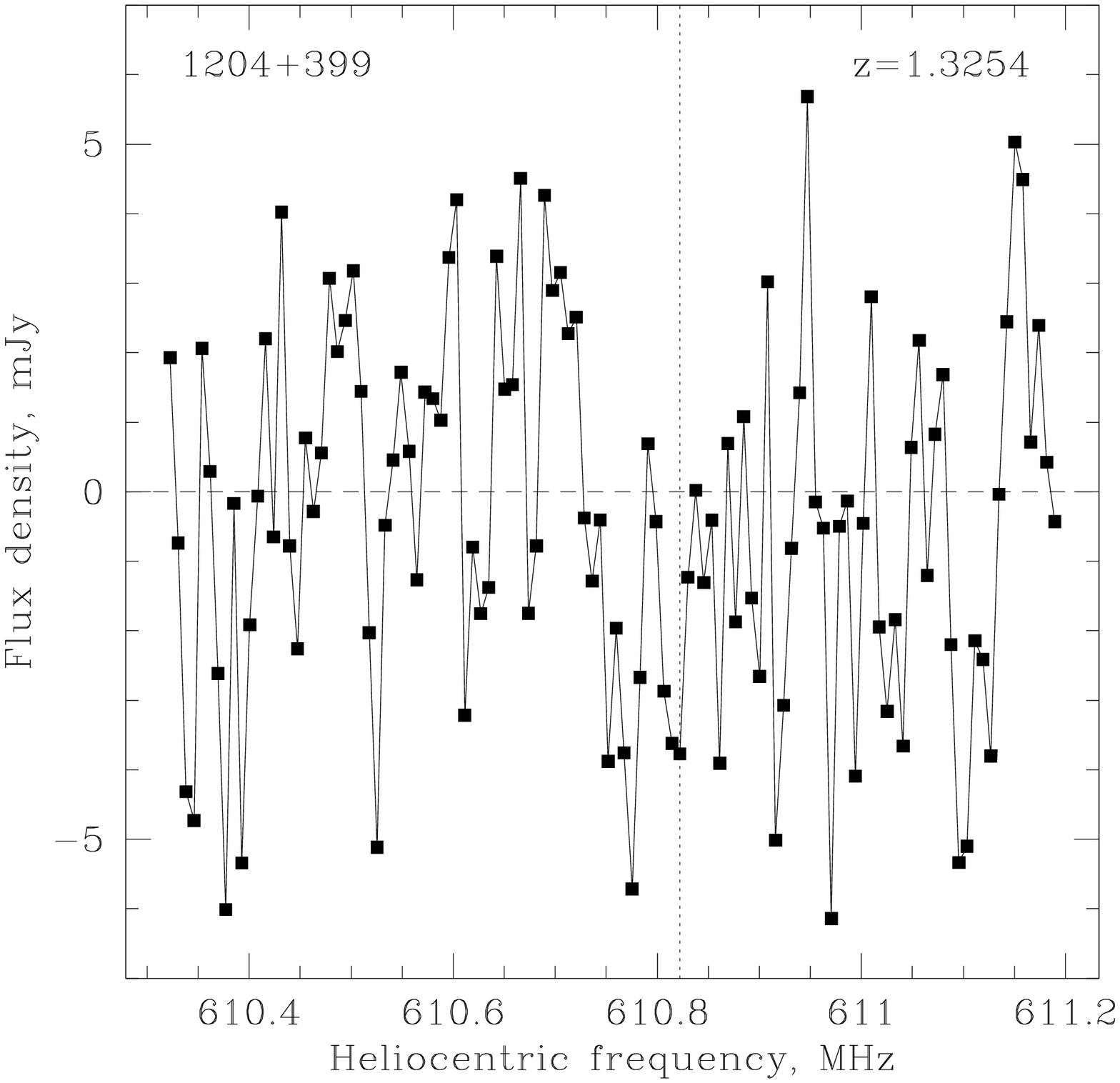,height=1.65in}
\epsfig{file=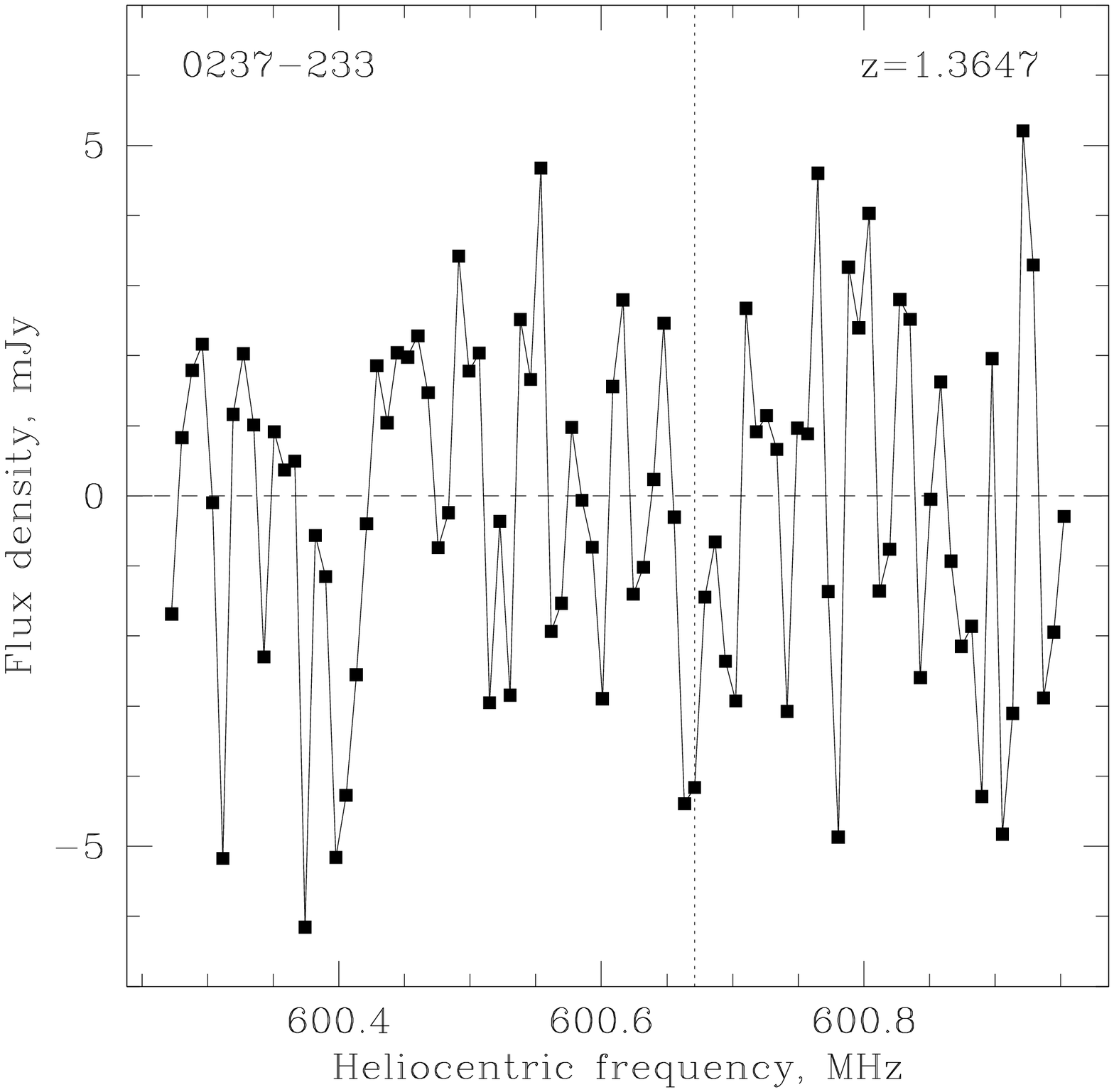,height=1.65in}
\epsfig{file=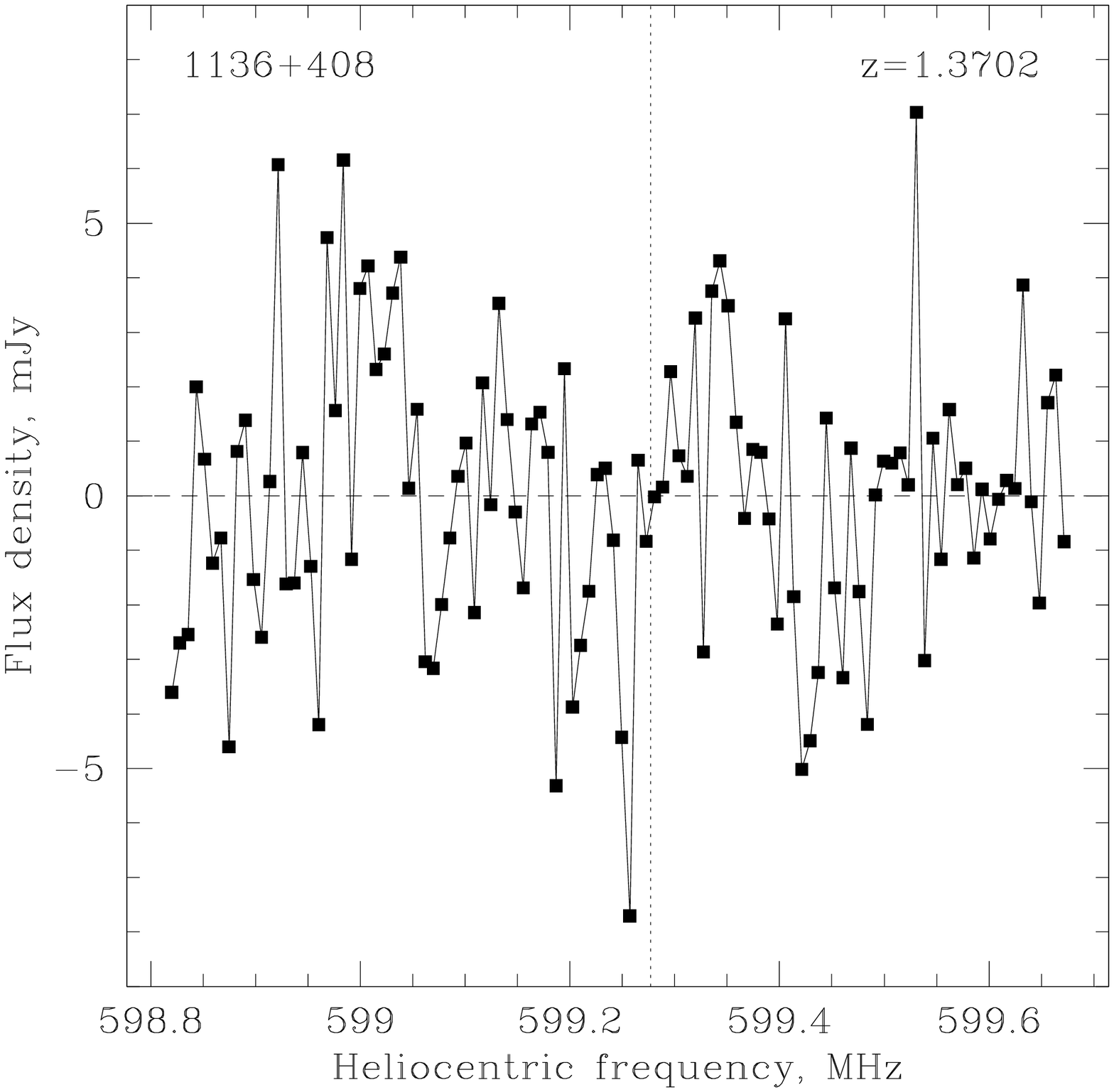,height=1.65in}
\epsfig{file=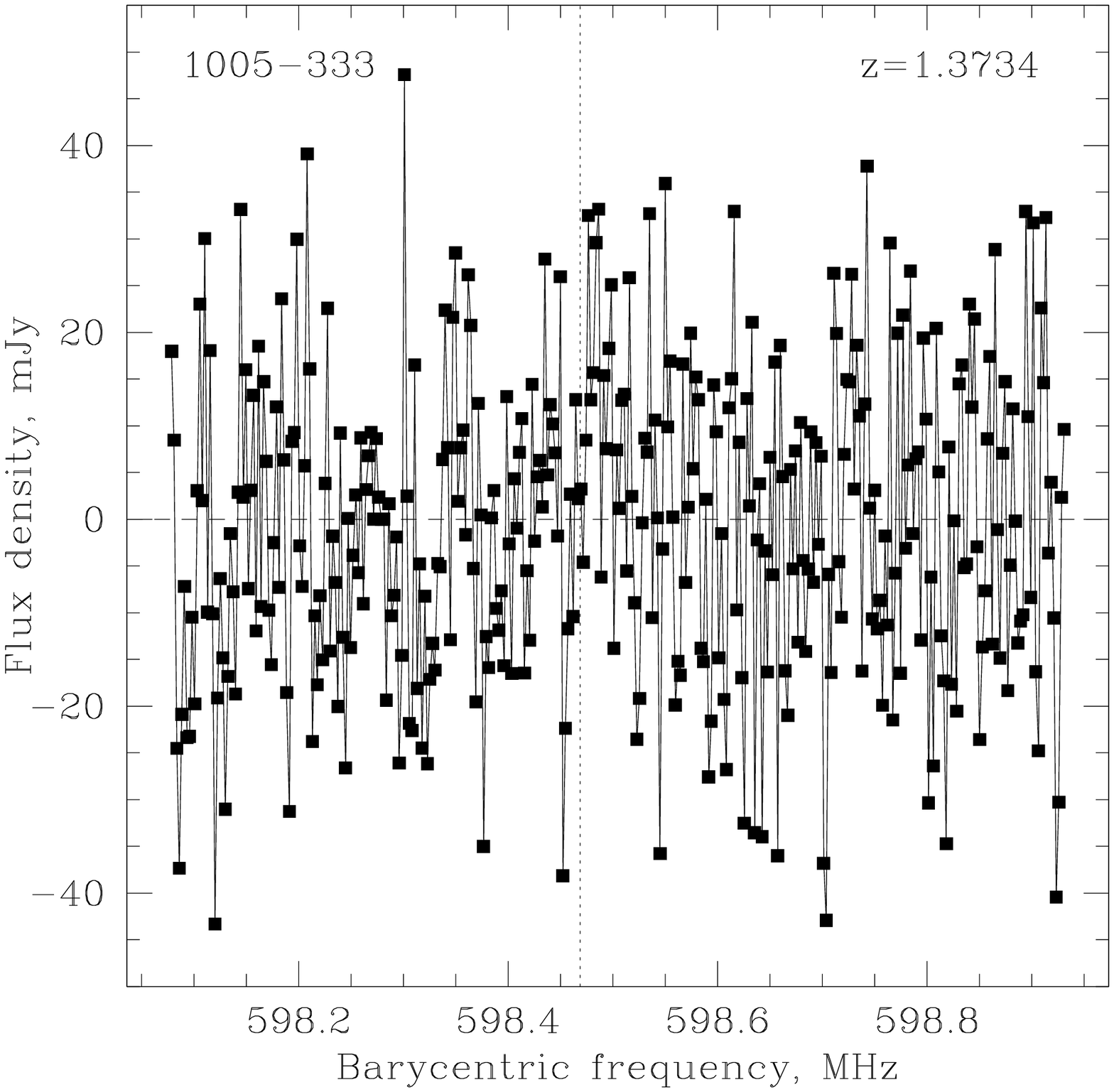,height=1.65in}
\epsfig{file=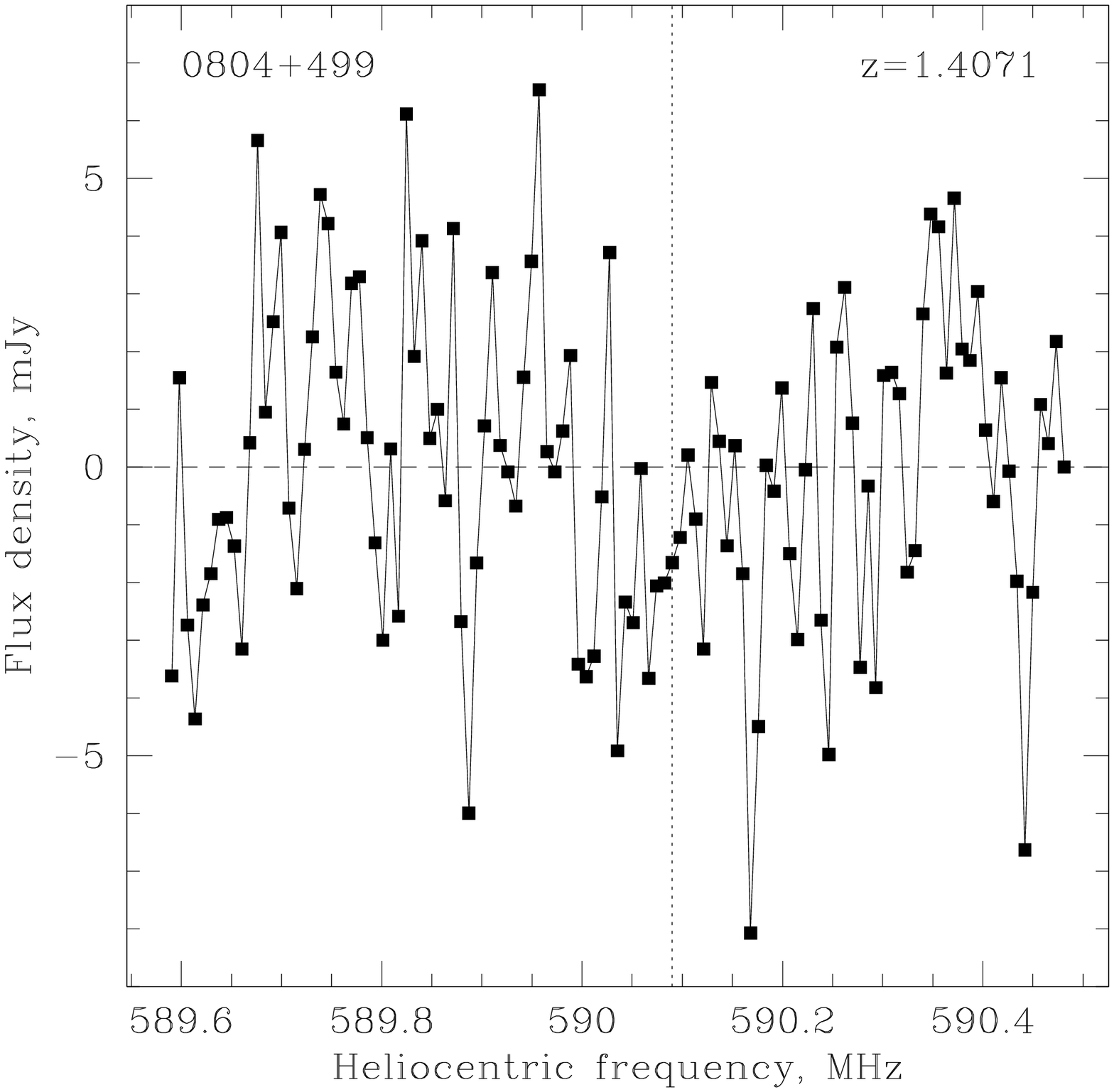,height=1.65in}
\epsfig{file=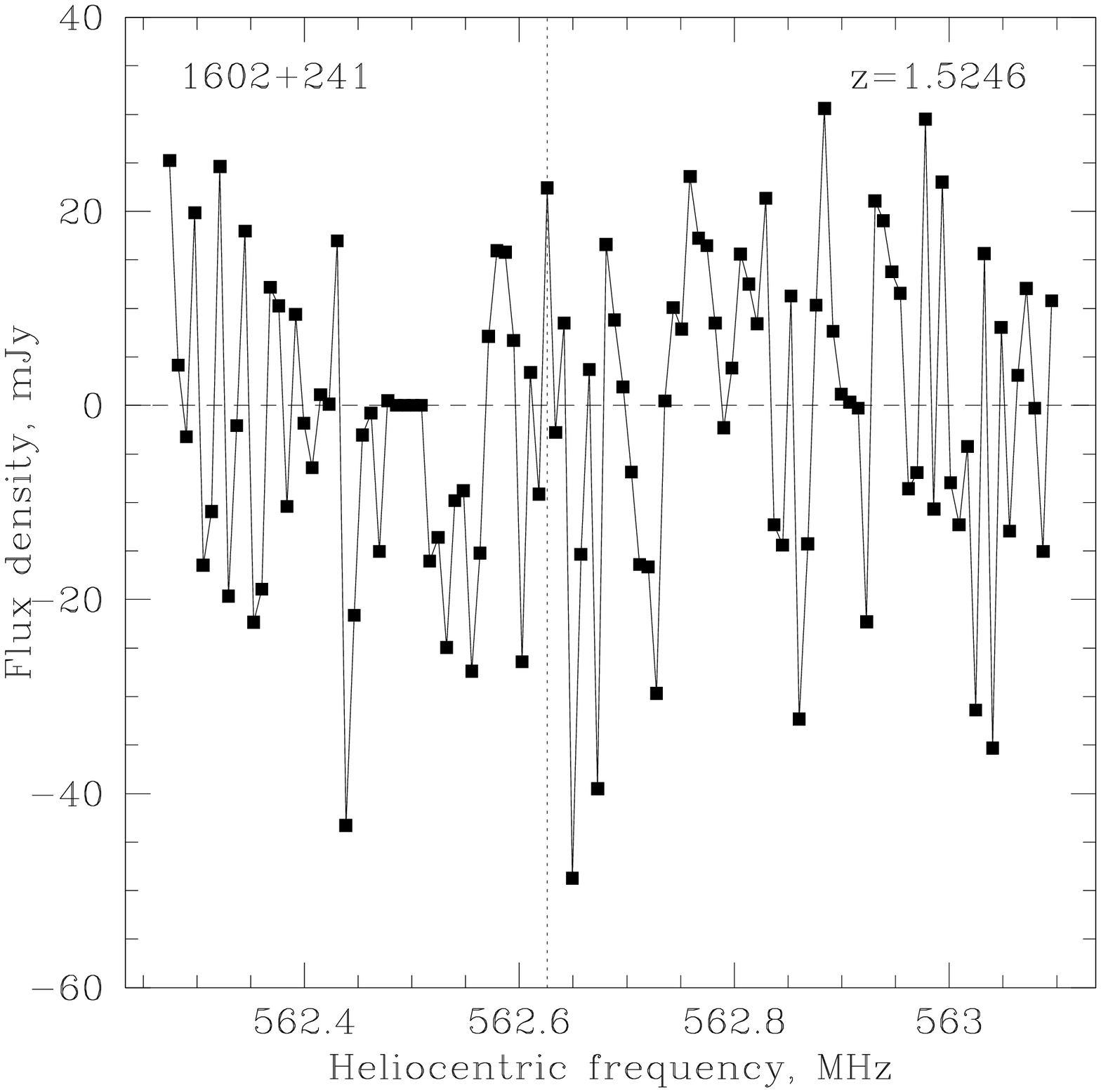,height=1.65in}
\epsfig{file=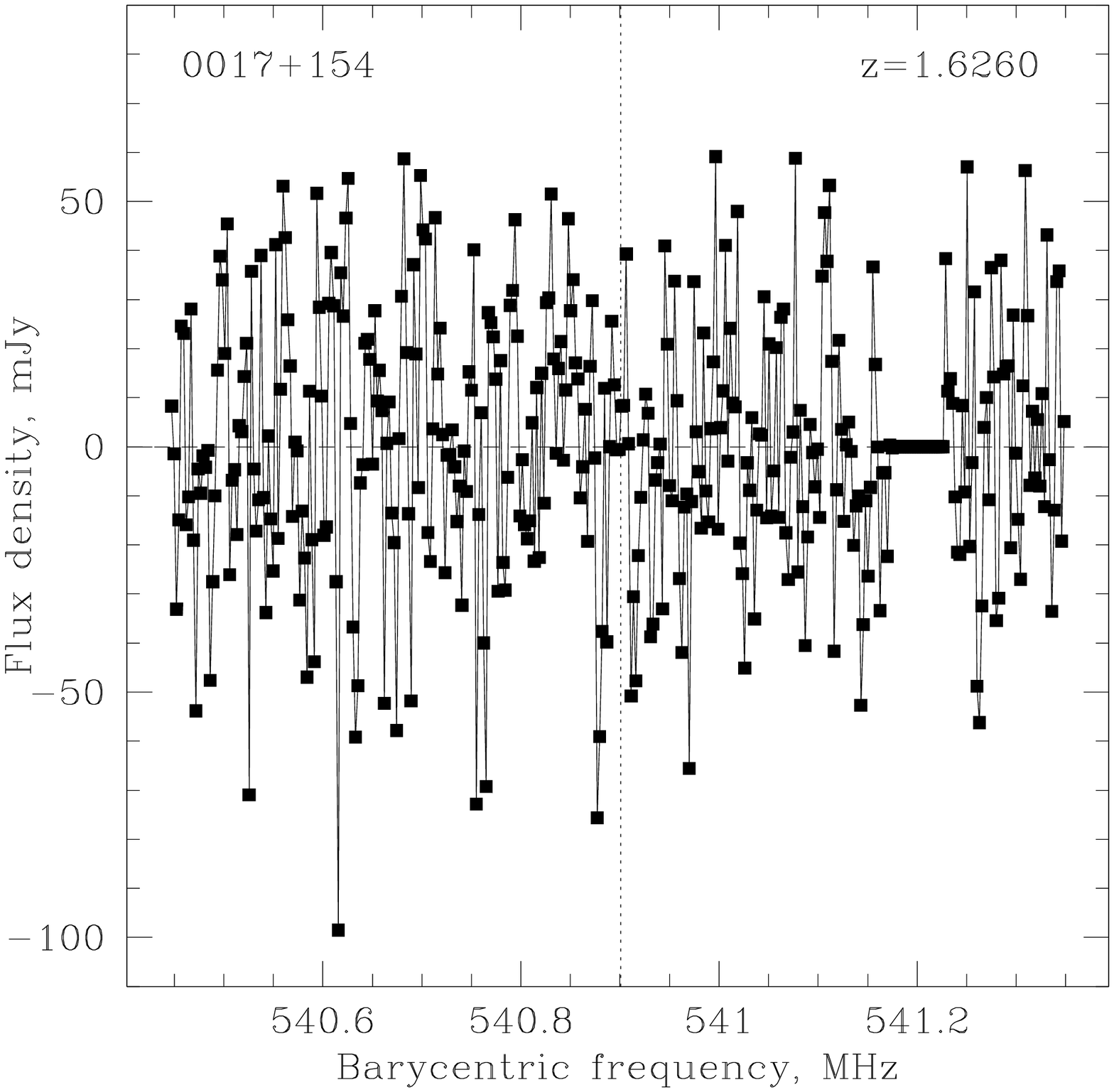,height=1.65in}
\vskip 0.00001in
\hskip 1.0in
\epsfig{file=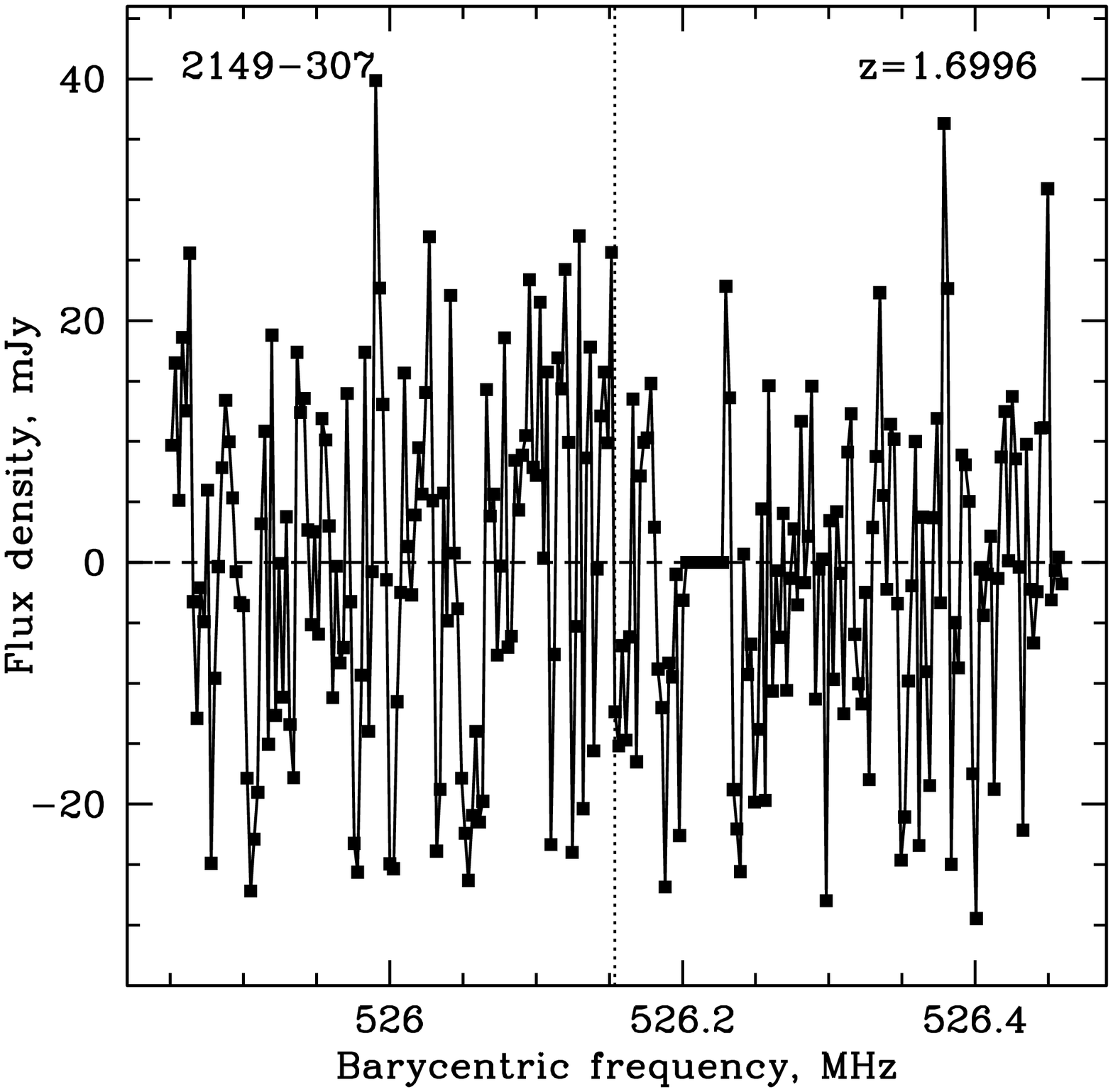,height=1.65in}
\epsfig{file=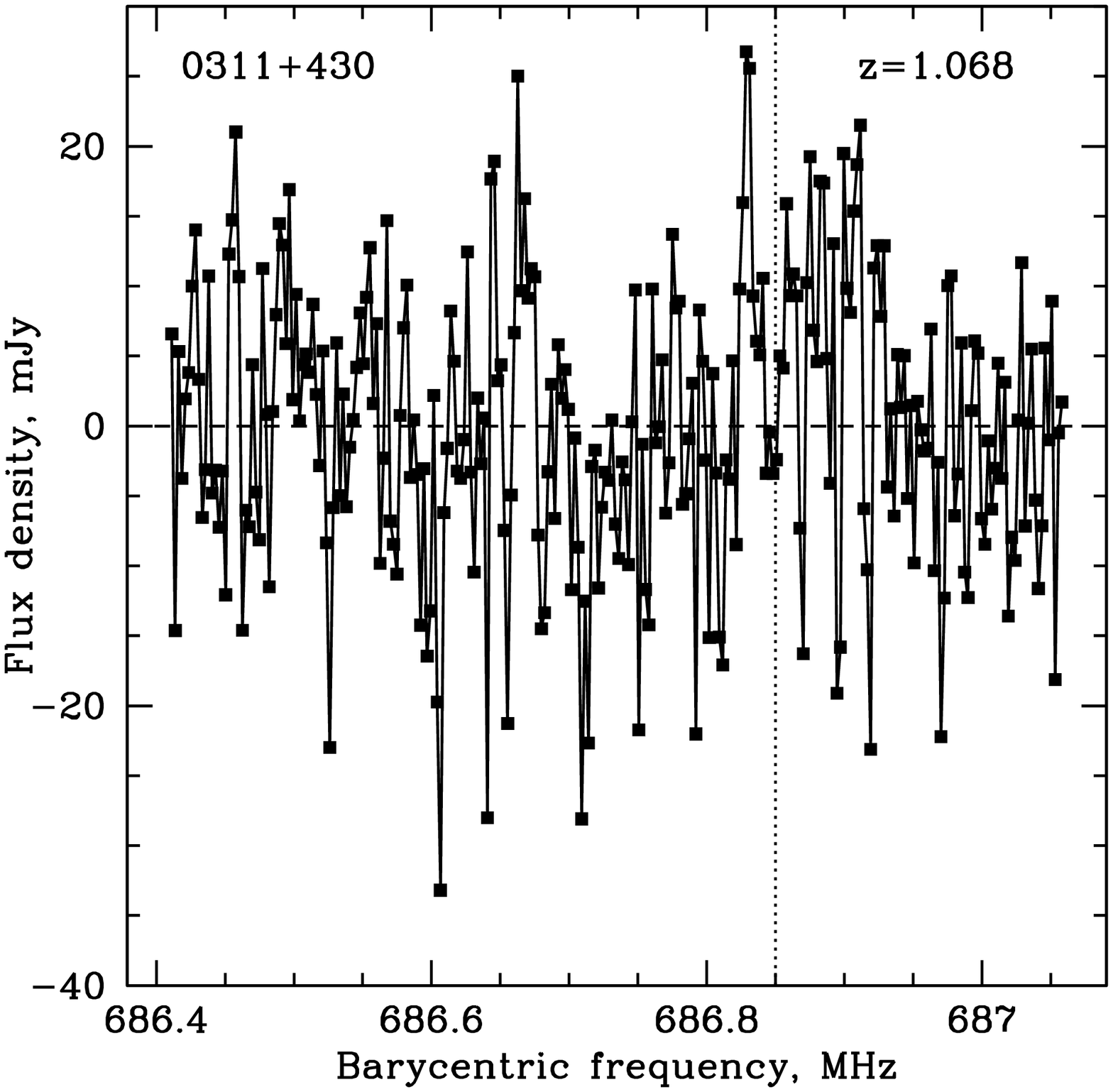,height=1.65in}
\label{fig:nondetect2}
\caption{(continued).}
\end{centering}
\end{figure*}

\section{Earlier searches for \hi~21cm absorption in \mgtwo-selected samples}
\label{sec:results}

The criterion of \mgtwo\ absorption has long been used as a criterion to select targets 
for searches for redshifted \hi~21cm absorption (e.g.  \citealp{roberts76,brown79,peterson80}), 
although these early studies typically observed few targets with relatively low sensitivity. 
The first survey for \hi~21cm absorption in a sizeable sample of \mgtwo\ absorbers was 
carried out by \citet{briggs83}, who used the Arecibo and NRAO~300-ft telescopes to target 
15~\mgtwo\ absorbers with $W_0^{\lambda 2796} \gtrsim 0.5 \AA$ and $0.4 \lesssim 
z_{\rm abs} \lesssim 1.8$. They obtained optical depth sensitivities of $\tau_{3\sigma} 
\sim 0.01 - 0.1$ per $\sim 10$~\kms, but with no new detections of \hi~21cm absorption.

More recently, \citet{lane00b} carried out a search for redshifted \hi~21cm absorption in a 
large sample of \mgtwo\ absorbers (62~systems, with $W_0^{\lambda 2796} > 0.3 \AA$ and at 
$0.2 < z_{\rm abs} < 1$), using the Westerbork Synthesis Radio Telescope. While this survey 
observed significantly more targets than earlier studies, the mean $3\sigma$ optical depth 
sensitivity of $\tau_{3\sigma} \sim 0.048$ was only sufficient to detect strong \hi~21cm absorbers, 
resulting in four new detections, all at $z_{\rm abs} < 0.44$ \citep{lane98,lane00b,lane01}. 
Indeed, four systems that were not detected in \hi~21cm absorption by \citet{lane00b} were 
later detected by deeper searches \citep{kanekar01a,kanekar03,curran07a}, that achieved 
sensitivities of $\tau_{3\sigma} \lesssim 0.01$ per $\sim 10$~\kms. The optical depth sensitivity 
of \citet{lane00b} was also typically lower for absorbers at $z_{\rm abs} \gtrsim 0.6$ than for 
the low-$z$ systems in their sample.

\citet{gupta07} used the GMRT to search for \hi~21cm absorption in 10~strong \mgtwo\ 
absorbers, at $1.11 < z_{\rm abs} < 1.45$, obtaining three detections and seven upper limits 
to the \hi~21cm optical depth. Their three absorption detections were observed during this 
survey and are included in our sample. Only one of their non-detections had an optical 
depth sensitivity $\tau_{3\sigma} \lesssim 0.01$ per $\sim 10$~\kms; the remaining six had 
$\tau_{3\sigma} \sim 0.02 - 0.04$ per $\sim 10$~\kms, in all cases worse than the sensitivity
of our statistical sample ($\tau_{3\sigma} \le 0.013$ per 10~\kms).

Finally, \citet{srianand08} report the detection of \hi~21cm absorption in two 
\mgtwo\ absorbers at $z_{\rm abs} \sim 1.3265$ towards SDSSJ0850+5159 and 
$z_{\rm abs} \sim 1.3095$ towards SDSSJ0852+3435 with the GMRT. These were targetted 
based on the criterion $\wmgtwo \ge 1 \AA$, in a continuation of the survey of 
\citet{gupta07}. Both absorbers have large metal line rest equivalent widths: the system 
at $z \sim 1.3265$ has $\wmgtwo = 4.89 \AA$, $\wmgone = 2.08 \AA$ and $\wfetwo = 2.27 \AA$, 
while that at $z \sim 1.3095$ has $\wmgtwo = 2.89 \AA$, $\wmgone = 1.11$ and $\wfetwo = 2.10 \AA$.

We investigated the possibility of augmenting our statistical sample with \hi~21cm 
results from these earlier studies. For this purpose, we required any such 
absorbers to satisfy the following constraints: (1)~$W_0^{\lambda 2796} \ge 0.5 \AA$, 
(2)~$0.58 \le z_{\rm abs} \le 1.70$, (3)~for non-detections, $\tau_{3\sigma} \le 0.013$ 
per $\sim 10$~\kms, and (4)~the velocity separation between the absorber and quasar redshifts 
should be $\ge 3000$~\kms. One system from the sample of \citet{briggs83}, at $z_{\rm abs} 
\sim 0.8596$ towards 0454+039, 
fulfils all four conditions and was hence added to our sample. The detections of 
\citet{gupta07} were all observed in this search and are included here. The only 
other system from their sample that satisfies the sensitivity constraint is the 
$z_{\rm abs} \sim 1.3647$ absorber towards 0237$-$233 \citep{srianand07}, for which 
we present a higher-sensitivity GMRT spectrum here. Both \hi~21cm detections of 
\citet{srianand08} satisfy all our criteria and were added to the sample. Finally, 
no system from the sample of \citet{lane00b} satisfies all three constraints; most 
are ruled out by the sensitivity criterion. 

\section{The incidence of \hi~21cm absorption in strong \mgtwo\ absorbers}
\label{sec:incidence}

\setcounter{figure}{2}
\begin{figure*}
\begin{centering}
\epsfig{file=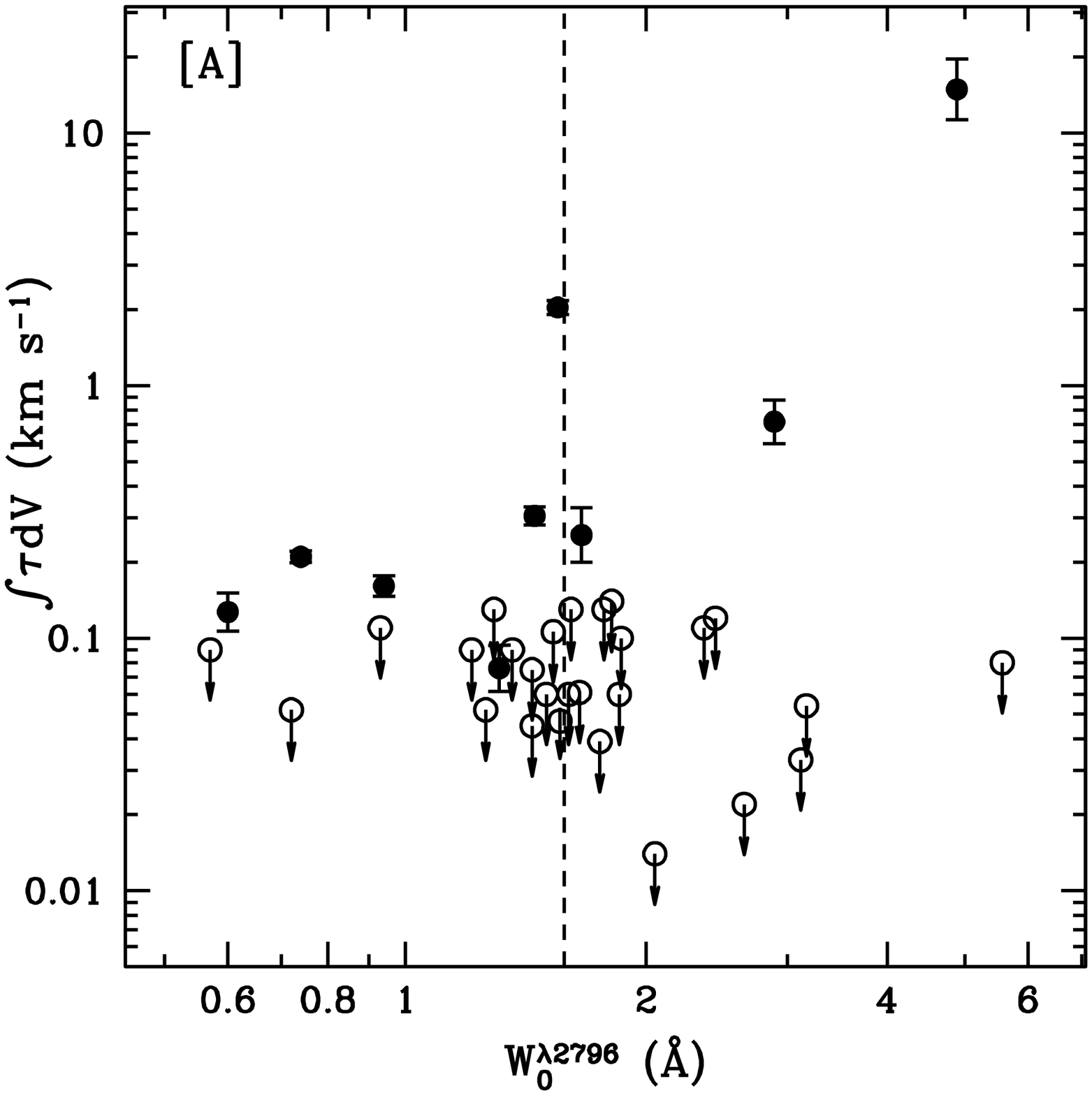,height=3.3truein,width=3.3truein}
\hskip 0.3in
\epsfig{file=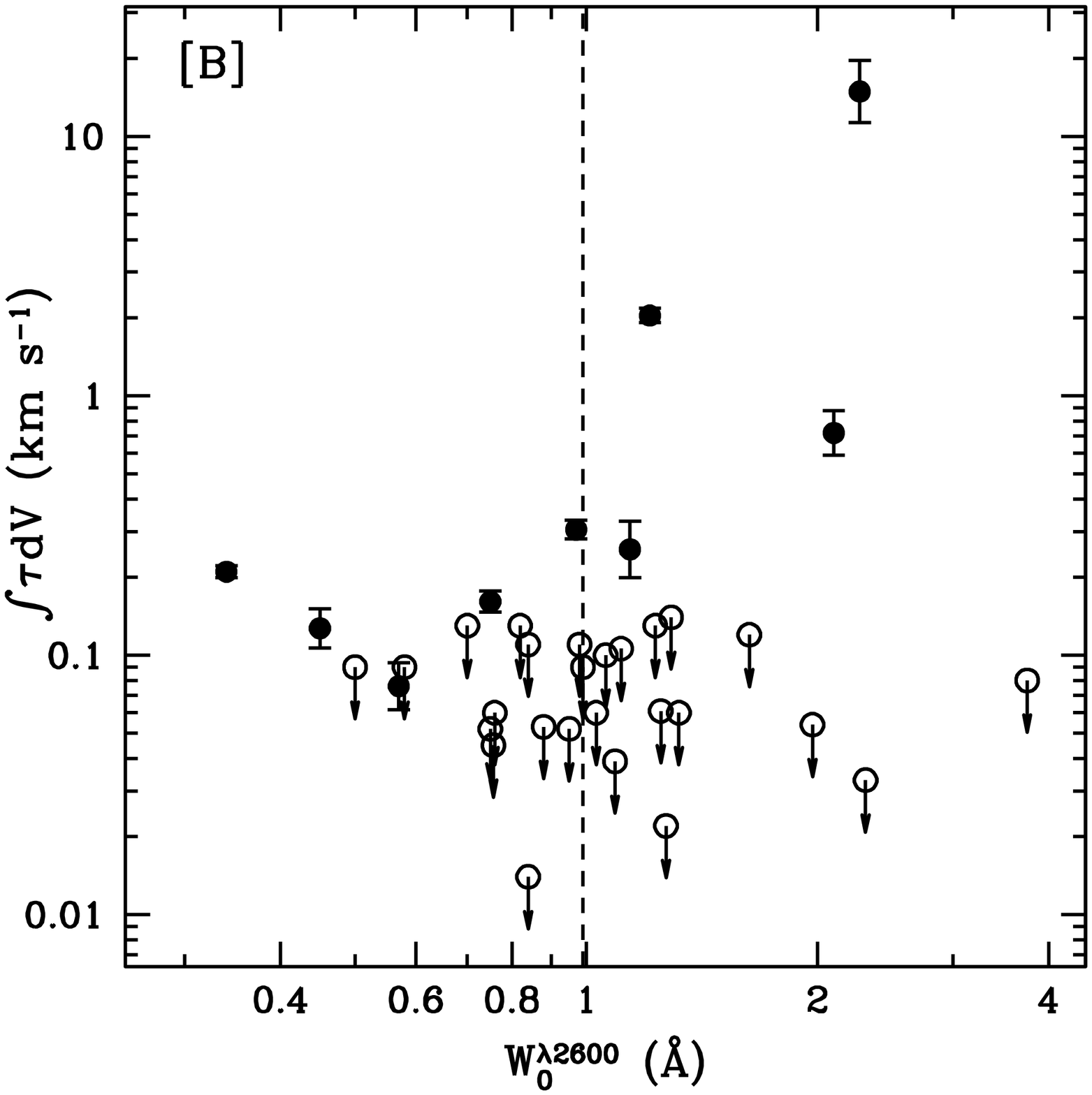,height=3.3truein,width=3.3truein}

\epsfig{file=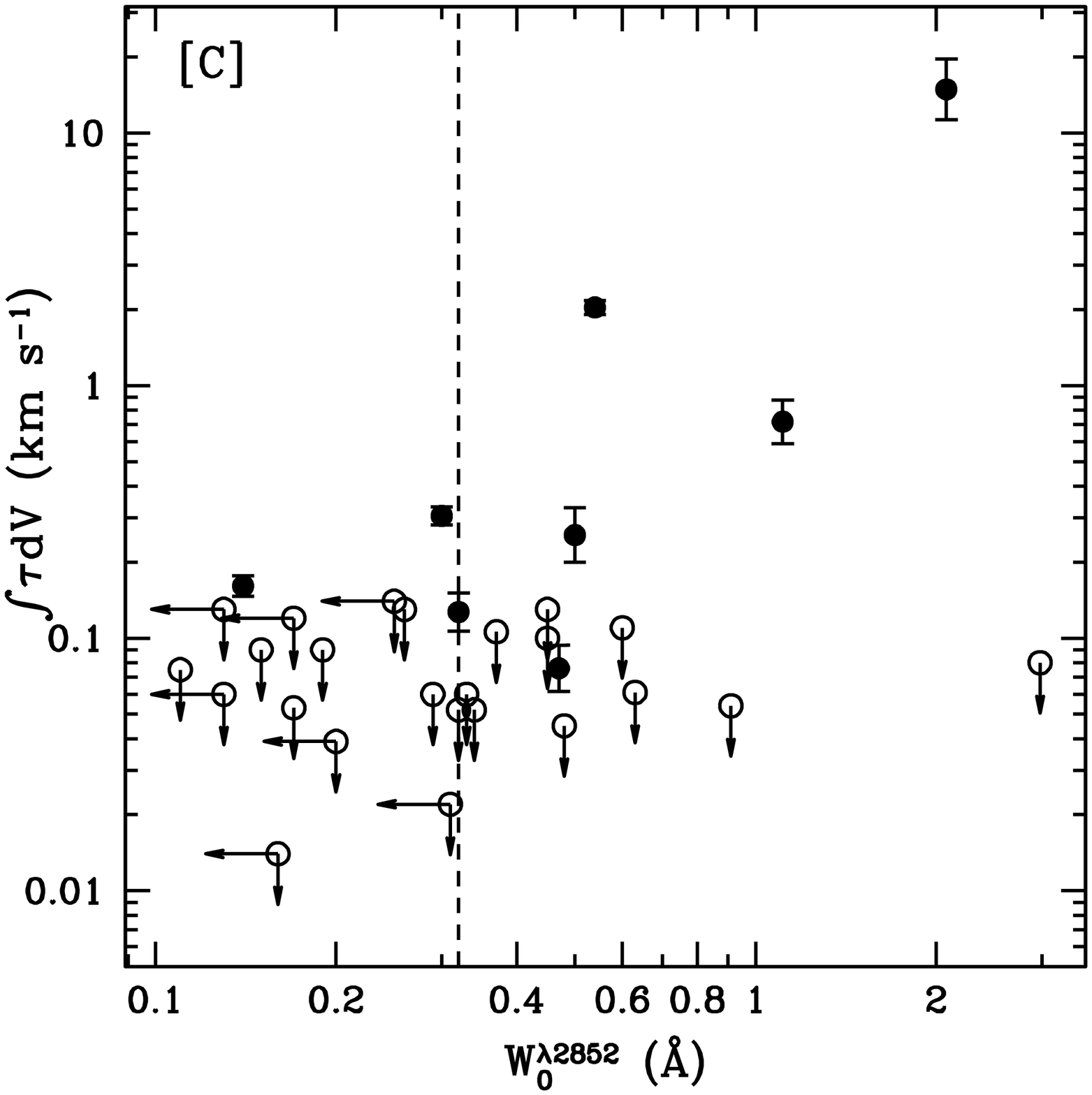,height=3.3truein,width=3.3truein}
\hskip 0.3in
\epsfig{file=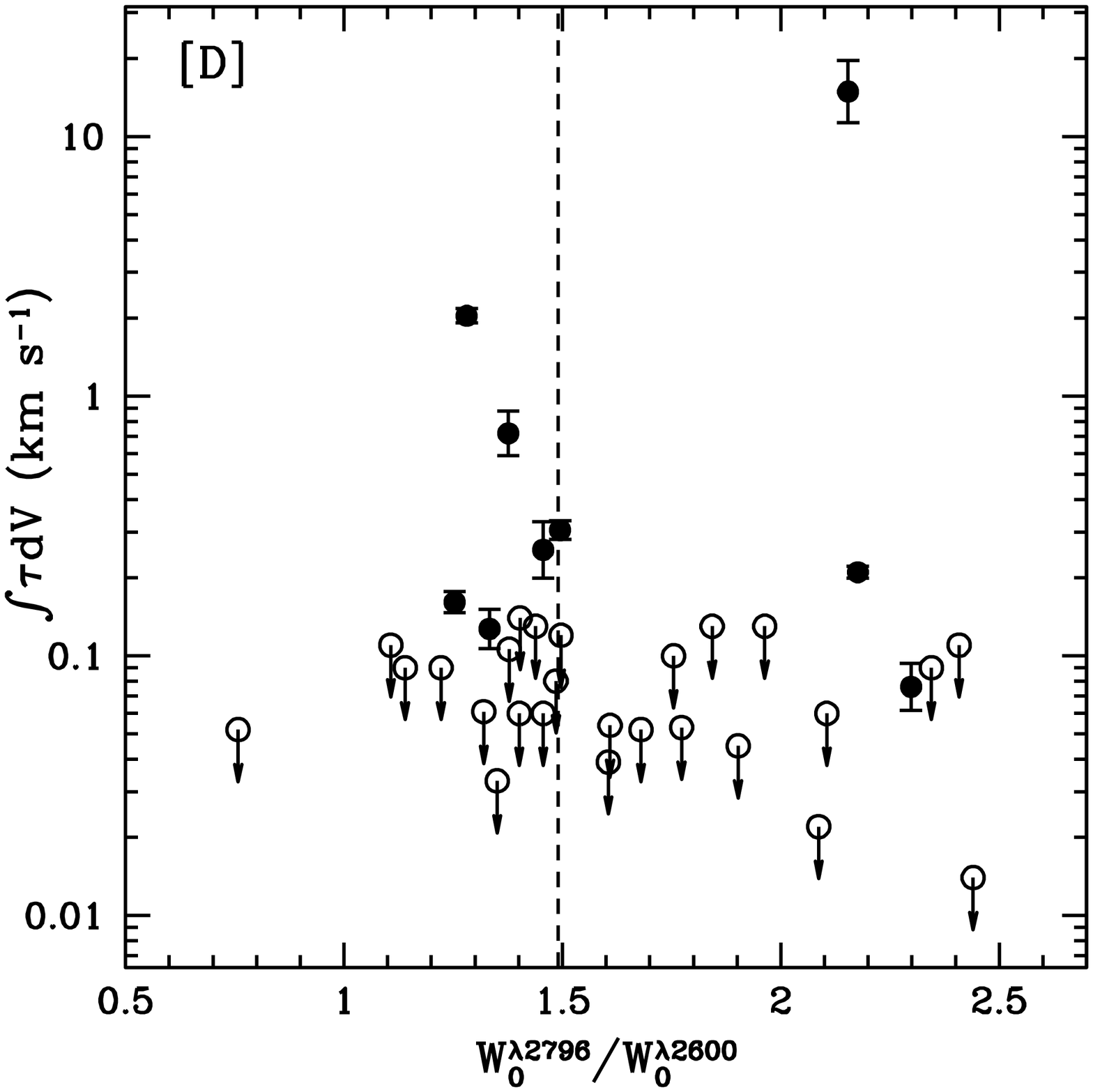,height=3.3truein,width=3.3truein}
\caption{The four panels show the integrated \hi~21cm optical depth plotted against (from top left)
[A]~the \mgtwo$\lambda$2796 rest equivalent width, [B]~the \fetwo$\lambda$2600 rest equivalent width 
(35~systems), [C]~the \mgone$\lambda$2852 rest equivalent width (33~systems), and 
[D]~$W_0^{\lambda 2796}/W_0^{\lambda 2600}$ (35~systems). Detections of \hi~21cm 
absorption are shown as filled circles. The dashed vertical lines mark the median value 
of the quantity plotted as the abscissa in each panel.}
\label{fig:ew21}
\end{centering}
\end{figure*}

\begin{figure*}
\begin{centering}
\epsfig{file=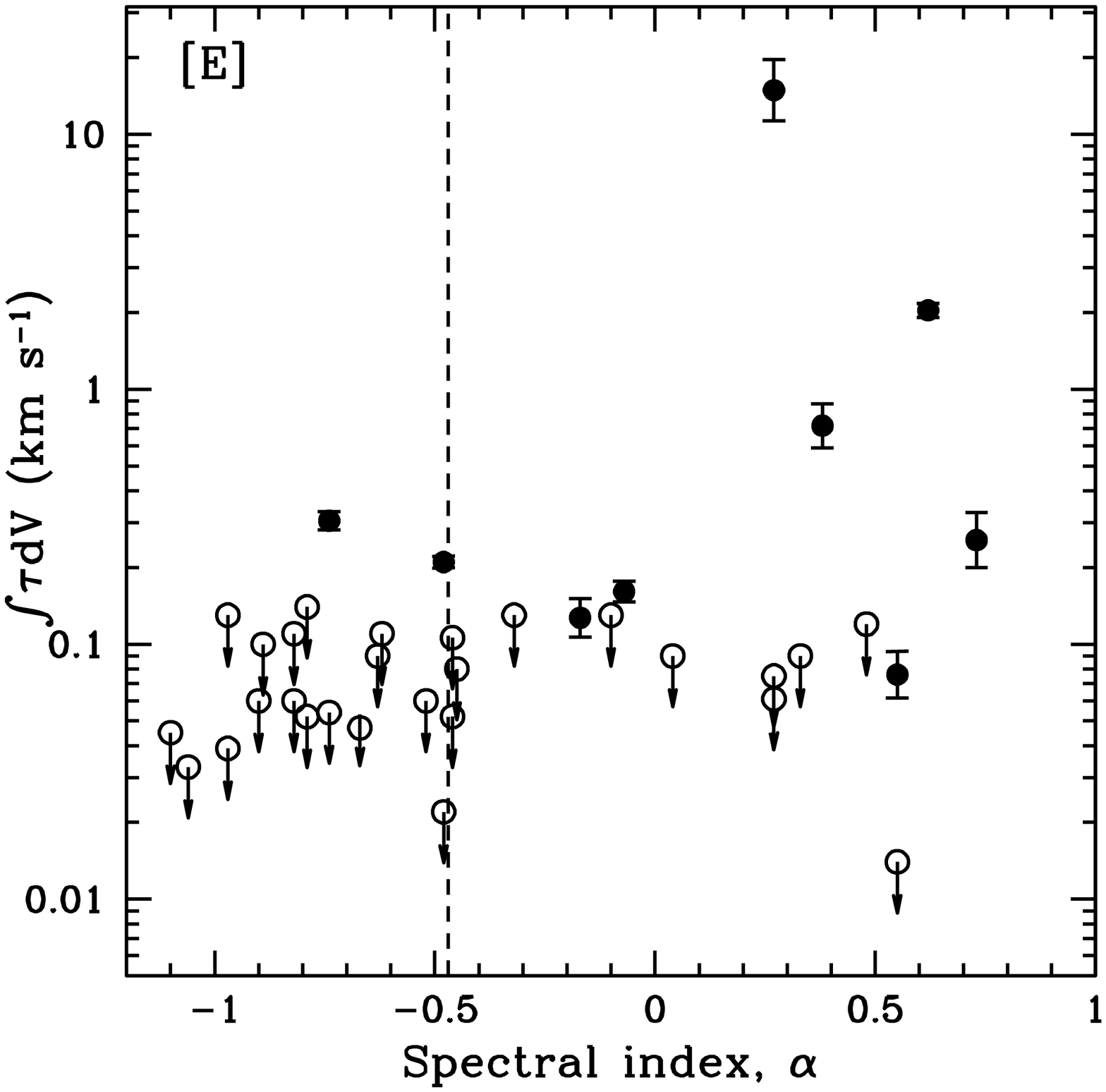,height=3.3truein,width=3.3truein}
\hskip 0.3in
\epsfig{file=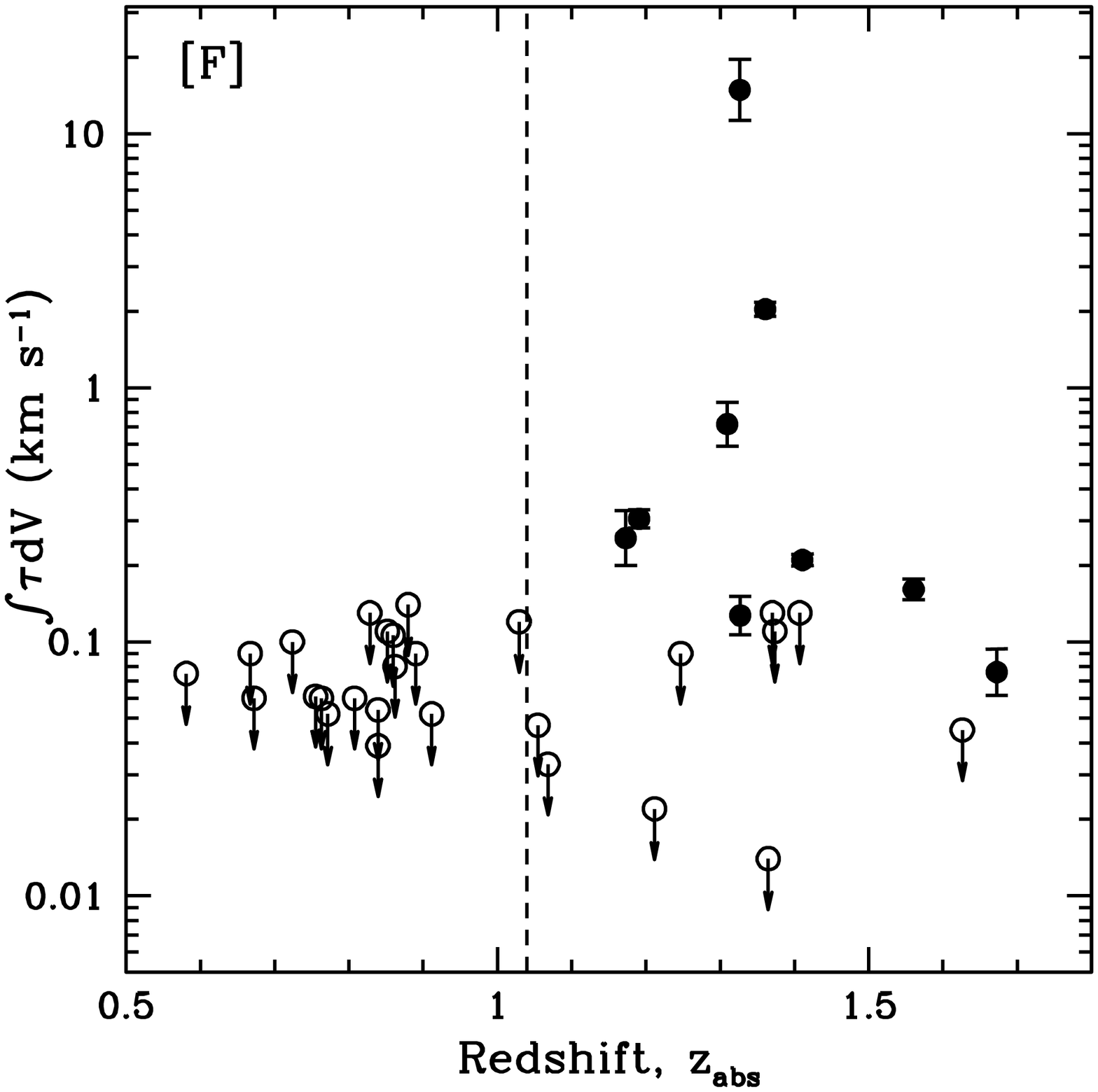,height=3.3truein,width=3.3truein}
\caption{The two panels of the figure show the integrated \hi~21cm optical depth plotted against 
[A]~the low-frequency spectral index $\alpha$ of the background quasar, and [B]~the absorber redshift,
$z_{\rm abs}$. Detections of \hi~21cm absorption are shown as filled circles. The dashed 
vertical lines mark the median value of the quantity plotted as the abscissa in each panel.}
\label{fig:ew21b}
\end{centering}
\end{figure*}

\begin{table}
\begin{center}
\setlength{\extrarowheight}{4pt}
\begin{tabular*}{0.45\textwidth}{@{\extracolsep{\fill}}|c|c|c|c|}
\hline
Attribute     & Median &  \multicolumn{2}{c|}{\hi~21cm detection rate, \% }\\
              &        &  ``High'' & ``Low'' \\
\hline
& & & \\
$\wmgtwo$     & $1.58 \AA$ & $17^{+16}_{-9}$  & $33^{+20}_{-13}$ \\
$\wfetwo$     & $0.99 \AA$ & $22^{+18}_{-11}$  & $28^{+19}_{-12}$ \\
$\wmgone$     & $0.32 \AA$ & $29^{+20}_{-13}$ & $18^{+17}_{-10}$  \\
$W_0^{\lambda 2796}/W_0^{\lambda 2600}$ & $1.49$ & $18^{+17}_{-10}$ & $33^{+20}_{-13}$  \\
$\alpha$      & $-0.47 $   & $39^{+21}_{-14}$ & $11^{+15}_{-7}$ \\
$z_{\rm abs}$ & $1.04$	   & $50^{+23}_{-16}$ & $0^{+10}$  \\
& & & \\
\hline
\end{tabular*}
\vskip 0.1in
\begin{tabular*}{0.45\textwidth}{@{\extracolsep{\fill}}|c|c|c|}
\hline
Telescope     & No. of systems & \hi~21cm detection rate, \% \\
\hline
GMRT          & 12     & $58^{+31}_{-22}$ \\
GBT/Arecibo   & 24     & $8^{+11}_{-5}$   \\
& & \\
\hline
\end{tabular*}
\end{center}
\caption{A comparison between the detection rates of \hi~21cm absorption in different 
sub-samples. In the upper table, for each attribute (e.g. $\wmgtwo$, $\wfetwo$, etc), 
the ``High'' and ``Low'' sub-samples consist of absorbers with attribute values 
higher and lower than the median, respectively.}
\label{tab:ew21}
\end{table}

Including systems from the literature, our final statistical sample consists 
of 36~strong \mgtwo\ absorbers at $0.58 < z_{\rm abs} < 1.68$, with 
$W_0^{\lambda 2796} \ge 0.57 \AA$; the median \mgtwo\ absorption redshift 
is $z_{\rm abs} \sim 1.04$. 33~systems have $\wfetwo \ge 0.5 \AA$, while 
three have either $\wfetwo < 0.5 \AA$ (two absorbers) or no information on the 
\fetwo\ transition (the $z \sim 0.5810$ absorber towards 0240$-$060). There 
are nine detections of \hi~21cm absorption (including the tentative detection towards 
0237$-$233) and 27~non-detections with $\tau_{3\sigma} \le 0.013$ per $\sim 10$~\kms. 
Seven of the detections and 26~of the non-detections are from the present survey, with two 
detections from \citet{srianand08} and one non-detection from \citet{briggs83}. The 
median $3\sigma$ optical depth sensitivity for the non-detections is $\tau_{3\sigma} 
\sim 0.0062$ per $\sim 10$~\kms. The net detection rate of \hi~21cm absorption is 
$25^{+11}_{-8}$\% at a mean absorption redshift of $\bar z \sim 1.1$.\footnote{All 
error bars correspond to $1\sigma$ Gaussian confidence 
levels, using small-number Poisson statistics \citep{gehrels86}.} 

We examined the data for correlations between the detection rate of \hi~21cm absorption 
and the following set of absorber attributes: (1)~$\wmgtwo$, (2)~$\wfetwo$, (3)~$\wmgone$, 
(4)~$W_0^{\lambda 2796}/W_0^{\lambda 2600}$ and (5)~\mgtwo\ absorption redshift. In order to
test for possible systematic effects, we also checked whether the \hi~21cm detection rate showed
a dependence on the spectral index of the background quasar (which is indicative of 
the absorber covering factor) or the telescope used for the observations (interferometer 
versus single-dish). The small size of the full sample implies that, for any given 
attribute, it is only meaningful to divide the sample into two sub-samples at 
the median value of the quantity in question (except, of course, for the comparison 
between telescope types).  This is the approach we will follow here.

Results of these comparisons are summarized in Table~\ref{tab:ew21}; the columns 
of the upper part of the table contain (1)~the attribute used for the division 
(e.g. the \mgtwo$\lambda$2796 rest equivalent width, the quasar spectral index $\alpha$, etc), 
(2)~the median value of this attribute, (3)~the detection rate of \hi~21cm absorption 
for the ``high'' sample, with attribute values higher than the median (e.g. $\alpha > 
\alpha_{\rm med}$), and (4)~the detection rate for the ``low'' sample, with values 
lower than the median. For the lower half of the table, the columns are (1)~the 
observing facility, divided into interferometers (GMRT) and single-dish telescopes 
(GBT and Arecibo), (2)~the number of absorbers, and (3)~the detection rates of 
\hi~21cm absorption. No statistically-significant difference ($\ge 3\sigma$) is 
found between the detection rates in the sub-samples for any criterion, largely 
due to the small size of each sub-sample. For example, all nine detections of \hi~21cm 
absorption were obtained at $z_{\rm abs} > 1.04$, but the difference in detection 
rates in the high-$z$ and low-$z$ sub-samples has only $\sim 2.6 \sigma$ significance. 
We conclude that the size of the present sample is too small to test the dependence 
of the detection rate on the properties of individual absorbers.

Finally, Figs.~\ref{fig:ew21} and \ref{fig:ew21b} show the integrated \hi~21cm 
optical depth ($EW_{\rm 21} \equiv \int \tau {\rm d}V$) plotted against 
$\wmgtwo$, $\wfetwo$, $\wmgone$, $W_0^{\lambda 2796}/W_0^{\lambda 2600}$, 
quasar spectral index $\alpha$ and \mgtwo\ absorption redshift, for absorbers
in the statistical sample. No 
statistically-significant correlation is found between $EW_{\rm 21}$ and any 
of these quantities. However, it can be seen in Figs.~\ref{fig:ew21}[A-C] that 
a trend may be present between $EW_{\rm 21}$ and each of $\wmgtwo$, $\wfetwo$ 
and $\wmgone$, for detections of \hi~21cm absorption. The possibility that a 
relation between $EW_{\rm 21}$ and $\wmgtwo$ might be present only in the 
highest \hi\ column density absorbers (i.e. DLAs) is explored further in the 
next section.

\section{\hi~21cm absorption in \dlas}
\label{sec:n21}

\begin{figure}
\begin{centering}
\epsfig{file=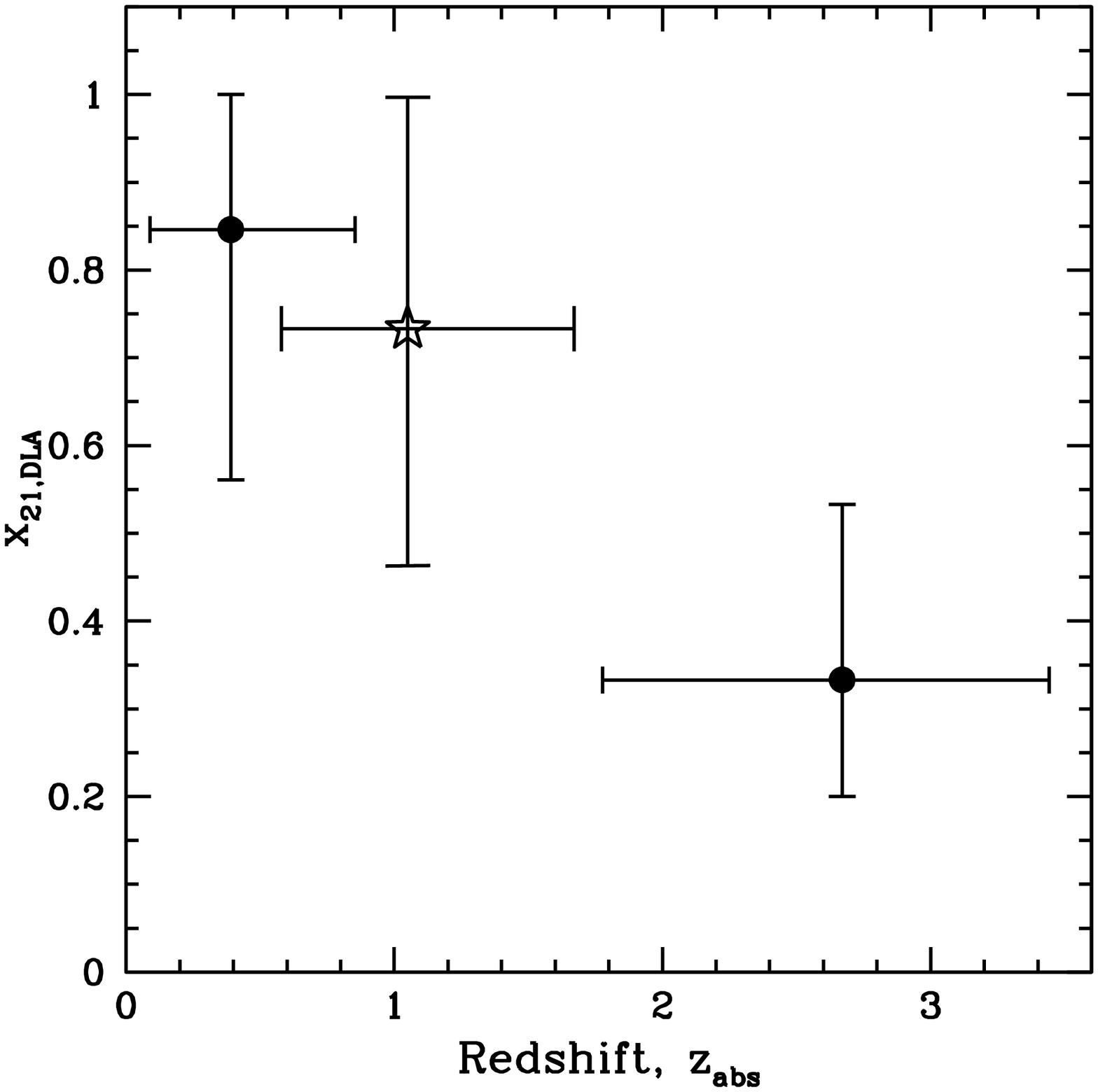,height=3.3truein,width=3.3truein}
\caption{The detection rate of \hi~21cm absorption in DLAs, plotted as a function 
of absorber redshift; the result from the present survey is shown as an open star. 
See the main text for discussion.}
\label{fig:det21}
\end{centering}
\end{figure}

\begin{figure}
\begin{centering}
\epsfig{file=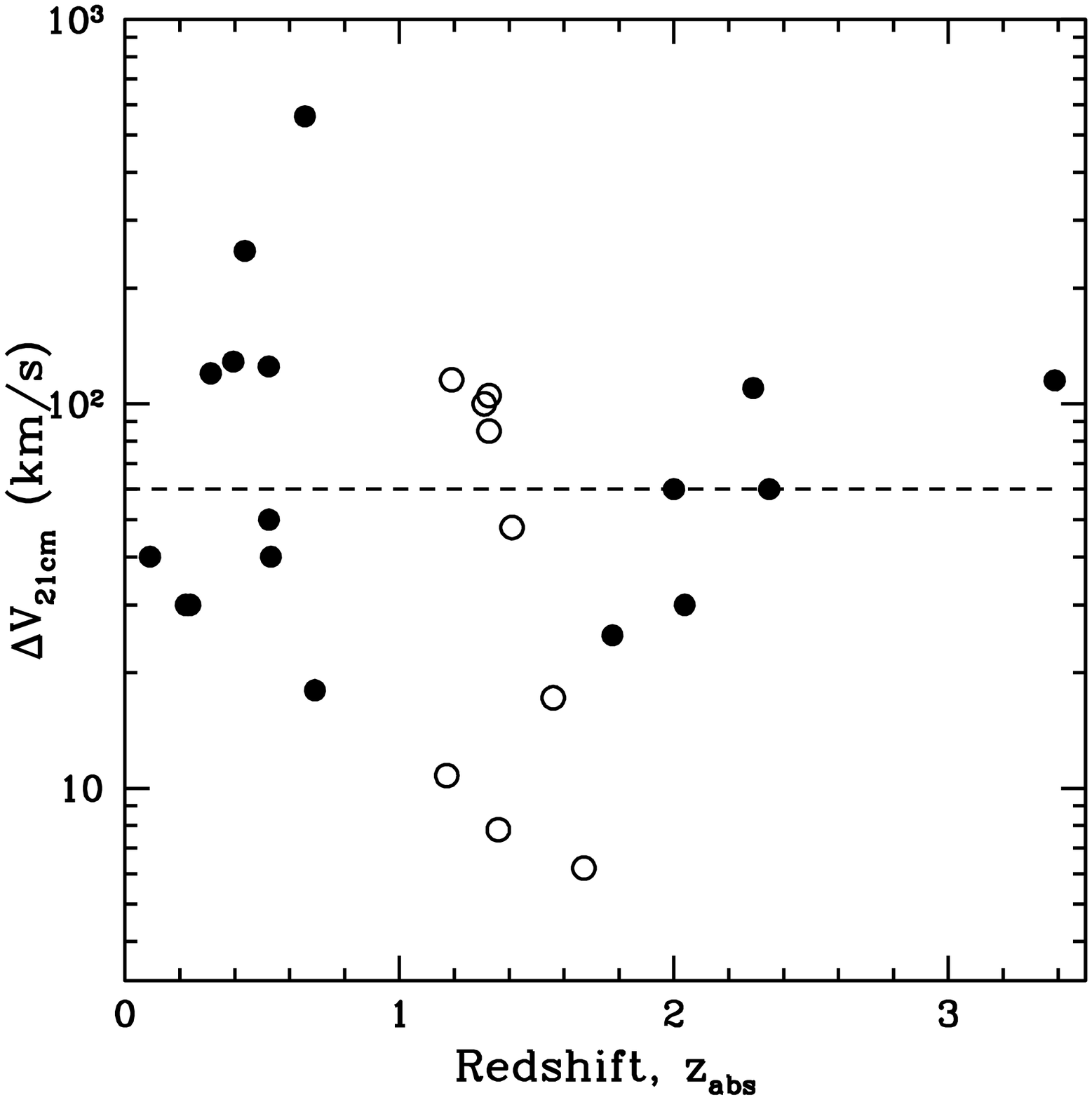,height=3.3truein,width=3.3truein}
\caption{The velocity spread of \hi~21cm absorption $\Delta V_{\rm 21}$ in 
strong \mgtwo\ absorbers (in open circles) and DLAs (filled circles), 
plotted as a function of absorber redshift; the dashed line is at 
$\Delta V_{\rm 21} = 60$~\kms.}
\label{fig:vel21}
\end{centering}
\end{figure}

\begin{figure}
\begin{centering}
\epsfig{file=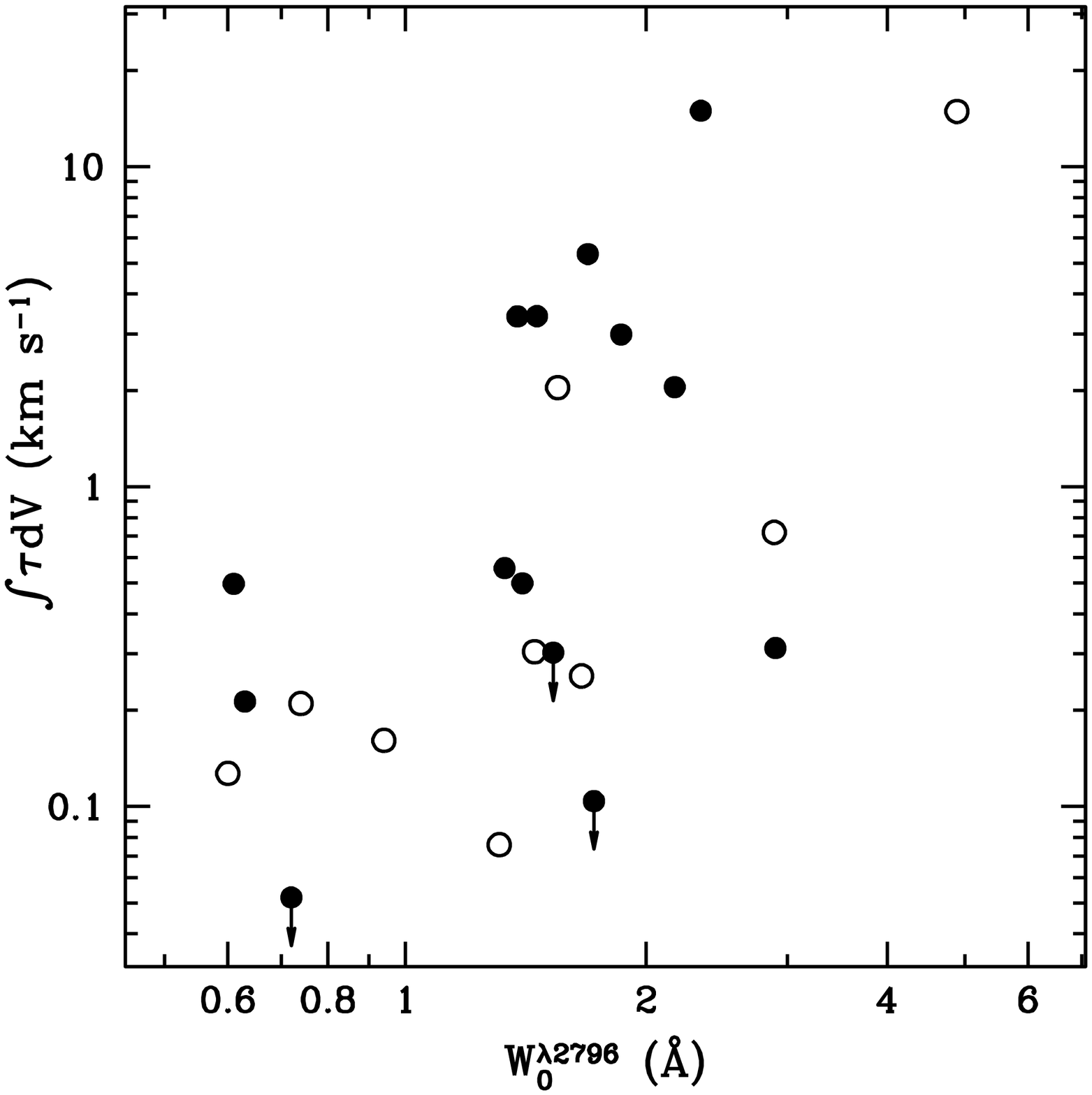,height=3.3truein,width=3.3truein}
\caption{The integrated \hi~21cm optical depth $EW_{\rm 21} \equiv \int \tau {\rm d}V$ 
plotted against the rest equivalent width in the \mgtwo$\lambda$2796 transition, $\wmgtwo$,
for \hi~21cm absorbers and DLAs. Strong \mgtwo\ absorbers with \hi~21cm detections are shown 
as open circles, while DLAs from the literature are shown as filled circles. A trend between 
$EW_{\rm 21}$ and $\wmgtwo$ is apparent in the figure, but has only $\sim 2.2\sigma$ 
significance in a non-parametric Kendall-tau test. See the main text for discussion.}
\label{fig:mgtwo}
\end{centering}
\end{figure}

\begin{figure}
\begin{centering}
\epsfig{file=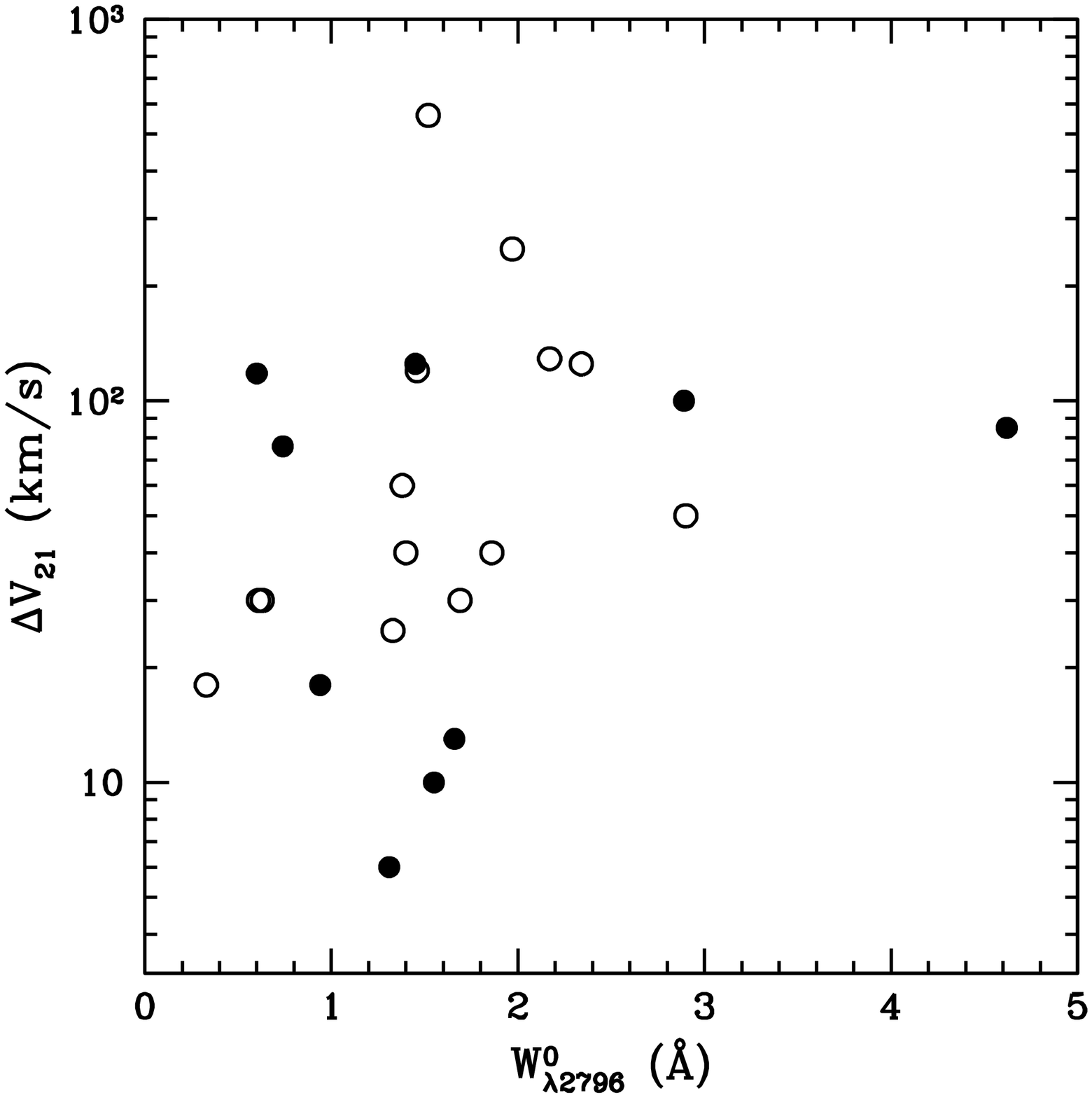,height=3.3truein,width=3.3truein}
\caption{The velocity spread of \hi~21cm absorption $\Delta V_{\rm 21}$ in 
strong \mgtwo\ absorbers (open circles) and DLAs (filled circles), 
plotted as a function of the \mgtwo$\lambda$2796 rest equivalent width, 
$\wmgtwo$. See the main text for discussion.}
\label{fig:vel21-mgII}
\end{centering}
\end{figure}

\begin{table}
\begin{center}
\begin{tabular}{ll}
\hline
&  \\
$x_{\rm 21,MgII} (z)$  & Detection rate of \hi~21cm absorption in \mgtwo$\lambda$2796 \\
& absorbers \\
$x_{\rm 21,DLA} (z)$   & Detection rate of \hi~21cm absorption in DLAs\\
$x_{\rm DLA,MgII} (z)$ & Detection rate of DLAs in \mgtwo$\lambda$2796 absorbers\\
$n_{\rm DLA}(z)$       & Number of DLAs per unit redshift \\
$n_{\rm MgII}(z,W_0)$  & Number of \mgtwo$\lambda$2796 absorbers with $\wmgtwo \ge W_0$, \\
& per unit redshift \\
& \\
\hline
\end{tabular}
\end{center}
\caption{Notation used in this paper.}
\label{tab:notation}
\end{table}

Very few DLAs with $0.8 \lesssim z_{\rm abs} \lesssim 1.7$ are known towards radio-loud quasars,
due to which there have been few \hi~21cm absorption studies of systems in this redshift range.
In fact, prior to this work, only two ``classical'' DLAs (i.e. with N$_{\rm HI} \ge 
2 \times 10^{20}$~\cm) at these redshifts had been searched for \hi~21cm absorption, 
with no detections \citep{kanekar03}. The situation is quite different at both lower and higher 
redshifts. So far, six \hi~21cm absorbers have been detected in ``non-associated'' 
DLAs at $z_{\rm abs} \gtrsim 1.7$, with 12~limits on the \hi~21cm optical depth (typically 
$\tau_{\rm 21} \sim 0.01$; e.g. 
\citealp{wolfe79,wolfe81,wolfe85,kanekar03,kanekar06,york07,kanekar07}). The 
detection rate of \hi~21cm absorption in high-$z$ DLAs is thus $x_{\rm 21,DLA}({\bar z} \sim 2.7) 
= 33^{+20}_{-13}$\%, where ${\bar z}$ is the mean redshift of the DLA sample (see 
Table~\ref{tab:notation} for the notation used in this paper). Conversely, \hi~21cm absorption 
has been detected in eleven DLAs at $z_{\rm abs} < 0.9$, with only two non-detections 
(e.g. \citealp{brown73,wolfe76,brown79,brown83,lane98,chengalur99,kanekar01a,curran07a}). The 
detection rate of \hi~21cm absorption in DLAs at $z_{\rm abs} < 0.9$ is thus 
$x_{\rm 21,DLA}({\bar z} \sim 0.4) = 85^{+15}_{-28}$\%.  Recent low-frequency VLBI studies 
have shown that DLAs in the $z_{\rm abs} > 1.7$ and $z_{\rm abs} < 0.9$ samples have very 
similar covering factors, $0.4 < f < 1$ \citep{kanekar07,kanekar09a}. The \hi\ column
density distributions of the high-$z$ and low-$z$ DLA samples (with \hi~21cm absorption studies) 
are also similar \citep{kanekar09a}.  The low detection rate of \hi~21cm absorption in the high-$z$ 
DLA sample is hence likely to be due to a higher fraction of warm \hi\ in these absorbers (e.g. 
\citealp{wolfe79,carilli96,kanekar03}). 

Our survey provides a direct estimate of the detection rate of \hi~21cm absorption in 
strong \mgtwo\ absorbers at $0.58 < z_{\rm abs} < 1.68$. We compute the 
\hi~21cm detection rate in DLAs at these redshifts by taking the ratio of the 
detection rate in our sample to the detection rate of DLAs in \mgtwo$\lambda$2796 
absorbers selected with the same criteria (RTN06).\footnote{This implicitly assumes that 
all systems with detected \hi~21cm absorption are DLAs, i.e. have $\nhi \ge 2 \times 
10^{20}$~\cm.}  35~of the 36~absorbers of our 
sample also have detected \fetwo$\lambda$2600 absorption, with $\wfetwo \ge 0.34 \AA$; 
we will use this 35-system sub-sample for the comparison with the DLAs of RTN06, 
applying the same selection criteria to their absorbers (i.e. $\wmgtwo \ge 0.57\AA$, 
$\wfetwo \ge 0.34 \AA$ and $0.58 \le z_{\rm abs} \le 1.68$). There are 97~absorbers 
in the RTN06 sample that satisfy these selection criteria, of which 34~were found to be DLAs. 

The probability that a strong \mgtwo\ absorber is a DLA has been found to depend on 
the rest equivalent width of the \mgtwo\ line, being higher for systems with higher values 
of $\wmgtwo$ (RTN06). We hence first tested whether our sub-sample of 35~absorbers with 
$\wmgtwo \ge 0.57 \AA$ and $\wfetwo \ge 0.34 \AA$ has a similar distribution of $\wmgtwo$ 
values as the corresponding sub-sample of RTN06.  A Kolmogorov-Smirnov rank-1 test finds 
a Gaussian probability of $\sim 24.5$\% that the two samples are drawn from the same distribution 
of $\wmgtwo$ values, consistent with the null hypothesis within $\sim 1.2\sigma$ 
significance. We will hence assume that the probability of detecting a DLA is the same 
in the two sub-samples.

Of the 35~systems of our sample with $\wmgtwo \ge 0.5 \AA$ and $\wfetwo \ge 0.34 \AA$, 
there are nine detections of \hi~21cm absorption, i.e. a detection rate 
of $x_{\rm 21,MgII} = 26^{+12}_{-8}$\% at a mean absorption redshift ${\bar z} = 1.1$. 
34~DLAs are present amongst the 97~absorbers in the RTN06 sample that satisfy our selection criteria;
 this yields a detection rate of $x_{\rm DLA,MgII} \sim 35^{+7}_{-6}$\% for
DLAs in strong \mgtwo\ systems, also at a mean redshift of ${\bar z} \sim 1.1$. The implied 
detection rate of \hi~21cm absorption in DLAs in the redshift desert is then 
$x_{\rm 21,DLA}({\bar z} = 1.1) = x_{\rm 21,MgII}/x_{\rm DLA,MgII} \sim (73 \pm 27)$\%,
where we have used equation~(26) of \citet{gehrels86} to determine the $1\sigma$ Gaussian 
confidence level intervals for the ratio of two quantities governed by small-number Poisson 
statistics. 

Fig.~\ref{fig:det21} plots the detection rate of \hi~21cm absorption in DLAs as a function 
of redshift. While the errors are large due to the small size of each sub-sample (especially
for the present \mgtwo\ sample), the detection rates increase with decreasing redshift, 
as would be expected if the cold gas fraction in galaxies is increasing with time. The 
detection rate measured in the present survey is consistent with that found in low-$z$ DLAs, 
suggesting that significant fractions of cold \hi\ have already been formed in gas-rich 
galaxies by $z \sim 1$.

The fraction of DLAs in the redshift desert that are undetected in \hi~21cm absorption is 
$1 - x_{\rm 21,DLA}(\bar z = 1.1) = (27 \pm 27)$\%; all of these non-detections must have
$[\ts/f] > 800$~K, the $3\sigma$ sensitivity limit of our survey. This implies that at least
one-fourth of the DLAs in the redshift desert have high spin temperatures and/or low covering 
factors; this is a lower limit because some of the DLAs detected in \hi~21cm absorption are 
also likely to have high $[\ts/f]$ ratios (see Section~\ref{sec:detect}). 

Fig.~\ref{fig:vel21} shows the total velocity spread (FWBN) of \hi~21cm absorption 
$\Delta V_{\rm 21}$ plotted against redshift, for the nine absorbers of this sample 
and the 17~DLAs with detected \hi~21cm lines. Only two systems (both at $z_{\rm abs} 
< 0.7$) have $\Delta V_{\rm 21} > 150$~\kms, while $\sim 60$\% of the absorbers 
in each redshift range have $\Delta V_{\rm 21} \lesssim 60$~\kms. Overall, the 
velocity spread of \hi~21cm absorption displays no significant evolution with redshift.

Next, no statistically-significant correlation was found in the previous section 
between the integrated \hi~21cm optical depth $EW_{\rm 21}$ and $\wmgtwo$, $\wfetwo$, 
or $\wmgone$, for the strong \mgtwo\ absorbers of the full statistical sample. However, 
Figs.~\ref{fig:ew21}[A-C] show that a trend may be present between $EW_{\rm 21}$ and the 
above rest equivalent widths for \hi~21cm detections alone. We hence explored the possibility of 
a relation between $EW_{\rm 21}$ and $\wmgtwo$ in high column density absorbers, considering 
all DLAs and \hi~21cm absorbers in the literature with observations of both the \hi~21cm 
and the \mgtwo$\lambda$2796 transitions.  We again excluded ``associated'' absorbers from 
this analysis, as well as four DLAs with \hi~21cm studies where the radio and optical 
sightlines are known to be different (the $z_{\rm abs} \sim 0.437$ DLA towards 3C196, the 
$z_{\rm abs} \sim 0.656$ DLA towards 3C336, and the $z_{\rm abs} \sim 1.4$ absorbers
towards QSO~0957+561A and PKS~1354+258; \citealp{briggs01,curran07a,kanekar03}).
Fig.~\ref{fig:mgtwo} plots $EW_{\rm 21}$ versus $\wmgtwo$ for the 24~systems of this sample.
The non-parametric generalized Kendall rank correlation statistic \citep{brown74,isobe86}, 
as implemented in the {\sc ASURV} package (the BHK test), was used to test for a 
correlation between the two quantities, treating $\wmgtwo$ as the independent variable. 
We find the putative correlation to have $\sim 2.2\sigma$ statistical significance, with 
a probability of chance occurrence of $\sim 3$\%. Restricting the sample to detections 
of \hi~21cm absorption (21~systems) increases the significance to $\sim 2.5\sigma$. 
This is, at most, weak evidence for a relation between the integrated \hi~21cm optical depth 
and the rest equivalent width of the \mgtwo$\lambda$2796 transition.

Finally, \citet{curran07c} suggest that the velocity spread of \hi~21cm absorption and 
$\wmgtwo$ are correlated, based on a sample of 13~redshifted \hi~21cm absorbers. 
Fig.~\ref{fig:vel21-mgII} plots the total velocity spread (FWBN) of \hi~21cm absorption 
$\Delta V_{\rm 21}$ against $\wmgtwo$, for all absorbers (23~systems) with detections 
of both \hi~21cm and \mgtwo$\lambda$2796 absorption (again excluding ``associated'' systems, 
but including the \hi~21cm detections towards 3C196 and 3C336). No clear trend 
can be seen between $\Delta V_{\rm 21}$ and $\wmgtwo$; the BHK test finds the possible 
correlation between these quantities to have only $\sim 1.7\sigma$ significance, 
with a probability of chance occurrence of $\sim 11$\%. Excluding the two absorbers 
towards 3C196 and 3C336 does not noticeably affect the result; the correlation then has 
$\sim 1.6\sigma$ significance.  We thus find no statistically-significant evidence 
for a correlation between the velocity spread of \hi~21cm absorption and the rest 
\mgtwo$\lambda$2796 equivalent width.

\section{The cosmological mass density of neutral gas }
\label{sec:omega21}

\begin{figure}
\begin{centering}
\epsfig{file=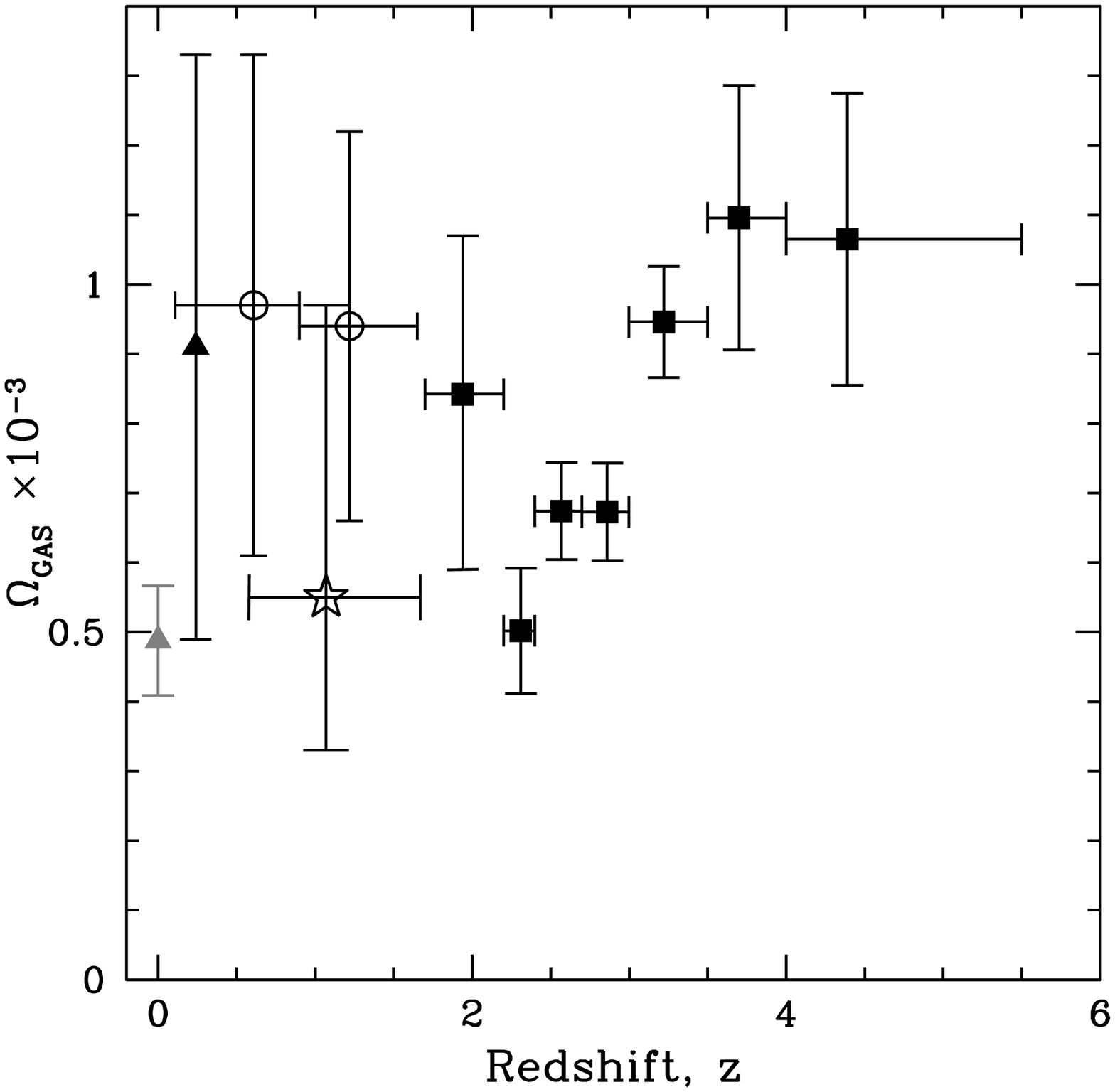,height=3.3truein,width=3.3truein}
\caption{The cosmological mass density of neutral gas $\Omega_{\rm GAS}$ plotted 
as a function of redshift, from the following techniques: (1)~the {\sc HIPASS} \hi~21cm
emission survey (grey triangle; \citealt{zwaan05b}), (2)~a GMRT measurement at 
$z \sim 0.24$, obtained by co-adding \hi~21cm emission spectra of field galaxies 
(filled black triangle; \citealt{lah07}), (3)~a survey for DLAs in strong \mgtwo\ 
absorbers (open circles; RTN06), (4)~this work (open star), and (5)~traditional ``blind'' 
DLA surveys, primarily based on the SDSS-DR5 (filled squares; \citealt{prochaska08c}). 
See the main text for discussion.}
\label{fig:omega}
\end{centering}
\end{figure}

The cosmological mass density of neutral gas in DLAs, $\Omega_{\rm GAS}(z)$, is related 
to their line density $n_{\rm DLA}(z)$ (i.e. the number of DLAs per unit redshift, 
also sometimes referred to as the redshift number density) by the following expression 
(e.g. RTN06):
\begin{equation}
\Omega_{\rm GAS}(z) = \frac{\mu m_{\rm H} H_0}{c \rho_c} n_{\rm DLA}(z) 
{\bar N}_{\rm HI} \frac{E(z)}{(1+z)^2} \;,
\label{eqn:omega}
\end{equation}
where $\rho_c \equiv {3H_0^2/8\pi G}$ is the critical mass density of the Universe, 
$\mu = 1.3$ is the correction factor for the 25\% contribution of He{\sc ii} to the 
neutral gas mass, ${\bar N}_{\rm HI}$ is the average \hi\ column density of the DLAs, 
$m_{\rm H}$ is the mass of the hydrogen atom and $E(z)$ is given by 
\begin{equation}
E(z) = \left[\Omega_m (1+z)^3 + (1 - \Omega_m - \Omega_\Lambda)(1+z)^2 + 
\Omega_\Lambda \right]^{1/2} \;.
\end{equation}

The present survey has measured the detection rate of \hi~21cm absorption in a sample of 
strong \mgtwo\ absorbers, finding $x_{\rm 21,MgII} = 25^{+11}_{-8}$\%, at the mean absorption 
redshift ${\bar z} = 1.1$. The fraction of DLAs in the \mgtwo\ absorber sample is then 
\begin{equation}
x_{\rm DLA,MgII}(z) = \frac{x_{\rm 21,MgII}(z)}{x_{\rm 21,DLA}(z)} \;\; ,
\end{equation}
where $x_{\rm 21,DLA}(z)$ is the detection rate of \hi~21cm absorption in DLAs. In turn, the 
line density of DLAs is related to the above DLA fraction by the expression
\begin{equation}
n_{\rm DLA}(z) = x_{\rm DLA,MgII}(z) \times n_{\rm MgII}(z, W_0) 
\label{eqn:ndla}
\end{equation}
where $n_{\rm MgII}(z,W_0)$ gives the number of \mgtwo\ absorbers per unit redshift 
with $\wmgtwo \ge W_0$; this has been determined using large samples of strong 
\mgtwo\ absorbers \citep{nestor05}. In other words, an estimate of the detection 
rate of \hi~21cm absorption in DLAs in the redshift desert would allow us to infer 
$\Omega_{\rm GAS}$ from our survey statistics [see also \citet{lane00b} for an estimate 
of the cosmological mass density of neutral hydrogen in \hi~21cm absorbers, based on 
an \hi~21cm absorption survey of \mgtwo\ absorbers].

Unfortunately, we do not have an {\it independent} estimate of $x_{\rm 21,DLA}(z)$ in the redshift
desert, due to the paucity of earlier \hi~21cm absorption studies here. Note that we cannot 
use the value of $x_{\rm 21,DLA}(z)$ derived in Section~\ref{sec:n21} as this already includes 
the detection rate of \hi~21cm absorption in strong \mgtwo\ absorbers. However, deep searches 
for \hi~21cm absorption (at comparable sensitivity to ours; e.g. \citealt{kanekar03}) have so 
far been carried out in 33 ``non-associated'' DLAs (at $0.09 < z_{\rm abs} < 3.4$), with 
17~detections, yielding a net \hi~21cm detection rate in DLAs of $x_{\rm 21,DLA} ({\bar z} = 1.7) 
= 52^{+16}_{-12}$\%. Assuming that $x_{\rm 21,DLA}({\bar z} = 1.1) 
\approx x_{\rm 21,DLA}(\bar z = 1.7)$, we obtain
\begin{equation}
x_{\rm DLA,MgII}(\bar z = 1.1) = 0.49^{+0.31}_{-0.19} \;.
\end{equation}

Next, following RTN06 [see their equations~(1--3)], we will use the parameterization 
of \citet{nestor05} to write
\begin{equation}
n_{\rm MgII}({\bar z}) = (1/36) \times \Sigma_i w_i [dn/dz]_i (z) 
\label{eqn:nmg2}
\end{equation}
where, for the $i$th absorber
\begin{equation}
[dn/dz]_i = N^* (1 + z)^\alpha e^{-(W_0/W^*)(1+z)^{-\beta}}  {\rm .}
\label{eqn:nmg2b}
\end{equation}
Here, $N^* = (1.001 \pm 0.132)$, $W^* = (0.443 \pm 0.032) \AA$, $\alpha = (0.226 \pm 0.170)$,
and $\beta = (0.634 \pm 0.097)$ are constants (see the appendix of \citealt{nestor05}) and
$w_i$ is the weight given to the $i$th absorber, based on the fractional abundance of
its class in the \mgtwo\ sample of \citet{nestor05}. All absorbers of our sample have 
$\wmgtwo \ge 0.6 \AA$, within the errors.\footnote{Note that one of our absorbers 
has $\wmgtwo = 0.57 \AA$, although consistent with $0.6 \AA$ within the errors. 
We estimate that using a threshold width of $\wmgtwo = 0.57 \AA$ in 
equation~\ref{eqn:nmg2b} would change our results by $< 5$\%, much below our actual 
errors. We hence choose to retain a threshold width of $\wmgtwo = 0.6 \AA$, for 
which the fraction of systems with $\wfetwo \ge 0.5 \AA$ is quoted by RTN06.} 
31~absorbers of our sample were selected with the additional criterion 
$\wfetwo \ge 0.5 \AA$; for these systems, $w_i = 0.54$, as 54\% of absorbers 
in the \mgtwo\ sample of \citet{nestor05} with $\wmgtwo \ge 0.6 \AA$ also have 
$\wfetwo \ge 0.5 \AA$ (RTN06). We also assume here that the fraction of strong 
\mgtwo\ absorbers that are also strong \fetwo\ absorbers does not depend on redshift.  
Conversely, $w_i = 1$ for the remaining five systems, which were selected on 
the basis of their \mgtwo\ rest equivalent width. Equation~\ref{eqn:nmg2} then yields, 
for ${\bar z} = 1.1$ and $W_0 = 0.6 \AA$,
\begin{equation}
n_{\rm MgII}(\bar z = 1.1) = (0.2833 \pm 0.0065)   \; .
\end{equation}

Replacing for $n_{\rm MgII}(\bar z = 1.1)$ and $x_{\rm DLA,MgII}(\bar z = 1.1)$ in 
equation~\ref{eqn:omega} gives a DLA line redshift density of $n_{\rm DLA}(\bar z = 1.1) 
= 0.137^{+0.088}_{-0.054}$.\footnote{For all cases of error propagation with asymmetric 
errors, we used $10^5$ Monte-Carlo runs to derive the $1\sigma$ confidence 
intervals on the final quantity.} The only remaining unknown in equation~\ref{eqn:omega} 
is ${\bar N}_{\rm HI}$, the average \hi\ column density in the nine \hi~21cm detections of 
the sample. Prior to this work, all redshifted \hi~21cm absorbers with observations of the 
Lyman-$\alpha$ transition have been found to be DLAs; it is thus likely that the \hi~21cm 
absorbers detected in this survey are also DLAs. If so, it is plausible that their 
average \hi\ column density is the same as that obtained by RTN06 for the DLAs detected in their 
\mgtwo\ survey. With this assumption, ${\bar N}_{\rm HI} = (1.07 \pm 0.23) \times 10^{21}$~\cm, where 
we have used the result of RTN06 from the redshift range $0.9 < z < 1.65$, where all of our \hi~21cm 
absorbers lie. We then obtain $\Omega_{\rm GAS}(\bar z = 1.1) = (0.55^{+0.42}_{-0.22}) \times 10^{-3}$.

Fig.~\ref{fig:omega} shows a comparison between estimates of the cosmological mass density of 
neutral gas $\Omega_{\rm GAS}(z)$, from the following techniques (in order of increasing redshift):
(1)~the {\sc HIPASS} survey for local \hi~21cm emission (grey triangle; \citealt{zwaan05b}), 
(2)~a GMRT survey for \hi~21cm emission at $z \sim 0.24$, obtained by co-adding spectra of
121~field galaxies (filled black triangle; \citealt{lah07}), (3)~a survey for DLAs in 
strong \mgtwo\ absorbers (open circles; RTN06), (4)~a survey for \hi~21cm absorbers 
in  strong \mgtwo\ absorbers (open star; this work), and (5)~traditional ``blind'' DLA surveys,
primarily based on the SDSS-DR3 (filled squares; \citealt{prochaska05}). The two measurements 
with the smallest errors, the HIPASS result at $z \sim 0$ \citep{zwaan05b} and the lowest-redshift 
SDSS-DR3 measurement at $z \sim 2.36$ \citep{prochaska05}, are in excellent agreement, suggesting a 
scenario wherein the conversion of gas to stars at $z \lesssim 2.3$ is balanced by other processes 
that replenish the 
neutral gas reservoir in galaxies. Curiously enough, however, earlier estimates of $\Omega_{\rm GAS}$ 
at $ 0.2 < z < 2$ (from multiple techniques, the GMRT observations, the RTN06 survey and a single 
non-SDSS DLA measurement) are all higher than the estimates at $z \sim 0$ and $z \sim 2.3$, although 
individually consistent with these measurements due to their far larger errors. While the estimate from 
the present survey, $\Omega_{\rm GAS} (\bar z = 1.1) = (0.55^{+0.42}_{-0.22}) \times 10^{-3}$ is in good
agreement with the two high-sensitivity results, the error bars on our result are also large 
and we cannot rule out the values obtained by \citet{rao06}.

We emphasize that the present result for $\Omega_{\rm GAS}$ assumes that the average \hi\ column 
density of our \hi~21cm absorbers is the same as that in the DLA sample of RTN06. Direct measurements 
of the \hi\ column density of these absorbers through spectroscopy in the Lyman-$\alpha$ 
line would allow a measurement of ${\bar N}_{\rm HI}$ for the sample, obviating the need for 
this assumption. Equally important, we have assumed $x_{\rm 21,DLA}({\bar z} = 1.1) 
\approx x_{\rm 21,DLA}(\bar z = 1.7)$. As seen in Section~\ref{sec:n21}, the detection 
rate of \hi~21cm absorption in DLAs appears to increase with decreasing redshift, 
implying that $x_{\rm 21,DLA}({\bar z} = 1.1)$ is likely to be higher than 
$x_{\rm 21,DLA}(\bar z = 1.7)$. In fact, the estimated average value 
$x_{\rm 21,DLA}(\bar z = 1.7) = (52^{+16}_{-12})$\% from the DLA sample is lower 
than the value $x_{\rm 21,DLA}({\bar z} = 1.1) = (73 \pm 27)$\% obtained 
in Section~\ref{sec:n21}. In other words, this assumption could result in 
an {\it upward} bias in our estimate of $\Omega_{\rm GAS}$.

Unlike direct surveys for DLAs in \mgtwo\ absorber samples (e.g. \citealp{rao00}; RTN06), 
observing time on space-based facilities is only needed for the last stage of the present 
survey, the measurements of the \hi\ column density for detections of \hi~21cm absorption. 
Deep \hi~21cm absorption surveys of large samples of \mgtwo-selected absorbers towards 
radio-loud quasars thus provide an interesting avenue to find new samples of DLAs, and 
determine $\Omega_{\rm GAS}$, at $z < 1.7$, with relatively little expenditure of observing 
time on space telescopes. 

\section{Summary}
\label{sec:summary}

We report results from a search for \hi~21cm absorption in 55 strong \mgtwo\ absorbers 
at $0.58 \lesssim z_{\rm abs} \lesssim 1.70$ towards radio-loud quasars, with 
$\wmgtwo \ge 0.5 \AA$. The survey yielded nine detections of \hi~21cm absorption 
(one of which is still tentative), all at $1.17 < z_{\rm abs} < 1.68$, and 27~strong
constraints on the \hi~21cm optical depth ($\tau_{3\sigma} \lesssim 0.013$, per $\sim 
10$~\kms). Six other systems had weaker constraints on the \hi~21cm optical depth 
($\tau_{3\sigma} \sim 0.02 - 0.06$, per $\sim 10$~\kms), while data on 13~targets 
were affected by RFI and were hence not useful. 

\hi\ column density estimates are available in the literature for a few of the \mgtwo\ 
absorbers of our sample, two of which, at $z \sim 0.9115$ towards 2149+212 
and $z \sim 1.4106$ towards 2003$-$025, are known damped Lyman-$\alpha$ systems.
Our detection of \hi~21cm absorption towards 2003$-$025 yields an absorber spin 
temperature of $\ts = (905 \pm 380) \times f$~K, while the non-detection towards
2149+212 gives the $3\sigma$ limit $\ts > 2700 \times f$~K. 

Including three systems from the literature and excluding all ``associated'' absorbers, 
our statistical sample of strong \mgtwo\ absorbers consists of 36~systems at 
$0.58 < z_{\rm abs} < 1.68$, with $\wmgtwo \ge 0.57 \AA$, and with either 
detections of \hi~21cm absorption or, for non-detections, $\tau_{\rm 21} 
\lesssim 0.013$ per $\sim 10$~\kms. This implies a detection rate of 
$x_{\rm 21,MgII}({\bar z} = 1.1) = 25^{+11}_{-8}$\% for \hi~21cm absorption in strong \mgtwo\ 
absorbers. Comparing the detection rates of \hi~21cm absorption and DLAs in similarly-selected 
samples of strong \mgtwo\ absorbers, we estimate that the detection rate of \hi~21cm absorption 
in DLAs in the redshift desert is $x_{\rm 21}({\bar z = 1.1}) \sim (73 \pm 27)$\%. 
This lies between the estimates of detection rates at lower and higher redshifts, 
$x_{\rm 21}({\bar z \sim 0.4}) \sim 85^{+15}_{-28}$~\% and 
$x_{\rm 21}({\bar z \sim 2.7}) \sim 33^{+20}_{-13}$~\%, 
although all three estimates have large errors. If the relatively-high detection rate 
of \hi~21cm absorption in DLAs in the redshift desert is confirmed in larger samples, it 
would imply that significant fractions of cold \hi\ are present in normal galaxies by 
$z \sim 1$. The $\sim (27 \pm 27)$\% of DLAs that are not detected in \hi~21cm absorption 
must have $[\ts/f] > 800$~K, implying that at least one-fourth of the DLAs in the redshift 
desert have high spin temperatures and/or low covering factors. 

Contrary to earlier results based on smaller samples, we find no statistically-significant
evidence for a correlation between the total velocity spread of \hi~21cm absorption
and the rest \mgtwo$\lambda$2796 equivalent width in DLAs and strong \mgtwo\ absorbers.
We find a weak ($2.2\sigma$) correlation between the integrated \hi~21cm optical depth 
and $\wmgtwo$, for a sample consisting of DLAs and \hi~21cm detections alone.


We have used the net \hi~21cm detection rate in strong \mgtwo\ absorbers to estimate 
the cosmological mass density in neutral gas in DLAs, obtaining 
$\Omega_{\rm GAS}({\bar z} = 1.1) = (0.55^{+0.42}_{-0.22}) \times 10^{-3}$. This assumes 
that (1)~the average \hi\ column density of the \hi~21cm absorbers is the same 
as that in the DLA sample of RTN06, and (2)~the detection rate of \hi~21cm absorption 
in DLAs at $z \sim 1.1$ is the same as that at $z \sim 1.7$. While this estimate 
has large errors, it is in good agreement with estimates of $\Omega_{\rm GAS}$ in 
both the local universe and at $z \sim 2.2$, and somewhat lower than earlier estimates in 
the redshift desert from \mgtwo\ absorber samples. Future deep searches for \hi~21cm 
absorption in large samples of strong \mgtwo\ absorbers should allow more accurate 
estimates of $\Omega_{\rm GAS}$ in the redshift desert, without significant requirements 
of observing time on space telescopes.

For the nine detections of \hi~21cm absorption, spectroscopy in the Lyman-$\alpha$ 
line to measure their \hi\ column densities, and low-frequency VLBI continuum 
observations of the background QSOs to determine the absorber covering factors, will 
allow estimates of the absorber spin temperatures. For systems with narrow \hi~21cm profiles 
(e.g. 2337$-$011, 2355$-$106 and 0105$-$008), high-resolution optical spectroscopy 
in the low-ionization metal lines will allow a probe of evolution in the fundamental 
constants. These observations are now in progress.

\section{Acknowledgments}

We thank Carl Bignell, Bob Garwood, Toney Minter, Frank Ghigo and Glen Langston 
for much help with the GBT observations and data analysis, and Ishwara-Chandra 
and Nimisha Kantharia for help with the scheduling of the GMRT observations. 
We thank the staff of the GMRT who made these observations possible; the GMRT 
is run by the National Centre for Radio Astrophysics of the Tata Institute of 
Fundamental Research. The National Radio 
Astronomy Observatory is operated by Associated Universities, Inc, under cooperative 
agreement with the National Science Foundation. This research has made use of the 
NASA/IPAC Extragalactic Database (NED) which is operated by the Jet Propulsion Laboratory, 
California Institute of Technology, under contract with the National Aeronautics and Space 
Administration. This research has also made use of the SAO/NASA Astrophysics Data System. 
NK acknowledges support from the Max-Planck Foundation and an NRAO Jansky Fellowship.
We also thank an anonymous referee for useful comments on an earlier draft of this
manuscript.

\bibliographystyle{mn2e}
\bibliography{ms}

\end{document}